\documentclass[12pt]{article}
\usepackage{helvet}
\usepackage{multicol}

\usepackage[usenames]{color}
\usepackage{url}
\usepackage{hyperref}

\usepackage{graphicx}
\usepackage{url}
\usepackage[font=small]{caption}    
\newcommand{\etal}{et al.}   

\newcommand{\blue}[1]{\textcolor{blue}{#1}}

\newcommand{\benum}{\begin{enumerate}}
\newcommand{\eenum}{\end{enumerate}}
\newcommand{\bitem}{\begin{itemize}}
\newcommand{\eitem}{\end{itemize}}
\newcommand{\bCitem}
	   {\begin{list}
	    {$\bullet$}
	    {\itemsep 0.01in \parsep -.01in \topsep .08in}}
\newcommand{\eCitem}{\end{list}}

\newcommand{\Ssection}[1]{\subsection{#1}}

\newcommand{\SSsection}[1]{\subsubsection{#1}}

\newlength{\oldbaselineskip}

\newcommand{\dictentry}[1]{\vspace{1 cm}}

\def\scaledpicture #1 by #2 (#3) indent #4 scale #5 {}

\newcommand{\coli}{{\em E. coli}}

\evensidemargin \oddsidemargin 
\usepackage{longtable}

\newcommand{\nLinesCodePTools}{867,000}		
\newcommand{\nPToolsLicensees}{15,567}		
\newcommand{\nLinesJavascriptPTools}{84,000}   
\newcommand{\NbiocycPGDBs}{20,000}             
\newcommand{\NworldWidePGDBs}{45,000}           

\begin{document}

\title{Pathway Tools version 28.0: Integrated Software for Pathway/Genome Informatics and Systems Biology}

\author{
{\bf Peter D. Karp$^{*}$, Suzanne M. Paley, Markus Krummenacker, } \\
{\bf Anamika Kothari, Peter E. Midford, Pallavi Subhraveti, } \\
{\bf Austin Swart, Lisa Moore, and Ron Caspi}\\
Bioinformatics Research Group, SRI International\\
333 Ravenswood Ave, Menlo Park, CA 94025 \\
pkarp@ai.sri.com \\
\ \\
$^{*}$To whom correspondence should be addressed.\\
}

\date{}

\maketitle

\newcommand{\biovelo}{{\sf BioVelo}}
\newcommand{\metacyc}{{\sf MetaCyc}}

\newpage

\section*{Abstract}

Pathway Tools is a bioinformatics software environment with a broad
set of capabilities.  The software provides genome-informatics tools
such as a genome browser, sequence alignments, a genome-variant
analyzer, and comparative-genomics operations.  It offers
metabolic-informatics tools, such as metabolic reconstruction,
quantitative metabolic modeling, prediction of reaction atom mappings,
and metabolic route search.  Pathway Tools also provides
regulatory-informatics tools, such as the ability to represent and
visualize a wide range of regulatory interactions.  The software
creates and manages a type of organism-specific database called a
Pathway/Genome Database (PGDB), which the software enables database
curators to interactively edit.  It supports web publishing of PGDBs
and provides a large number of query, visualization, and omics-data
analysis tools.  Scientists around the world have created more than
\NworldWidePGDBs{} PGDBs by using Pathway Tools, many of which are curated databases
for important model organisms.  Those PGDBs can be exchanged using a
peer-to-peer database-sharing system called the PGDB Registry.

\vspace{1 in}

\section*{Biographical Note}

Dr. Peter Karp is the Director of the Bioinformatics Research Group at
SRI International.  He received the PhD degree in Computer Science
from Stanford University.

\newpage

\section{Introduction}

Pathway Tools \cite{PTools10,PToolsOverview06,PTools02,EcoCycJCB96} is a
software environment for management, analysis, simulation, and visualization of
integrated collections of genome, pathway, and regulatory
data.  This article provides an overview of all current capabilities of
Pathway Tools; previous such overview publications were \cite{PTools10,PTools15Overview}.  
A shorter version of this article describes recent
enhancements to  Pathway Tools \cite{PTools19BiB}.
New software capabilities developed since the previous (2020) version of this
article are typeset in \blue{blue} in this article.\footnote{This article incorporates significant text from
\cite{PTools19Overview, PTools19BiB,PTools15BiB,PTools10} by permission of the publisher.  
}

Pathway Tools handles many types of information beyond pathways
and offers extensive capabilities.  The software has been under continuous
development within the Bioinformatics Research Group (BRG) within SRI
International since the early 1990s.  Pathway Tools serves several
different use cases in bioinformatics and systems biology:

\bitem

\item It supports development of organism-specific databases
(also called model-organism databases) that integrate many
bioinformatics data types.

\item It supports scientific visualization, web publishing, and dissemination of those
organism-specific databases.

\item It performs computational inferences from sequenced genomes including prediction of an
organism's metabolic network, prediction of metabolic pathway hole fillers,
and prediction of operons.

\item It enables creation of steady-state quantitative metabolic flux
models for individual organisms and for organism communities.

\item It provides tools for graph-based analysis of biological networks, such as
  for identification of metabolic choke points, dead-end metabolites,
  and blocked reactions.

\item It provides tools for analysis of gene expression, metabolomics,
proteomics, and multi-omics data sets.

\item It provides comparative analyses of organism-specific databases.

\item It supports metabolic engineering.

\eitem

Pathway Tools is focused around a type of model-organism database
called a Pathway/Genome Database (PGDB).  A PGDB integrates
information about an organism's genes, proteins, metabolic network, and 
regulatory network.

Pathway Tools has several major components.  The {\bf PathoLogic} component
enables users to create a new PGDB from the annotated genome of an
organism, containing the genes, proteins, biochemical reactions,
and predicted metabolic pathways and operons of the organism.  

The {\bf Pathway/Genome Editors} let PGDB developers interactively
refine the contents of a PGDB, such as editing a metabolic pathway or
an operon, or defining the function of a newly characterized gene.

The {\bf Pathway/Genome Navigator} supports querying, visualization,
and analysis of PGDBs.  Whereas all other Pathway Tools components run
as desktop applications only, the Navigator can run as both a desktop
application and as a web server.  The Navigator enables scientists to
quickly find information, to display that information in familiar
graphical forms, and to publish a PGDB to the scientific community via
the web.  The Navigator provides a platform for systems-level analysis
of high-throughput data by providing tools for painting
combinations of gene expression, protein expression, and metabolomics
data onto a full metabolic map of the cell, onto the full genome, and
onto a diagram of the regulatory network of the cell.

The {\bf MetaFlux} component enables construction and execution of
steady-state metabolic flux models from PGDBs.  MetaFlux has modes to
accelerate development of metabolic models, and to use metabolic
models to simulate both gene and reaction knockouts.  Pathway Tools
provides a unique environment for metabolic flux modeling: by
combining a tool for reconstructing metabolic networks from genome
annotations with metabolic-model debugging tools such as a reaction
gap-filler, the software enables rapid development of metabolic models
from sequenced genomes.  And by tightly coupling the metabolic model
with other enriching information such as the sequenced genome,
chemical structures, and regulatory information, Pathway Tools-based
metabolic models are easier to understand, validate, reuse,
extend, and learn from.

\blue{These components contain more than 60 individual tools that are
summarized in Figure~\ref{fig:tools}.}

Pathway Tools includes a sophisticated ontology and database
application programming interface (API) that enables programs to
perform complex queries, symbolic computations, and data mining on the
contents of a PGDB.  For example, the software has been used for
global studies of the \coli\ metabolic network
\cite{KarpVcell00} and genetic network \cite{KarpSymTheory01}.

Pathway Tools is seeing widespread use across the bioinformatics
community to create Pathway/Genome Databases in all domains of
life. More than
\nPToolsLicensees\ groups to date have licensed the software.  As well as supporting the
development of the EcoCyc \cite{EcoCycNAR13} and MetaCyc
\cite{MetaCycNAR14} databases (DBs) at SRI, and SRI's BioCyc
collection of \NbiocycPGDBs\ PGDBs \cite{BioCyc17}, the software is used
by genome centers, experimental biologists, and groups that are
creating curated DBs for a number of different
organisms (see Section~\ref{sec:pgdbs} for a more detailed listing of
available PGDBs).

This article provides a comprehensive description of Pathway Tools.
Where possible, it references earlier publications that provide more algorithmic
details.  However, in some cases, those earlier publications are
outdated by new developments in the software that are described here.
This article also emphasizes new aspects of the software that have not
been reported in earlier publications.

The organization of this article is as follows.  
\bitem
\item Section~\ref{sec:use-cases}: Use cases for which Pathway Tools was designed

\item Section~\ref{sec:creating}:  PGDB creation, Pathway Tools
  computational inferences,  interactive editing, associated author-crediting system, tools for automatic
upgrading of a PGDB schema and for bulk updating of a PGDB genome annotation

\item Section~\ref{sec:schema}: PGDB schema and ontologies

\item Section~\ref{sec:nav}: Search and visualization facilities

\item Section~\ref{sec:omics}: Omics data analysis

\item Section~\ref{sec:apis}: Mechanisms for importing and
exporting data from Pathway Tools, and for accessing and updating
PGDB data via  APIs

\item Section~\ref{sec:analyses}  Pathway Tools modules for network analyses
of PGDBs, such as for identifying dead-end metabolites

\item Section~\ref{sec:comparative}: Comparative analysis tools

\item Section~\ref{sec:arch}: Software architecture of Pathway Tools

\item Section~\ref{sec:metaflux}: Metabolic modeling capabilities

\item Section~\ref{sec:pgdbs} PGDBs that have been
created by Pathway Tools users outside SRI International; describes
a peer-to-peer data-sharing facility within Pathway Tools that enables
users to easily exchange PGDBs

\item Section~\ref{sec:related}: Compares Pathway Tools to related efforts
\eitem

Supplemental Figures are available at \cite{PT24SuppFigs}.

\begin{figure}
\begin{center}

\begin{multicols}{2}
{\small

{\bf \noindent Genome Informatics}\\
\vspace{-.07in}
\bCitem
{\footnotesize
\item  Gene/protein/RNA/site searches
\item  Gene and genome site information pages
\item  Genome browser
\item  BLAST search, sequence pattern search
\item  Multiple sequence alignment
\item  Sequence retrieval
}
\eCitem

\vspace{.1in}
{\bf \noindent Pathway Informatics}
\vspace{-.07in}
\bCitem
{\footnotesize
\item Reaction/metabolite/pathway searches 
\item Reaction/metabolite/pathway pages
\item Single and multi-pathway diagrams
\item Zoomable metabolic network diagrams
\item Pathway inference from annotated genome
\item Flux-balance analysis, dFBA, FVA
\item Gap filling, blocked metabolites, chokepoints
\item Find metabolic routes to goal metabolite
\item Metabolic network explorer
}
\eCitem

\vspace{.1in}
{\bf {\noindent Regulatory Informatics}}
\bCitem
{\footnotesize
\item  Operon inference
\item  Operon information page includes promoters, TF binding sites
\item  Regulatory network viewer
\item  Visualize regulon of transcription factor
}
\eCitem

\vspace{.1in}
{\noindent \bf Transcriptomics Data Analysis}
\bCitem
{\footnotesize
\item Enrichment analysis
\item Paint transcriptomics data onto single pathways, zoomable
  metabolic chart
\item Omics Dashboard
\item Sort pathways by pathway activation score
}
\eCitem
\ \\

\ \\
\columnbreak

{\noindent \bf Metabolomics Data Analysis}
\bCitem
{\footnotesize
\item Enrichment analysis
\item Paint metabolomics data onto single pathways, zoomable
  metabolic chart
\item Omics Dashboard
\item Sort pathways by pathway activation score
\item Pathway covering analysis
}
\eCitem

\vspace{.1in}
{\noindent \bf Microbiome Informatics}
\bCitem
{\footnotesize
\item Calculate pathway abundances across metagenome samples
\item Create PGDBs for organisms in a community
\item Search across multiple PGDBs
\item Analysis of meta-omics data
}
\eCitem

\vspace{.1in}
{\noindent \bf Advanced Data Manipulation}
\bCitem
{\footnotesize
\item SmartTables
\item Interactive editing tools
\item Advanced Query interface
\item Comparative genomics and pathways
}
\eCitem

\vspace{.1in}
{\noindent \bf APIs and Data Import/Export}
\bCitem
{\footnotesize
\item APIs: Web services, Python, Java, Perl, Lisp
\item Formats: GFF, GenBank, BioPAX, SBML, tab-delimited files
}
\eCitem

}
\end{multicols}

\caption{The individual software tools within Pathway Tools.
}
\label{fig:tools}
\end{center}
\end{figure}

\section{Pathway Tools Use Cases}
\label{sec:use-cases}

This section articulates the objectives for which Pathway Tools was
designed.  Please note that when we assert that Pathway Tools supports
a given type of use case, it does not mean that Pathway Tools provides
every type of computational tool needed in that area.  For example,
omics data analysis is a huge field, and although Pathway Tools
contributes novel and useful omics data analysis capabilities, it does
not provide every omics data analysis method: in fact, it is
intended to be used in conjunction with other omics analysis tools
(such as for data normalization).  Section~\ref{sec:limitations}
summarizes the limitations of Pathway Tools.

\Ssection{Development of Organism-Specific Databases}

Organism-specific DBs (also known as model-organism DBs) describe the
genome and other information about an organism
\cite{EcoCycNAR17,SGD14,ApiDB07,WormBase14,MGD15}.
We posit that every organism with a completely sequenced genome and an
experimental community of significant size requires an
organism-specific DB to fully exploit the genome sequence.  Such DBs
should provide a central information resource that integrates
information dispersed through the scientific literature about the
genome, molecular parts, and cellular networks of the organism.  Such
DBs both direct and accelerate further scientific investigations.

Pathway Tools facilitates rapid initial computational construction of
organism-specific DBs, followed by manual refinement of the PGDB, to
produce an extremely rich and accurate DB in minimal time.  Our
approach differentiates experimental versus computationally inferred
information whenever possible.  Rapid construction of PGDBs is
achieved by importing an annotated genome into a PGDB in the form of a
GenBank or GFF file, and by applying several computational inference tools to
infer new information within the PGDB, such as metabolic pathways.
Scientists can then employ the Pathway/Genome Editors to correct and
supplement computational inferences when necessary, and to perform
ongoing manual curation of the PGDB if desired.  

The Pathway Tools DB schema (see
Section~\ref{sec:schema}) is significant in both its breadth and its
depth: it models an unusually broad set of bioinformatics data types
ranging from genomes to pathways to regulatory networks, and it
provides high-fidelity representations of those data types that allow
PGDBs to accurately capture complex biology.

\Ssection{Web Visualization and Querying of Organism-Specific Databases}
\label{sec:viz}
To speed user comprehension of the complex information within PGDBs,
the Pathway/Genome Navigator provides many scientific visualization
services, including a genome browser, visualization of single metabolic
pathways and entire metabolic maps, visualization of single operons
and of entire regulatory networks, and visualization of chemical
compounds and reactions (see Section~\ref{sec:nav} for more
details).

These visualizations are generated programmatically for each PGDB and
are also available on the desktop as well as the web.  Hence, all
developers of PGDBs get the same high-quality visualization tools
available for their PGDBs. Additionally, most of the graphics on the
web are rendered using a protocol we call ``web graphics''. A
browser-based engine renders ``web graphics'' using HTML5 canvas
technology, enabling high-resolution, zoom-able graphics on all
web-browsers. These improvements are only for rendering; our core,
rapid, and robust layout algorithms remain in place.

These visualization tools operate within a web server,
permitting developers of PGDBs to publish their PGDBs to the
scientific community through a website.  This form of PGDB publishing
supports interactive querying and browsing
using a three-tiered series of web query interfaces (see Section~\ref{sec:query}),
including a quick search, a set of object-specific query tools, and
a tool for interactively constructing queries whose power is
comparable to that of SQL.

We have developed other publishing paradigms to support computational
analysis and dissemination of PGDBs.  Pathway Tools APIs exist in
four languages \cite{PTools05} and as web services.  PGDBs can be exported in several formats and
imported into the BioWarehouse DB integration system
\cite{Biowarehouse06}.  Finally, users can easily share and exchange
PGDBs using a peer-to-peer DB-sharing system that we have
developed.

\Ssection{Extend Genome Annotations with Additional Computational Inferences}

Pathway Tools extends the paradigm of genome analysis.  After
traditional analyses such as gene calling and gene function prediction
are  performed by external software packages, Pathway Tools provides additional computational genome
analyses that layer additional information above the traditional
genome annotation.  Pathway Tools predicts the operons of the
organism.  It predicts the metabolic pathways of the organism.  It
also predicts which genes in the organism code for missing enzymes in
the predicted metabolic pathways, thus using pathway information to
predict additional gene functions.  See Section~\ref{sec:creating} for
more details.

\Ssection{Analysis of Omics Data}

Pathway Tools provides a variety of tools for visualization and
analysis of omics data sets, including the Omics Dashboard~\cite{DashGene17} and three
genome-scale viewers for animated visualization of omics data sets in
the context of the full metabolic network \cite{PToolsOverview06},
full transcriptional regulatory network, and full genome.  It also
provides enrichment analysis and SmartTable-based analysis of omics
data, as well as several pathway-based tools.  See
Section~\ref{sec:omics} for more details.

\Ssection{Quantitative Metabolic Flux Modeling}

The MetaFlux module of Pathway Tools supports development and execution
of steady-state metabolic flux models for individual organisms and
organism communities from PGDBs.  MetaFlux supports a
{\em literate modeling} approach that makes metabolic flux models
highly accessible to and understandable by scientists.

\Ssection{Analysis of Biological Networks}

Pathway Tools includes programs for symbolic analysis of biological
networks (see Section~\ref{sec:analyses} for more details) that rely
on the detailed biological network ontology underlying Pathway Tools.
The software identifies dead-end metabolites
and blocked reactions, both of which usually reflect errors or
incompleteness of our knowledge of a metabolic network.

Pathway Tools indirectly supports a two-phased, pathway-based paradigm for drug
discovery.  Phase I is the search for essential in vivo metabolic
pathways: pathways whose function is essential for microbial growth in
the host. Phase II is the search for targets within essential in vivo
pathways.  Both phases are supported by a Pathway Tools module that
predicts choke-point reactions within the metabolic network as
likely drug targets \cite{Yeh04}.

\Ssection{Comparative Analyses of Organism-Specific Databases}

Pathway Tools provides a suite of comparative analysis operations that
can be applied to multiple user-selected PGDBs (see
Section~\ref{sec:comparative} for more details).  Pathway Tools
emphasizes comparisons at the functional level, rather than the
sequence level.  Example comparisons include \blue{(1) the Comparative
Genome Dashboard, which aggregates metabolic, transport, and other functional
capabilities into a set of interactive charts,} (2) highlighting on the
Cellular Overview of one organism the reactions that it shares (or
does not share) with one or more other organisms; (3) a tabular
comparison of the reaction complements of several organisms,
organized by substrate type (e.g., small molecules, RNAs, proteins) or
by the number of isozymes per reaction; (4) a comparison of the pathway
complements of several organisms, where the tabular pathway comparison
is organized by a pathway ontology; (5) a table showing which genes have
orthologs in which PGDBs; and (6) a comparison of the genome organization of
orthologs using the genome browser.

\Ssection{Metabolic Engineering}

Metabolic engineering is a discipline that seeks to modify the
metabolic network of an organism in a desired fashion, such as to
achieve overproduction of desired end products, or degradation of
specified compounds \cite{Stephan98}.  Pathway Tools is designed to
assist metabolic engineers in several respects.  Its
metabolic-reconstruction capabilities aid in rapid characterization of
a host organism for metabolic engineering.  Its editing tools permit
refinement of that pathway database.  Its omics analysis capabilities
aid metabolic engineers in understanding the activity levels of
different portions of the metabolic network under different growth
conditions.  Its RouteSearch tool supports design of novel reaction
pathways from a feedstock compound to a desired product compound, and its
metabolic modeling capabilities enable computational exploration of
modified flux routes.

\section{Creating and Curating a PGDB}
\label{sec:creating}

The life cycle of a PGDB typically includes the following three 
procedures.

{\bf 1. Initial creation of the PGDB}  starts with
one or more input files describing the functionally annotated genome of an
organism.  The PathoLogic component of Pathway Tools transforms the
genome into an Ocelot \cite{Karp-JIIS-97a} DB structured according to the Pathway Tools
schema.  Next, the user applies one or more computational inference
tools within PathoLogic to the genome to infer new information such as
metabolic pathways.  For several of the PathoLogic inference tools, we
have created graphical user interfaces (GUIs) that enable
the user to review the inferences made by these tools, and to accept,
reject, or modify those inferences.

{\bf 2. PGDB curation.}  Manual refinement and updating of a PGDB is
performed using the Pathway/Genome Editors.  This phase can last for
years, or for decades, as in the case of EcoCyc \cite{EcoCycNAR17}.
Curation can be based on information found about the organism in the
experimental literature, information from in-house experiments, or
information inferred by the curator, perhaps with help from 
computational tools.  PGDB curation is multidimensional, involving addition and/or deletion of genes or
metabolic pathways to/from the PGDB; changing gene functions; altering
the structure of metabolic pathways; authoring of summary comments for
genes or pathways; attachment of Gene Ontology terms to
genes and gene products; entry of chemical structures for small
molecules; defining regulatory relationships; and entry of data
into many different PGDB fields including protein molecular weights,
pIs, and cellular locations.

{\bf 3.  Bulk updating of a PGDB.}  A PGDB developer might run an
external program that predicts cellular locations for hundreds of
genes, and want to load those predictions into the
PGDB.  Some groups store the
authoritative genome annotation for an organism in another genome-data management
system, and want to periodically import the latest genome annotation
into Pathway Tools; the software provides a tool for this operation.
In addition, some of the individual
components within PathoLogic that were used to initially create a PGDB
can be run again at a later date to take advantage of updated
information.

The following subsections describe these procedures in more detail.

\Ssection{PathoLogic PGDB Creation}

\begin{figure}
\begin{center}
\includegraphics[width=6in]{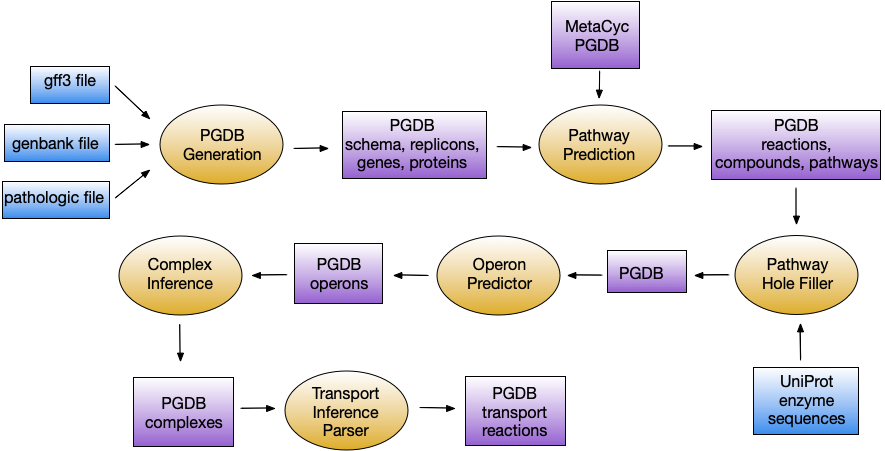}
\caption{Inputs and outputs of the computational inference modules within PathoLogic.
The initial input to PathoLogic is a file in GFF3, GenBank, or PathoLogic format.
Blue boxes describe input data files.  Yellow ellipses indicate
processing modules.
Purple boxes indicate where a PGDB is an input to or an output from some processing
step; the notations at the bottom of the purple boxes indicate what types of data have
been added by the previous processing step (for example, the Transport Inference Parser
adds transport reactions to a PGDB).
}
\label{fig:pathologic}
\end{center}
\end{figure}

PathoLogic performs a series of computational inferences that are
summarized in Figure~\ref{fig:pathologic}.  These inferences can be
performed in an interactive mode, in which the user guides the system
through each step, and can review and modify the inferences made by
the system using interactive tools.  PathoLogic can also execute in
a batch mode, in which all processing is automated, to process hundreds
or thousands of genomes.

The input to PathoLogic is
the annotated genome of an organism.  PathoLogic does not perform
genome annotation; its input must supply the genome sequence, the
locations of genes, and functions of gene products.  
The annotation is supplied as a set of files in GenBank format, GFF format, or PathoLogic format,
each of which describes the annotation of one replicon (chromosome or plasmid),
or of one contig for genomes that are not fully assembled.
When the annotation is provided in PathoLogic format, the sequence is
provided as one or more separate FASTA files.

The annotation specified in a GenBank or PathoLogic file can include
the start and stop positions of the coding region for each gene, and
intron positions.  It can also include a description of the function
of the gene product as a text string, one or more EC numbers, and one
or more Gene Ontology terms.  The annotation can also include a gene
name, synonyms for the gene name and the product name, links to other
bioinformatics databases, and comments.

PathoLogic initializes the schema of the new PGDB by copying from
MetaCyc into the new PGDB the definitions of the ontology
classes and the 350 slots (DB attributes) that define the schema of a
PGDB.  

PathoLogic next creates a PGDB object for every replicon and contig
defined by the input files, and for every gene and gene product
defined in the input files.  It populates these new objects with data
from the input files, such as gene names and their sequence
coordinates, and gene product names.  As a result of these operations,
the new PGDB now mirrors the information in the input files.

\Ssection{PathoLogic Inference of Metabolic Pathways}
\label{sec:pwy-inference}

Pathway Tools predicts the metabolic pathway complement of an organism
by assessing what known pathways from the MetaCyc PGDB \cite{MetaCycNAR18} are present in
the annotated genome of that organism's PGDB.  This inference is
performed in two steps that are described and evaluated in
\cite{KarpHpy02,KarpPathPred10,KarpPathPred11}.  

{\bf Step 1:} The enzymes in the PGDB are assigned to their
corresponding reactions in MetaCyc, thus defining the reactome of the
organism.  PathoLogic performs this assignment by matching to MetaCyc
reactions the gene-product names (enzyme names), the EC numbers, and
the Gene Ontology terms assigned to genes in the genome.  The program
can use whatever combination of these three information types is
available in a given genome.  For example, the {\em fabD} gene in {\em
  Bacillus anthracis} was annotated with the function ``malonyl
CoA-acyl carrier protein transacylase.''  That product name was
recognized by PathoLogic as corresponding to the MetaCyc reaction
whose EC number is 2.3.1.39.  PathoLogic therefore imported that
reaction and its substrates into the {\em B. anthracis} PGDB, and
created an enzymatic-reaction object linking that reaction to that
{\em B. anthracis} protein.  If the product name is not recognized or
is ambiguous, and there is no EC number or GO term, then PathoLogic
will look up the gene name to try to identify the reaction catalyzed,
if any.

{\bf Step 2:} Once the reactome of the organism has been established
by Step 1, PathoLogic predicts what metabolic pathways are likely
present based on the reactome.  PathoLogic considers every pathway in
MetaCyc, and computes a score that indicates the likelihood that the
pathway is present. The pathway score is computed from the sum of the
scores of the reactions in the pathway, divided by the number of
reactions in the pathway (excluding spontaneous reactions). A reaction
score is computed by summing these factors:

\bitem
\item Is an enzyme catalyzing the reaction present in the organism?
\item How unique is the reaction to this pathway? Is it found
  only in this pathway, or in other pathways as well? The less unique a reaction, the lower its score. 
\item Some reactions in a pathway are designated as key reactions,
  meaning the reaction distinguishes the pathway from other similar
  pathways; the presence of an enzyme that catalyzes a key reaction
  boosts the score of that reaction.
\eitem
  
A rule-based expert system makes the final determination of whether the pathway is inferred as present by considering the following factors:
\bitem
\item The pathway score
\item The presence of designated key non-reactions for the pathway --– reactions whose presence inhibits inference of the pathway
\item Was some other variant of this pathway assigned a superior score?
\item Is the pathway outside its taxonomic range as specified in
  MetaCyc, with at most one reaction having no enzymes present?
\eitem

\SSsection{Calculation of Pathway Abundance for Metagenomics Analysis}

PathoLogic computes abundances of metabolic pathways based on gene
abundances, which is useful for comparing the metabolic profiles of
different microbial communities. Gene abundances are specified in the
annotated genome file (PathoLogic format only).

No preprocessing of the gene abundances (such as outlier removal) is done by PathoLogic. 
The abundance of a pathway is computed based on the gene abundances
involved in the pathway. More precisely, assume that $R$ is the set of
reactions in pathway $P$ for which gene abundances are specified, $|R|$ is
the size of $R$, and $g_a$ is the given abundance of gene $g$. The
abundance of a pathway $P$ is

    $$\sum_{r \in P} r_a/|R|\;{\rm where\ } r_a = \sum_{g\ {\rm catalyzes}\; r} g_a$$

That is, the abundance of a pathway is the sum of the abundances of
the genes catalyzing the reactions of the pathway, divided by the
number of reactions of the pathway for which gene abundances are
given.  Notice that this formula does take into account all the known
isozymes catalyzing a reaction and the spontaneous reactions do not
take part in the computation. The abundances are provided, among other
results, in the file \texttt{pathways-report.txt}.

\Ssection{PathoLogic Inference of Operons}

The Pathway Tools operon predictor identifies operon boundaries by
scoring pairs of adjacent genes, $A$ and $B$, on the
basis of a set of features, including the intergenic distance in addition to
features that predict a functional relationship between $A$ and $B$.  Other
features include membership in the same pathway \cite{KarpOperon04}, membership in
the same multimeric protein complex, \blue{and overlapping sets of Gene Ontology function
annotations.  The predictor can also use correlated levels of gene expression for
$A$ and $B$ across RNASeq experiments.  The features are combined in a logistic regression
model to predict whether the pair is either within an operon or straddles a boundary between
operons.}

\Ssection{PathoLogic Inference of Pathway Holes}

A pathway hole is a reaction in a metabolic pathway for
which no enzyme has been identified in the genome that catalyzes that
reaction.  Typical microbial genomes contain 200--300 pathway holes.
Although some pathway holes are probably genuine, we believe that the
majority are likely to result from the failure of the genome-annotation
process to identify the genes corresponding to those
pathway holes.  For example, genome-annotation systems systematically
under-annotate genes with multiple functions, and we believe that the
enzyme functions for many pathway holes are unidentified second
functions for genes that have one assigned function.
Erroneous pathway holes can result from the prediction of pathways
that are not actually present in the organism.

The pathway hole-filling program PHFiller \cite{Green04} (a component
of PathoLogic) generates hypotheses as to which genes code for these
missing enzymes by using the following method.  Given a reaction that
is a pathway hole, the program first queries the UniProt database to
find all known sequences for enzymes that catalyze that same reaction
in other organisms.  The program then uses the BLAST tool to compare
that set of sequences against the full proteome of the organism in
which we are seeking hole fillers.  It scores the resulting BLAST hits
using a Bayesian classifier that considers information such as genome
localization (that is, is a potential hole filler in the same operon
as another gene in the same metabolic pathway?).  At a stringent
probability-score cutoff, our method finds potential hole fillers for
approximately 45\% of the pathway holes in a microbial genome
\cite{Green04}.

PHFiller includes a graphical interface that optionally presents each
inferred hole filler to the user along with information that helps the 
user evaluate the hole fillers, and enables the user to
accept or reject the hole fillers that it has proposed.

\Ssection{PathoLogic Inference of Transport Reactions}

Membrane transport proteins typically make up 5--15\% of the gene
content of organisms sequenced to date.  Transporters import nutrients
into the cell, thus determining the environments in which cell growth
is possible.  The development of the PathoLogic Transport Inference
Parser (TIP) \cite{KarpTIP08} was motivated by the need to perform
symbolic inferences on cellular transport systems, and by the need to
include transporters on the Cellular Overview diagram.  The motivating
symbolic inferences include the problems of computing answers to the
following queries: What chemicals can the organism import or export?
For which cellular metabolites that are consumed by metabolic
reactions, but never produced by any reaction, does no known
transporter exist? This would mean that the origin of such metabolites is a mystery,
and would indicate missing knowledge about transporters or reactions that
produce the compound.

To answer such queries, Pathway Tools uses an ontology-based representation of
transporter function in which transport events are
represented as reactions in which the transported compound(s) are
substrates.  Each substrate is labeled with the cellular compartment
in which it resides, and each substrate is a controlled-vocabulary
term from the extensive set of chemical compounds in MetaCyc
\cite{MetaCycNAR18}.  The TIP program converts the free-text
descriptions of transporter functions found in genome annotations
(examples: ``phosphate ABC transporter'' and ``sodium/proline
symporter'') into computable transport reactions.

\Ssection{\blue{PathoLogic Inference of Protein Complexes}}

\blue{
Many protein gene products form complexes, to enable various cellular
functions, such as complicated reactions, transport through membranes,
or mechanical components in cell architecture.  Therefore, PathoLogic
aims to predict such complexes, in part for supporting simulations of
gene knockout experiments.  When a gene for a subunit in a metabolic
complex is knocked out, the corresponding reaction or function is usually
disabled.
}

\blue{
Complex Inference is accomplished by several different approaches.
One method collects all protein monomers that are connected to the
same reaction, but only if the monomer genes are in mutual vicinity to
each other on a chromosome and contain a word like ``subunit'' in
their names.
}

\blue{
However, there are many complexes that contain some subunits that are
not directly attached to a reaction, potentially due to annotation
pipeline flaws, or where a complex plays a non-metabolic role
altogether.  To catch these cases, a name-based analysis of the
subunit names was developed.  First, all protein monomer name strings
are searched for words like ``subunit'', which indicate the monomer
might be part of a complex.  Second, the names of these subunit
candidates are trimmed to what we call the ``core-string'', by
removing all extraneous words such as the (now redundant) subunit
words, and indicators of the identity of a subunit, such as "alpha",
"beta", Roman numerals, etc.  This reduction tries to reveal the
essence of the complex' function and purpose.  Third, subunits with
identical core-strings are grouped together as candidates that should
form a complex.  If there are subunits with exact duplicates of their
full names, then possibly more than one complex needs to be formed, as
separate iso-complexes.  This is done by also grouping according to
mutual proximity on the chromosome, to provide further evidence for
which group of candidate subunits should be joined into one complex.
}

\blue{
Two types of transport complexes receive special analysis, namely PTS
and ABC transporters.  Some common patterns can be taken advantage of,
for these cases.
}

\blue{
At the end of running Complex Inference, a detailed report is recorded
in a file, detailing which method created a given complex, along with
its detailed subunits.
}

\Ssection{Atom Mappings}

The atom mapping of a reaction specifies for each non-hydrogen atom in
each reactant
its corresponding atom in a product compound.
Pathway Tools contains an algorithm for computing atom mappings,
described in~\cite{latendresse2012}. Essentially, this approach computes atom
mappings that minimize the overall cost of bonds broken and made in
the reaction, given assigned propensities for bond creation and
breakage.  This algorithm has been applied to compute atom mappings
for almost all of the reactions in 
the MetaCyc database.

Atom mappings are used in two other parts of Pathway Tools.  Atom
mappings are used in the rendering of Pathway Tools
reaction pages, to depict the conserved chemical moieties in a reaction.
Conserved moieties are depicted by using the same color on the reactant
and product sides.  The bonds made or broken by a reaction are identified
from the atom mapping for the reaction, and are colored black.
Atom mappings are also used in the RouteSearch module of Pathway Tools
described in Section~\ref{sec:routesearch}.

Atom mappings are typically stored in the MetaCyc PGDB only, except
for the reactions unique to other PGDBs.

\Ssection{Computation of Metabolite Gibbs Free Energies}

The MetaCyc database provides the standard Gibbs free energy
of formation for its compounds, and the change in Gibbs free energy for
its reactions. 
These data were calculated by an algorithm within Pathway Tools.  The
algorithm first calculates the free energy of formation at pH 0 and
ionic strength 0 ($\Delta_f G^0$) by using a technique based on the
decomposition of the compounds into chemical groups with known
free-energy contributions to the overall energy, based on the method
of~\cite{Jankowski08}. Then, the standard Gibbs free energy at pH 7.3
and ionic strength 0.25 ($\Delta_f G^{'0}$) is computed based on a
technique developed by Robert A. Alberty~\cite{Alberty03}. In his
technique, Alberty proposes to use several protonation states for some
compounds, but we simplified the technique by always using only one
protonation state, the state stored in MetaCyc.
We use pH 7.3 because this is a common cellular pH, and computation of
the protonation state of all compounds in MetaCyc were performed at
that pH.

The change in standard Gibbs free energy of reactions, $\Delta_r G^{'0}$,
is computed based on the $\Delta_f G^{'0}$ values of the
compounds involved in the reaction.
The $\Delta_f G^{'0}$ could not be computed for some of the compounds
in MetaCyc due to the impossibility of decomposing them into the
groups provided by the technique
of~\cite{Jankowski08}. Consequently, the $\Delta_r G^{'0}$ is not
computed for any reaction which has a substrate for which its
$\Delta_f G^{'0}$ is not stored in MetaCyc.

\Ssection{Pathway/Genome Editors}

The Editors support PGDB curation through interactive modification and
updating of all the major data types supported by Pathway Tools.
The editing tools included in Pathway Tools are as follows:

\bitem
\item {\bf Gene Editor:}  Supports editing of gene name, synonyms, database
links, and start and stop position within the sequence.

\item {\bf Isoform/Coding Segment Editor:} Supports specifying
multiple coding regions within a gene, including mRNA splicing
and ribosomal slippage.

\item {\bf Protein Editor:} Supports editing of protein attributes,
subunit structure, and protein complexes (see \href{http://www.ai.sri.com/pkarp/pubs/pt20suppfigs4.pdf}{Supplemental Figures}~12-14).  Enables users to assign terms from the
Gene Ontology controlled vocabulary.  Pathway Tools can store,
and display features of interest on a protein; see
Section~\ref{sec:prot-schema} for more details.  When editing a
protein feature the user selects a feature type (e.g., phosphorylation
site), defines the location
of the feature on the sequence, a bound or attached moiety where
appropriate, a textual label, an optional comment, citations, and
sequence motif.
\item {\bf Protein Subunit Structure Editor:} Supports editing of
protein subunit structure and protein complexes.

\item {\bf Reaction Editor:} Supports editing of metabolic reactions,
  transport reactions, binding reactions and redox reactions.  This
  editor checks reactions for elemental balance and charge balance.

\item {\bf Atom Mapping Editor:} Supports editing of the mapping of atoms
  from reactants to products for chemical reactions.

\item {\bf Pathway Editor:} Enables users to interactively construct and edit
a metabolic pathway from its component reactions.  (See
\href{http://www.ai.sri.com/pkarp/pubs/pt20suppfigs4.pdf}{Supplemental Figure}~11.)

\item {\bf Signaling Pathway Editor:} Enables users to interactively
  construct and edit a signaling-pathway diagram by using a toolkit of
  icons and operations inspired by CellDesigner \cite{Funahashi03}
  (See \href{http://www.ai.sri.com/pkarp/pubs/pt20suppfigs4.pdf}{Supplemental Figure}~16).  Updates to the visual representation
  are automatically translated back to changes to component reactions
  and proteins.

\item {\bf Regulation Editor:} Enables definition of regulatory interactions
including regulation of gene expression by control of transcription
initiation, attenuation, and control of translation by proteins
and small RNAs (see \href{http://www.ai.sri.com/pkarp/pubs/pt20suppfigs4.pdf}{Supplemental Figure}~15).  This editor also allows
creation of operons and definition of
their member genes, as well as specifying the positions of promoters
and transcription-factor binding sites.

\item {\bf Transcription Unit Editor:} This editor allows creation of
  operons and definition of their member genes, as well as specifying
  the positions of transcription start sites and transcription-factor
  binding sites.

\item {\bf Compound Editor:} Supports editing of compound names, citations,
  and database links.  For specifying a compound structure, Pathway
  Tools has been interfaced to an
  external chemical structure editor, Marvin \cite{MarvinURL}  (both
  the JAVA applet and the JavaScript versions), and it can import and export
  MOL files.  A chemical compound duplicate checker runs
  whenever chemical structures are entered or modified, to inform the
  user if the resulting structure is identical to another compound in that
  user's PGDB or in MetaCyc.  Additionally, Pathway Tools can display
  glycan structures in an icon-based style that follows the
  conventions of CFG (Consortium for Functional Glycomics).  To edit
  these structures, the software can communicate directly with a
  modified
  version of the GlycanBuilder editor \cite{Ceroni07,Damerell12}, to
  which we have added functionality for better integration with Pathway Tools.

\item {\bf Publication Editor:}  Supports entry of bibliographic references,
  allowing automatic import of data using a DOI URL.

\item {\bf Organism Editor:} Supports editing information about an organism
including species name, strain name, and synonyms, and
taxonomic rank within the NCBI Taxonomy.

\item {\bf PGDB Info Editor:} This editor allows entering data about
  the PGDB, including additional information about the organism it
  describes such as sample collection data (e.g., date, geographic location,
  host, body site); and phenotypic information, such as pathogenicity
  and relationship to oxygen.
  
\item {\bf Cellular Architecture Editor:} Enables users to specify exactly
  which set of cellular components are present in an organism or cell
  type, with appropriate defaults derived from the organism's
  taxonomy.

\item {\bf Sequence Editor:} Supports interactive, visual editing of
    the nucleotide sequence for a replicon, allowing insertion,
    deletion, and replacement of arbitrary sections of sequence.
    Coordinates of all objects affected by the edits are updated
    automatically.

\eitem

\SSsection{Author-Crediting System}

Often, multiple curators collaborate on development of a given PGDB.
It is desirable to attribute their contributions accordingly, both
to identify whom to ask if questions about particular entries arise,
and to provide an incentive for high-quality contributions, because
contributors will be able to clearly demonstrate their
accomplishments.

Most Pathway Tools editors thus create {\em credits} of several kinds.
When an object such as a pathway is first created, by default, a
``created'' credit is attached to the object, along with a timestamp.
A credit for an object can refer to curators, to organizations, or to
both.  Pathway Tools provides a generated web page for every curator
and organization that lists all the objects for which they are
credited.

Other kinds of credits are ``revised'', which is used when a curator substantially
edits an object that was created some time ago, and a ``last curated''
flag that can be set to indicate when a curator has last researched the
literature available for a given object.  The last-curated flag is useful 
for those objects about which almost nothing is known, to distinguish
between the case where no curator has ever looked into that the object,
versus the case where despite 
an extensive search no new information was found.
The ``reviewed'' credit is used to attribute reviews of DB objects by
external domain experts.

\Ssection{Incorporation of Genome-Annotation Revisions}
\label{sec:bulk-update}

Some groups choose to store the authoritative version of the
organism's genome annotation in a database external to the PGDB.  Such
users need the ability to incorporate revisions to the genome
annotation into their PGDB without overwriting or otherwise losing any
manual curation they added to the PGDB.  Pathway Tools
provides an interface for incorporating annotation revisions that takes as input one or
more update files, either in GenBank format or as a PathoLogic Format file.
The files can contain either a complete revised annotation for the
organism, or they can contain just the information that has changed.
The software will parse the update files and determine all differences
between the new data and the old.  Types of changes that are detected
include new genes, as well as updated gene positions, names, synonyms,
comments, links to external databases, and updated functional
assignments.  A graphical interface summarizes different classes of
changes to the user, and gives the user the  option of either accepting all updates (e.g., creating
database objects for all the new genes) or of accepting/rejecting individual
updates.  Once this phase is complete and any changes to functional
assignments have been made, the software will re-run the pathway-inference
procedure described in Section~\ref{sec:pwy-inference},
identify any new pathways that are inferred to be present and any
existing pathways that no longer have sufficient evidence, and allow
the curator to review those changes.

\Ssection{Sequence Coordinate Mapping Service (Web Only)}
\label{sec:seq-coord-mapping}

The authoritative or reference DNA sequence of a replicon is sometimes updated to
fix sequencing errors. Because some of the errors can involve
insertions or deletions, the base-pair coordinates further downstream
will shift, compared to the uncorrected sequence. Such updates affect the
positions of genes, promoter sites, and other regions of importance.

Pathway Tools provides a web form that will map user data files
containing older coordinate information, to coordinates appropriate
for the latest genome version.  Note
that this operation will not change the coordinates used in the PGDB
itself, which always correspond to the latest version.  However, it
will enable updating older data files, so the revised file can be used
for analyses against the latest sequence.  The service can be invoked
on the BioCyc website from the menu item
{\bf Genome $\rightarrow$ Map Sequence Coordinates} 

Pathway Tools provides a sequence editor that can be used to update the
nucleotide sequence of a replicon.  These individual edits are recorded
in the replicon frame in the PGDB.  The coordinate mapping service works by
chaining together these edits to produce the overall mapping between
two specified versions of the replicon.

\Ssection{Consistency Checker and Aggregate Statistics}
\label{sec:ccheck}

Pathway Tools contains an extensive set of programs for performing
consistency checking of a PGDB to detect structural errors and
inconsistencies that sometimes arise within PGDBs. Such problems
are caused by either user data-entry errors or errors in Pathway Tools itself.
Also included in this component are tools for computing certain types
of data and statistics for a PGDB, such
as computing the molecular weights of all proteins from their amino
acid sequences.

Roughly half of the programs can automatically repair the problems that
they find.  Such problems could be caused by either user data-entry errors
or errors in Pathway Tools itself.
Example checks include: ensuring that inverse relationship
links are set properly (e.g., that if a gene is linked to its gene
product, and that the product links back to the gene); making sure
pathways do not refer to reactions they no longer contain; validating
and updating GO terms with respect to the latest version of the GO Ontology;
validating the format of hypertext within summaries and names; and
removing redundant bonds from chemical structures. The other half
of the programs can identify problems and point to their location but
leave it to a curator to solve the issues.

\Ssection{Schema Upgrading and Propagation of MetaCyc Updates}

Most new releases of Pathway Tools include additions or modifications to
the Pathway Tools schema that are made to model the
underlying biology more accurately (such as adding support for introns
and exons), and to extend the data types within Pathway Tools (such as
adding support for features on protein sequences).
Because each new version of the software
depends on finding data within the fields defined by the associated
version of the schema, existing user PGDBs created by older versions
of the software will be incompatible with these new software versions.

Therefore, every release of Pathway Tools contains a program to
upgrade PGDBs whose schema corresponds to the previous version of the
software, to the new schema version.  For users who have not upgraded the
software for several releases, several upgrade operations will be
performed consecutively.  Example upgrade operations include adding
new classes to the PGDB from the MetaCyc PGDB; adding new slots to
PGDB classes; deleting PGDB classes; moving data values from one slot
to another; and moving objects from one class to another.  The schema
upgrade leaves the user's curated data intact.

Every new release of Pathway Tools includes a new
version of the MetaCyc database, which, in addition to providing new
data content, typically contains updates and corrections to existing
pathways, reactions, and compounds.  Pathway Tools includes an option
to propagate such updates and corrections to an existing organism
PGDB.  However, because we do not want to override any manual edits
made to a PGDB, this tool does not run automatically.  Much like the
tool for incorporating a revised genome annotation, described in
Section~\ref{sec:bulk-update}, this tool organizes the changes into
logical groups (such as all compounds with newly added structures, or
all reactions with changed reaction equations), and allows the user to
either accept an entire group of changes, or to examine and confirm
individual members of a group.

\section{The Pathway Tools Schema}
\label{sec:schema}

\begin{figure}
\includegraphics[width=6in]{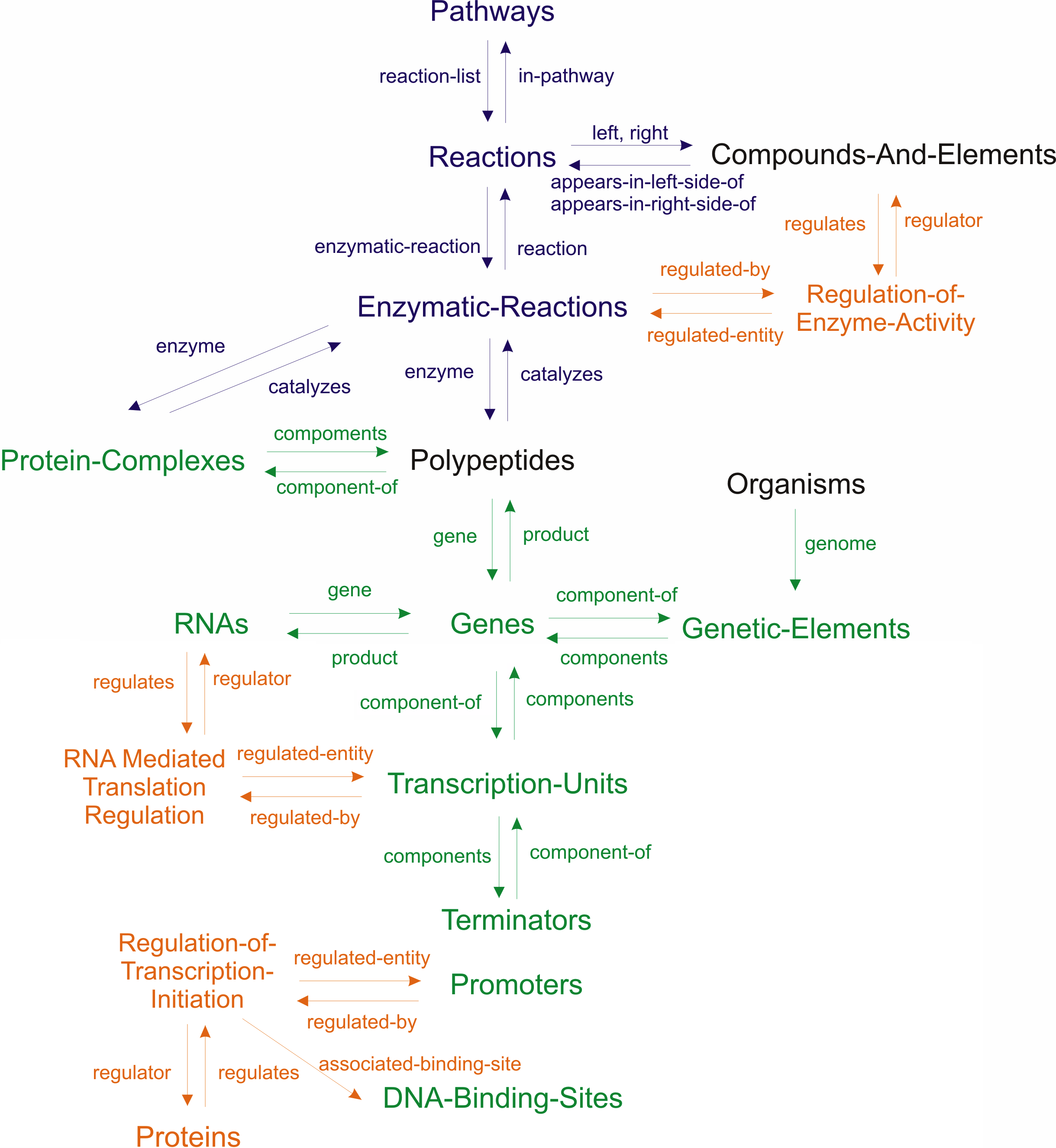}
\caption{Major relationships among the major classes of the Pathway
Tools schema.  Colors indicate biological areas: blue for reaction and
pathway information, green for genome and protein information, orange
for regulation.}
\label{fig:relationships}
\end{figure}

The Pathway Tools schema defines structured representations of a broad
range of biological data types to enable computational analyses and
integration of many types of data.
The schema consists of a set of classes and a set of slots.
Classes describe types of biological entities, such as genes and
pathways, and are arranged in a class--subclass hierarchy.  Slots
define attributes of PGDB objects and 
relationships between PGDB objects.  Figure~\ref{fig:relationships}
provides an overview of the relationships among PGDB classes.  For
example, user queries can follow the relationship from a gene to the
protein that it codes for, from a protein to a reaction that it
catalyzes, and from a reaction to a metabolic pathway in which it is a
component, to answer questions such as ``find all metabolic pathways, in
which the products of a given gene play a role''.

Every PGDB object has a stable unique identifier (ID)---a
symbol that uniquely identifies that object within the PGDB.  Example
unique IDs include TRP (an identifier for a metabolite), RXN0-2382 (an
identifier for a reaction), and PWY0-1280 (an identifier for a
pathway).  Relationships within a PGDB are implemented by storing
object IDs within slots.  For example, to state that the TRP
(L-tryptophan) object is a reactant in the reaction RXN0-2382, a slot
of RXN0-2382 called LEFT (meaning reactants) contains the value TRP.
Many PGDB relationships exist in both forward and backward directions
(for example, the TRP object contains a field called
APPEARS-IN-LEFT-SIDE-OF that lists all reactions in which TRP is a
reactant).  The slots LEFT and APPEARS-IN-LEFT-SIDE-OF are called
inverses.

\Ssection{Metabolites, Reactions, and Pathways}
\label{sec:schema-metab}

There are two alternative ways in which one might choose to represent
the metabolic network in a computer: as a simple listing of all
metabolic reactions that occur in the cell, or by partitioning the
reaction list into a carefully delineated set of metabolic pathways
that describe small, functionally linked subsets of reactions.  Which
approach is preferred?  Both approaches have value,
and they are not mutually exclusive; therefore, Pathway Tools supports
both views of metabolism in a PGDB.

Pathway Tools conceptualizes the metabolic network in three layers.
The first layer consists of the small-molecule substrates upon which
metabolism operates.  The second layer consists of the reactions
that interconvert the small-molecule metabolites.  The third layer consists of
the metabolic pathways whose components are the metabolic reactions of
the second layer.  Note that not all reactions in the second layer are
included in pathways in the third layer, because some metabolic
reactions have not been assigned to any metabolic pathway by biologists.
Scientists who choose to view the metabolic network within a PGDB solely as a
reaction list can operate on the second layer directly without
interference from the third layer.  
The compounds, reactions, and pathways in levels 1--3 are each
represented as distinct database objects within a PGDB.  The
relationships among the metabolic data types in a PGDB are depicted by
the blue region of Figure~\ref{fig:relationships}.

The representation of reactions can capture atom-mapping information
that records, for each atom in a reactant compound, its terminus atom
in a product compound.  The representation of metabolites can capture
stereochemical structural information as well as glycan structures.

The pathways in PGDBs are modules of the metabolic network of a single
organism.  Pathway boundaries are defined by considering the following
factors.  Pathways are often regulated as a unit (based on
substrate-level regulation of key enzymes, on regulation of gene
expression, and on other types of regulation).  Pathway boundaries are
often defined at high-connectivity, stable metabolites
\cite{GreenK06}.  Pathway conservation across multiple species is also
considered, as are pathway definitions from the 
experimental literature \cite{MetaCycNAR18}.

\Ssection{The Proteome and the Genome}

PGDBs define the proteome and the genome of an organism in
the following manner, as depicted by the green region of
Figure~\ref{fig:relationships}.
The proteome of the organism is described as a set of PGDB objects,
one for each gene product in the organism, and one for each complex
formed from two or more (identical or nonidentical) polypeptides.
Furthermore, every chemically modified form of a monomer or of a
multimer is encoded by a distinct PGDB object.  For example, we might
create one object representing an unmodified protein and another
representing the phosphorylated form.  Each protein object is in turn
linked, through a slot in the object, to the metabolic reactions that
it catalyzes.  Proteins can also be substrates of reactions.
Additional PGDB objects define features on proteins, as described in
Section~\ref{sec:prot-schema}.
Each protein product resulting from alternatively spliced forms of a
gene is also represented by a distinct protein object.  Each protein
object records the exons of the gene that encodes it.

Protein objects are linked to gene objects that define the gene
encoding each protein.  Each gene in the genome is defined by a
distinct PGDB object, as is every replicon (chromosome or plasmid) in
the genome.  Genes are linked to the replicon on which they reside.
In addition, other features on the genome, such as operons, promoters,
and transcription-factor binding sites, are described by PGDB objects.

The associations between enzymes and the reactions they catalyze are
implemented by using an intermediary object called an enzymatic reaction,
as shown in Figure~\ref{fig:relationships}.  This arrangement enables us to
capture the many-to-many relationship that exists between enzymes and
reactions---one reaction can be catalyzed by multiple enzymes, and
multifunctional enzymes catalyze multiple reactions.  The purpose of
the enzymatic reaction is to encode information that is specific to
the pairing of the enzyme with the reaction, such as cofactors,
activators, and inhibitors.  Consider a bifunctional enzyme with two
active sites, where one of the active sites is inhibited by pyruvate,
and the second active site is inhibited by lactate.  We would
represent this situation with two enzymatic reactions linking the
enzyme to the two reactions it catalyzes, and each enzymatic reaction
would specify a different inhibitor.

\Ssection{Pathway Tools Regulation Ontology}
\label{sec:schema-reg}

The Pathway Tools schema can represent a wide range of regulatory
interaction types.  A regulation object within a PGDB captures information about 
each type of regulatory interaction.  The available regulation types are as
follows:
(a) Substrate-level
regulation of enzyme catalytic activity, such as the allosteric
activation or competitive inhibition of an enzyme by a small molecule.
Slots of this class identify the regulator molecule, the
regulated enzymatic reaction object, encode the polarity of
regulation (activation or inhibition), and the mechanism of regulation
(allosteric, competitive, or noncompetitive).
(b) Regulation of a bacterial promoter by a transcription-factor protein.
The slots of this regulation class describe the
transcription factor, the promoter that is regulated, and the binding
site to which the regulator binds.  
(c) Regulation via premature termination of transcription
(attenuation).  This class of regulation is divided
into six subclasses, each describing a different attenuation mechanism
(e.g., ribosome-mediated, protein-mediated, or RNA-mediated).  The slots
of these classes identify the regulated terminator region, the
regulator (a protein, RNA or small-molecule, depending on the type of
attenuation), and the regulator binding site if one exists.
(d) Regulation of the
translation of an mRNA transcript to the corresponding protein.  This
regulation class is divided into two subclasses to distinguish between
regulation by a protein and regulation by a small RNA.  The slots of
these classes identify the regulated transcription-unit (which
corresponds to a single transcript), the regulator protein or RNA,
and the mRNA binding site where the regulator binds.  An additional
slot indicates whether regulation is by direct interference with the
translation machinery, by processing of the mRNA transcript to promote or
inhibit its degradation before translation, or both.
(e) Regulation of protein activity by chemical modification, such as by
phosphorylation, is represented by a reaction that converts the
unmodified form of the protein to the modified form.

\Ssection{Conditions of Cellular Growth}
\label{sec:growth-conditions}

 The Pathway Tools schema supports representation of
  conditions of cellular growth that include the chemical composition
  of the growth medium, pH, temperature, and aerobicity.  This
  representation enables us to capture low-throughput information about
  conditions of cellular growth, and high-throughput information such
  as Phenotype Microarray \cite{Bochner-etal01,Bochner09} data sets.

\Ssection{Gene Essentiality}

The Pathway Tools schema supports representation of
gene-essentiality experiments.  Our representation links growth
phenotype (no growth, limited growth, or growth) under a given gene
knockout with the conditions of cellular growth expressed as per Section~\ref{sec:growth-conditions}.
Multiple phenotypic observations can be recorded for a given gene
knockout and growth condition to express conflicting experimental outcomes.

\Ssection{Organism Phenotype Data and Genome Metadata}

The Pathway Tools schema supports representation of
microbial phenotypic data to enable users to query among the many
genomes stored within a Pathway Tools website to find organisms
pertinent to their research.  Our representation adapts the MIGS
\cite{MIGS} standard to incorporate metadata about the
sample from which the organism was derived (e.g., geographic location,
depth, health-or-disease state of host, human microbiome site), phenotypic
information about the organism itself (e.g., relationship to oxygen,
temperature range, and pathogenicity), and genome-annotation metadata (e.g.,
annotation date, provider, pipeline).

\Ssection{Pathway Tools Evidence Ontology}

Database users want to know the type(s) of evidence that support assertions
within a DB, and they want to know the strength of that evidence.
We have developed an evidence ontology \cite{KarpEv04} that can encode
information about {\em why} we believe certain assertions in a PGDB,
the {\em sources} of those assertions, and the {\em degree of
confidence} scientists hold in those assertions (although in practice
the latter field is rarely populated).  An example assertion is the
existence of a gene in a PGDB.  Was the
gene predicted by using computational gene finding?  Is it
supported by wet-lab experiments?  The Pathway Tools evidence ontology builds upon
and substantially extends the Gene Ontology evidence ontology, which
applies only to gene products.

Evidence about object existence in PGDBs is recorded as a structured
{\em evidence tuple}.  An evidence tuple enables us to associate
several types of information within one piece of evidence.  Each {\em
evidence tuple} is of the form

\begin{verbatim}
   Evidence-code : Citation : Curator : Timestamp : Probability
\end{verbatim}

\ \\
where
{\tt Evidence-code} is a unique ID for the type of evidence, 
within a hierarchy of 54 evidence types described in \cite{PToolsEvOntologyURL,KarpEv04}.
{\tt Citation} is an optional citation identifier such as a PubMed ID that 
indicates the source of the evidence.  For computational evidence, the
citation refers to an article describing 
the algorithm used.
{\tt Curator} identifies the curator who created this
evidence tuple. {\tt Timestamp}  encodes when this evidence tuple was created.
{\tt Probability} is an optional real number indicating the
probability that the assertion supported by this evidence is correct,
such as a probability provided by an algorithm.

The Pathway Tools editors enable users to manually enter evidence
codes, and the PathoLogic pathway and operon predictors annotate
objects that they create with appropriate computational evidence
codes.  The Navigator supports display and querying of evidence codes.

\Ssection{Pathway Tools Cell Component Ontology}

The Cell Component Ontology (CCO) is a controlled vocabulary of terms
describing cellular components and compartments, and relationships
between these terms \cite{PToolsCCOURL}.  It was developed to provide
a controlled vocabulary of terms for annotating the subcellular
locations of enzymes, and the compartments involved in transport
reactions, in PGDBs.  CCO spans all domains of life, and includes
terms such as cytoplasm, cell wall, and chloroplast. The ontology
currently contains 170 terms.  CCO includes many terms and their
definitions from the Gene Ontology \cite{GO2008}, but substantially
extends Gene Ontology.  A recent extension to CCO enables
any metabolic reaction to be annotated to one or more CCO compartments,
to allow metabolic models to span multiple compartments.

\Ssection{Pathway Tools Protein Feature Ontology}
\label{sec:prot-schema}

We have developed an ontology of protein features to identify
and represent post-translational modifications, binding sites, active
sites, conserved regions, and other regions of interest on a protein.
Starting from the list of feature types described in the UniProt User
Manual \cite{SprotManual09}, with some suggested additions from the
SRI EcoCyc and MetaCyc database
curators, we created an ontology of 40 feature classes.  

Features fall into two major classes.  For amino acid site features,  the
feature location is a list of one or more amino acid residue numbers
(or residue types, if the feature is associated with a generic protein
whose precise sequence is unspecified).  For protein segment features, 
the feature location is a range defined by its starting and
ending residue numbers.  

Feature types that are classified as binding features (either covalent
or non-covalent) permit specification of an attached group.  The
attached group could be a compound or compound fragment, as in the
case of a protein that binds a small molecule.  The attached group can
also be another protein feature, as in the case of a disulfide bond or
other cross-link between two features on different proteins, or any
other type of molecule or binding site (such as a DNA binding site).

A different protein object is created in a PGDB for each biologically
relevant modified form of a protein, and a single feature may be
linked to multiple forms of the same protein.  Some feature types are
capable of existing in multiple states.  For example, an amino acid
modification feature can  be in either the modified or the unmodified
state (as in the case of a phosphorylation feature, which will be in
the modified state when associated with the phosphorylated protein and
the unmodified state when associated with the unphosphorylated
protein), and a binding feature can  be in either the bound or unbound
state (as in the case of a metal-binding feature whose state indicates
whether or not the metal ion is bound to the protein).  We consider
the state to be not an attribute of the feature, but rather an
attribute of the pairing between a particular form of a protein and
the feature.  

\section{Visualization and Querying of PGDBs}
\label{sec:nav}

The Pathway/Genome Navigator component of Pathway Tools provides
mechanisms for interrogating PGDBs and for visualizing the results of
those queries.  We begin by describing the query tools.  We then
describe visualization tools for individual biological entities
(such as genes and pathways).  We next describe the SmartTables
system, which enables the user to manipulate groups of PGDB objects,
such as gene sets and metabolite sets.  Finally, we describe system-level visualization tools that 
graphically display the entire metabolic network, entire regulatory 
network, and entire genome map of an organism.

The Navigator runs as both a desktop application and a web server,
whereas the Pathway Tools components described in some other sections run as
desktop applications only (e.g., PathoLogic does not run as a web server).
The desktop mode has more overall functionality (see
\cite{BioCycDesktopVsWebURL} for details), but the web mode 
has some functionality that is not present in the desktop mode.

\Ssection{Query Tools}
\label{sec:query}

Pathway Tools provides a three-level query paradigm, meaning that three different types of
query tools are available, each of which represents a different
trade-off between ease of use and query power.

{\noindent \bf Level 1: QuickSearch:} Quick
Search is designed to provide a fast and simple way for new or casual
users to find general information in the site.  
The ``Quick Search'' box (top-right of most web pages; bottom-left of
the Desktop window) is extremely easy to use.  The
user selects the organism whose PGDB the user wants to query and
enters one or more search terms.  Pathway Tools searches that PGDB for objects
whose primary name or synonyms contain the search term as a substring,
and presents the list of results, organized by object type (e.g.,
gene, metabolite, pathway).  The user can click on an object name to
navigate to the display page for that object.

{\noindent \bf Level 2: Single Object Searches:} 
A set of intermediate-level query tools provides the ability to
construct more powerful and precise searches against objects of a
single class.  Web mode: The Search menu contains a specialized search page for genes, proteins, and
RNAs (see \href{http://www.ai.sri.com/pkarp/pubs/pt20suppfigs4.pdf}{Supplemental Figure}~1), for example, the user can search
for proteins with a specified pI, molecular weight, and sequence length.
Additional web-based query pages exist for
pathways, reactions, chemical compounds, DNA or mRNA sites, and growth
media. Desktop mode: See the Advanced Search commands under the Gene,
Protein, and Compound menus, for example, the Compound Advanced Search
enables a user to search for a metabolite with a specified full or
partial chemical formula, chemical substructure, and/or (monoisotopic)
molecular weight.

{\noindent \bf Level 3: Multi-Object Searches:} 
The web-based Structured Advanced Query Page (SAQP) described below enables advanced users to construct
extremely powerful searches (that are approximately as powerful as those
provided by the SQL language).  The graphical interactive nature of
this web form makes these searches much easier to construct than those
using the SQL language.  

\SSsection{Organism Selection}

PTools web-mode searches are conducted against a single PGDB, thus the first step
is to select the desired organism or PGDB.  The web-based organism
selection dialog allows for searching organisms by name or taxonomic
grouping, and includes quick links to a user's recently accessed
organisms.  In addition, this dialog now offers a search option for
finding PGDBs based on properties of the organism.  Example searches
include searching for organisms with a given growth phenotype (e.g.,
presence of oxygen or temperature); the geographic region where the
organism was collected; the human microbiome body site at which the
organism was collected; or properties of the PGDB, such as the
software used to annotate the genome, or the number of regulatory
interactions present in the PGDB.  This type of search will become
more valuable as more metagenomics projects generate data tied to
specific body sites (e.g., the Human Microbiome Project) or specific
geographic regions or built environments (MetaSub\cite{MetaSub16}).

In PTools desktop mode, some searches can be conducted against a single
PGDB only; other searches can be conducted  multiple PGDBs (see next section).

\SSsection{Cross-Organism Search}

It is also possible to search across all organisms within a
Pathway Tools web server, such as for the \NbiocycPGDBs\ organisms at
BioCyc.org. This is only implemented for the BioCyc website, those using
our software to instantiate their own website won't have this
available. The cross-organism search tool
\cite{BioCycCrossOrgSearchURL} searches for user-specified
combinations of words in the Common-Name/Synonyms attributes, and/or
the Summary attribute.  It can search all types of objects in a given
PGDB, or in user-specified object types, such as genes and/or pathways.
It can search all organism databases present in the Pathway Tools web
server, or it can search user-specified sets of organisms, such as all
organisms within a selected taxonomic group.  Indexing and searching
is implemented using SOLR \cite{SOLRURL}.

In desktop mode, many searches for
entities such as pathways, compounds, and genes can be run against a
collection of PGDBs, instead of just against a single PGDB.
The user can define and save a named collection of PGDBs, which will
become the subject of searches until changed.  For example, this
facility can be used to search multiple strains of the same species
for the presence of a specified gene, or to search all organisms
within a microbial community for the presence of a metabolite or a
pathway.  Searches return answer sets containing all objects that
match the search across all the PGDBs in the collection.

\SSsection{Structured Advanced Query Page}


The SAQP enables a biologist to construct precise structured
searches.  A query
can be as simple as looking up a gene given a name, or as complex as
searching several databases and several object types interconnected by
several relations. The SAQP enables biologists to formulate queries
whose power and expressiveness closely approach SQL, but without
having to learn SQL.  The SAQP translates a formulated query into
BioVelo, an SQL-like language \cite{88909}, before sending it to the web server.

\begin{figure*}
 \begin{center}

 \includegraphics[scale=0.4]{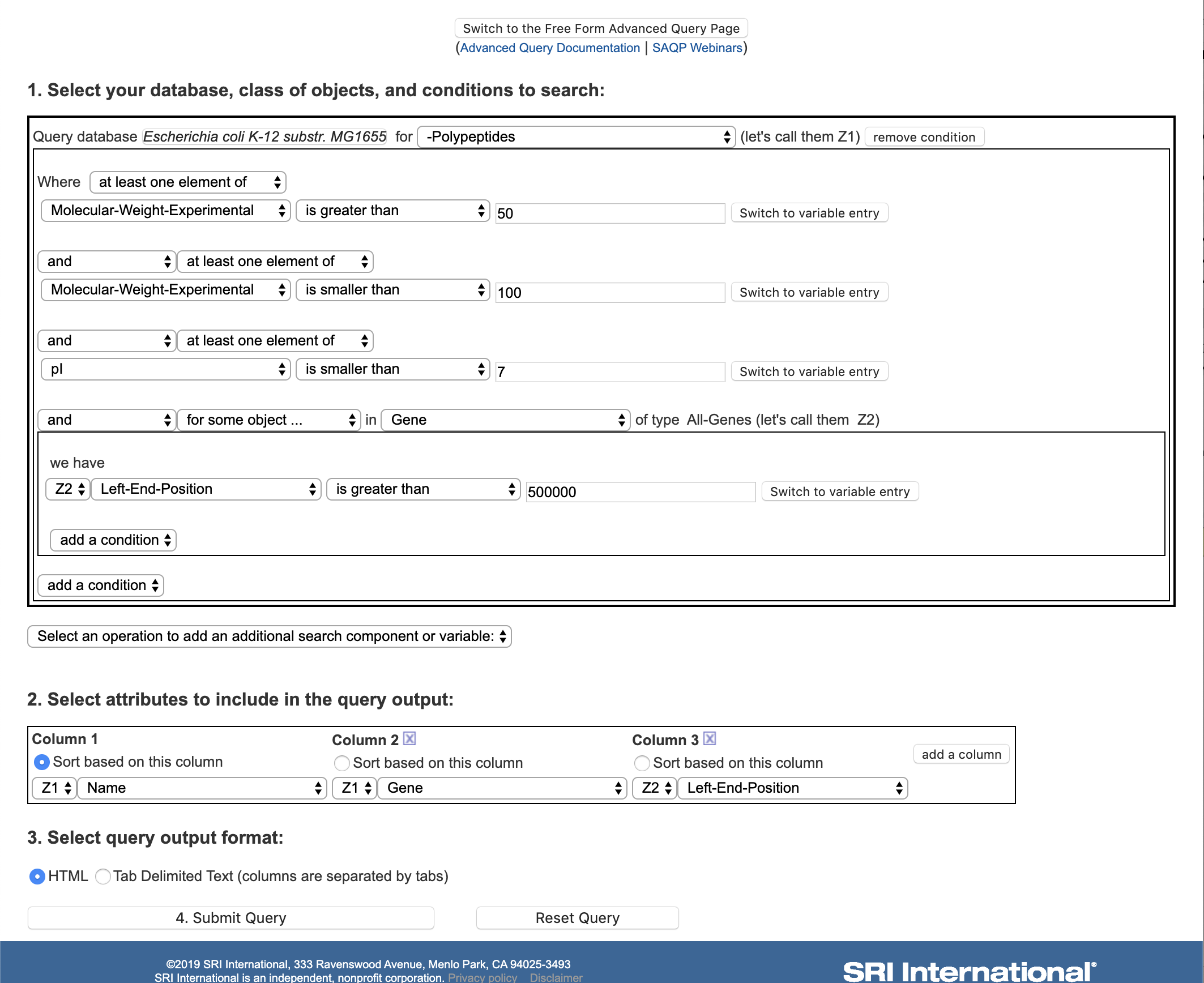}

 \end{center} 
\caption{\label{fig:polypeptides2SAQP} 
     A query for the \coli\ polypeptides whose experimental molecular
     weight lies between 50 and 100 kilodaltons, whose pI is smaller than 7, 
     and whose gene is located after
     the first 500 kb of the genome. An 
     output column is used to include the gene (or sometimes
     genes) producing each polypeptide using the second 
     variable~\texttt{Z2}.
     }
 \end{figure*}

The following explanation presents the elements of
this web user interface using the
example shown in Figure~\ref{fig:polypeptides2SAQP}, involving
a query against the class of protein monomers (Polypeptides) in the
EcoCyc DB.

{\bf \noindent Step 1: Select Database and Class.}
The first step in building a query is to specify at least one database
(DB) and the class of objects to search.

{\bf \noindent Step 2: Specify Conditions.}
Most queries include one or more conditions on the desired objects
within the class.  By clicking the button labeled \texttt{add a
condition} in the initial blank SAQP, a {\it where} clause is
added---visually boxed---in the search component.  This operation adds a
selector for an {\it attribute} (e.g., \verb+name+) of the objects and
a selector for a relational operator (e.g., {\tt contains the
substring}).  It also adds a {\it free text box} to enter a number or
string. Several other relational operators are provided, such as {\tt
is equal to}, and {\tt is a substring of}.
Regular expression matching is also available.
This new field forms an {\it atomic condition}.  Additional atomic
conditions can be added to the query by using the button labeled ``add
a condition''.

When selecting a relational operator, the
list of relational operators provided is compatible with the type of
the selected attribute.  In the case of the attribute {\tt name}, the
selectable operators are for strings, because the {\it type} of the
attribute {\tt name} is string.  The query in
Figure~\ref{fig:polypeptides2SAQP} has three atomic conditions
to filter the selected polypeptides.

Quantifiers on relations within the SAQP enable a join-like
capability.  For example, imagine that we want to extend the query
with an additional restriction that depends on the {\it gene encoding
the polypeptide,} not on the polypeptide itself.  
To do so, the user would add an \verb+and+ condition and then select
the \verb+gene+ attribute, which represents the gene encoding the
polypeptide. We then select
the quantifier operator {\tt for some object...}, meaning that we want
to define a condition that applies to some of the genes in the
\verb+gene+ attribute of this polypeptide (although in the majority of
cases only one gene will be present).

At this point, the SAQP adds a new indented query clause, to enable a
condition to be defined on the gene.  We have specified a constraint
that its nucleotide coordinate must lie after the first 500 kb of the
genome.  Because several attributes and logical connectors can be
specified in this new clause, forming a complex condition by itself,
the web interface draws a box around this condition and introduces it
with the {\tt we have} keyword. A new unique variable, named
\verb+Z2+, is also introduced. This variable represents every value of
the \verb+gene+ attribute.  

{\bf \noindent Step 3: Define Query Results.}
The section titled \texttt{Select attributes to include in the query
output} enables the user to describe the contents of the query results
by selecting the attributes to display for each result object.  The
result of a query is a table containing zero or more rows, one for each
query result.  Each column in the table is
a user-selected attribute.

\SSsection{Notification of Database Updates}

Users of the BioCyc website can register interest in sets of
genes, pathways, and/or Gene Ontology (GO) terms.  These interest
areas are specified using Pathway Tools ontologies, such as using the
pathway ontology to specify interest in cofactor biosynthesis, or
using Gene Ontology to specify interest in genes involved in the
cellular process Cell Adhesion.  When new information relevant to
their interest areas is curated in BioCyc, the user is notified by
email~\cite{Paley17}.  Email notifications are concise and targeted,
with brief descriptions of what in the user's interest area has
changed, and links to the updated BioCyc web pages.

\Ssection{Information Pages for Individual Biological Entities}

Pathway Tools provides information pages for a number of different
types of biological data types, such as genes, pathways, and
metabolites.  These pages summarize the information present
in a PGDB for that datatype.

{\bf Genes/Proteins/RNA Page:} The web version of the combined
Gene/Protein/RNA information page uses a tab-based interface.  Each
tab contains a subset of the extensive information previously listed
on one large gene page.  The tab-based redesign reduces the amount of
user scrolling required, makes the types of information available more
apparent, and speeds the loading of the initial gene page (the Summary
tab).  These tabbed sections display data in tables for easier reading
and faster loading.  The set of tabs available depend on the
information available for the gene and its product(s).  A ``Show All''
tab replicates the previous all-in-one display.  The full set of
possible tabs are: the initial summary tab; a listing of reactions
catalyzed by the gene product(s) and the pathways containing those
reactions; the GO terms associated with a gene; essentiality
information for the gene; protein features known for the gene; the
operon(s) containing the gene; the regulon (set of operons) controlled
by the gene if it is a regulator; the references for the gene.

The information available across the tabs is quite extensive,
including the map position of the gene on the chromosome, a graphical
depiction of the chromosomal region containing the gene, and available
gene-essentiality information.  The regulation-summary diagram
(see Supplemental Figure~21)
available on the summary tab integrates all known regulatory
influences on the gene and gene product into a single figure.  The
diagram summarizes regulation of transcription, regulation of
translation, post-translational regulation, and formation of
multimeric complexes and chemically modified forms of a protein.

Some types of proteins have additional information available, as follows.
Enzymes: The software displays the reaction catalyzed by the
enzyme and the name of the pathway that contains that reaction, if
any; the activators, inhibitors, and
cofactors required by the enzyme; and comments and citations for the
enzyme (See \href{http://www.ai.sri.com/pkarp/pubs/pt20suppfigs4.pdf}{Supplemental Figures}~3-4).  Transporters: The
software displays the transport
reaction catalyzed by the transporter.
Transcription factors: The software displays diagrams for all
operons controlled by the transcription factor (the regulon for the
transcription factor) (see \href{http://www.ai.sri.com/pkarp/pubs/pt20suppfigs4.pdf}{Supplemental Figure}~6).

{\bf Reaction Page:} Reaction pages apply to metabolic, transport, and
protein signaling reactions (see
\href{http://www.ai.sri.com/pkarp/pubs/pt20suppfigs4.pdf}{Supplemental
  Figures}~2 and 5).  The reaction page shows the one or more enzymes
that catalyze the reaction, the gene(s) that code for the enzymes, and
the pathway(s) that contain the reaction.  The display shows the EC
number for the reaction and the reaction equation.  \blue{Like the
  gene/protein page, the web version of the page includes multiple
  tabs. The summary tab shows the reaction equation with chemical
  structures (for a metabolic reaction) or compartments (for a
  transport reaction).} If atom-mapping data is available, the user
can display mappings for each of the atoms in a reaction using
color-coding (whereby conserved chemical moieties in the reaction
substrates are drawn in different colors) and/or numbering. If there
is more than one possible atom-mapping for a given reaction, the user
has the option to show them all. \blue{Other tabs include a more
  detailed description of the reaction enzymes (previously accessible
  only from the reactions tab of the enzyme pages), the ontology
  viewer (see below), and the references for the reaction and its
  enzymes.}

{\bf Pathway Pages:} All metabolic pathway diagrams are computed automatically
using pathway-layout algorithms.  Pathway Tools can draw metabolic pathways at
multiple levels of detail, ranging from a skeletal view of a pathway
that depicts the compounds only at the periphery of the pathway and at
internal branch points, to a detailed view that shows full structures
for every compound, and EC numbers, enzyme names, and gene names associated
each reaction.
Depending on the graph structure of
the pathway, it can be drawn using a linear layout
(see \href{http://www.ai.sri.com/pkarp/pubs/pt20suppfigs4.pdf}{Supplemental Figure}~22 --- 
this pathway diagram is drawn at the detail level that includes
chemical structures), 
tree layout (see \href{http://www.ai.sri.com/pkarp/pubs/pt20suppfigs4.pdf}{Supplemental Figure}~23),
or circular layout (see
\href{http://www.ai.sri.com/pkarp/pubs/pt20suppfigs4.pdf}{Supplemental
  Figure}~24 --- this diagram shows only the backbone metabolites of
the pathway, and includes links to other related pathways).
Pathway diagrams can also include cellular compartments
(see
\href{http://www.ai.sri.com/pkarp/pubs/pt20suppfigs4.pdf}{Supplemental
  Figure}~25).

\begin{figure}
\begin{center}
\includegraphics[width=7in]{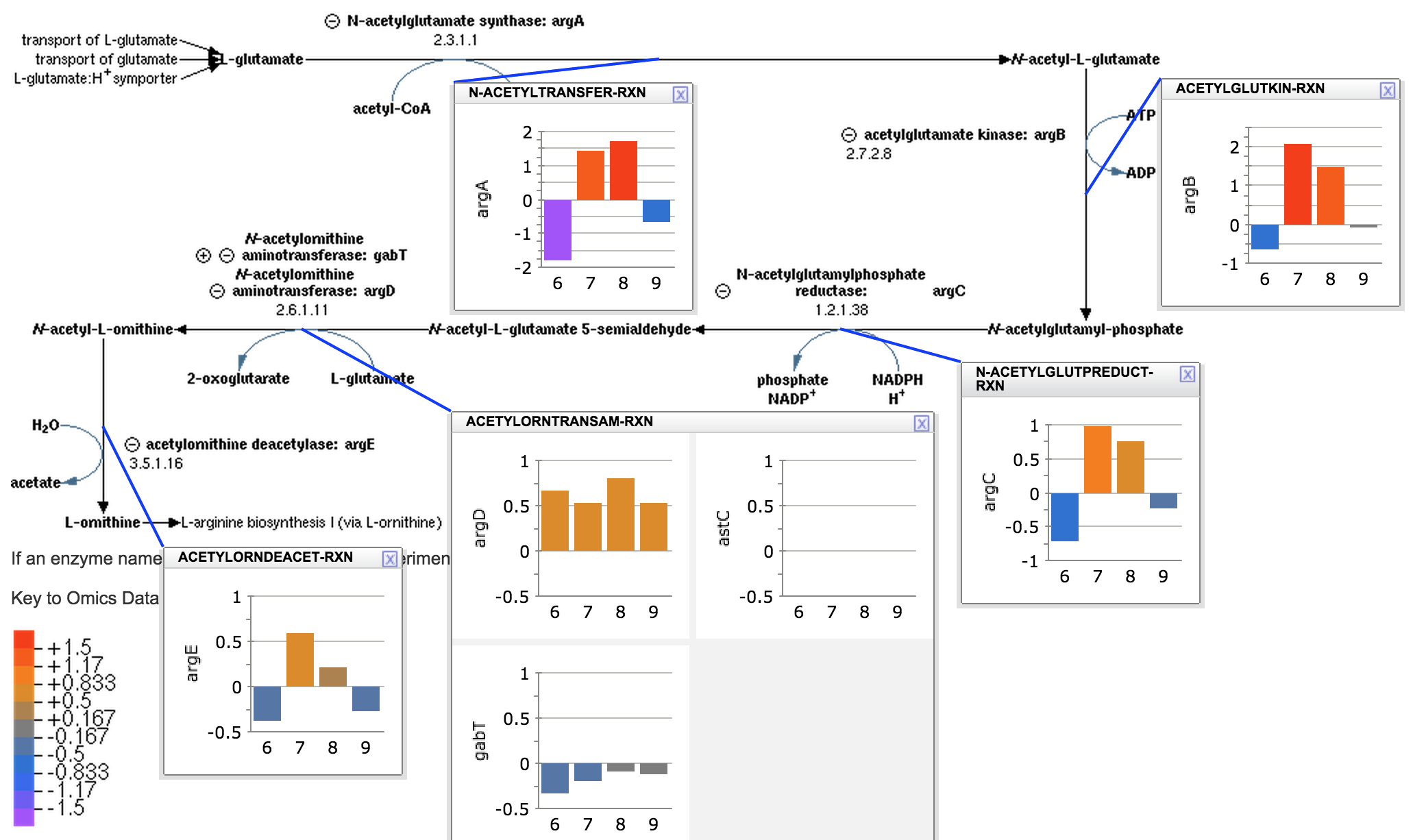}
\caption{The EcoCyc L-ornithine biosynthesis pathway shown with
  omics pop-ups containing time-series data from a gene-expression experiment.}
\label{fig:pathomics}
\end{center}
\end{figure}

\begin{figure}
\begin{center}
\includegraphics[width=3in]{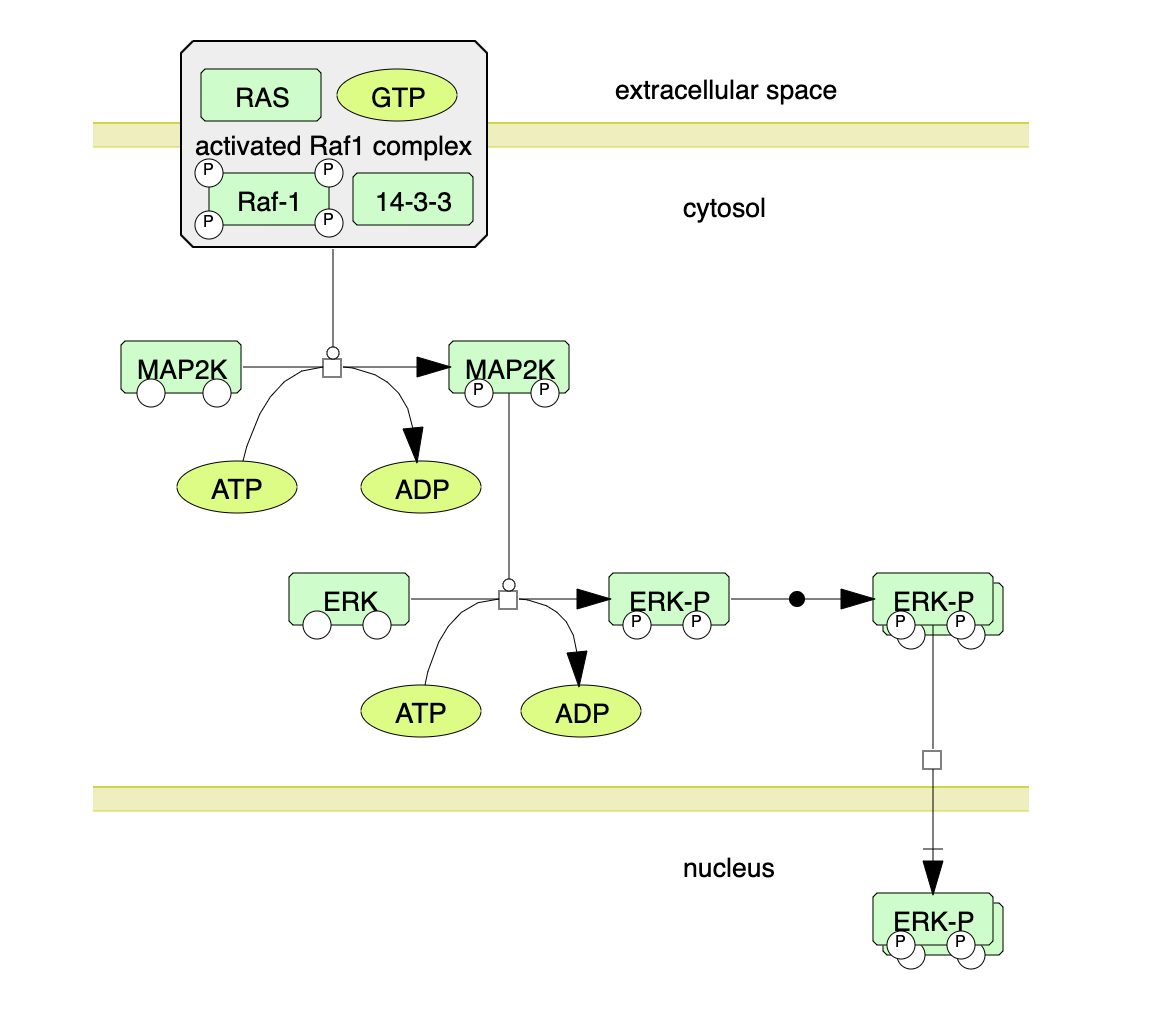}
\caption{The HumanCyc MAP kinase cascade, a signaling pathway diagram.}
\label{fig:sigpath}
\end{center}
\end{figure}

Pathway diagrams can also be shown at a detail level that includes the
transcriptional, translational and substrate-level regulators for
each step.  The user can customize a pathway drawing to
include desired elements only, and to include superimposed omics data
using ``omics pop-ups'' (see Figure~\ref{fig:pathomics}).  Display of signaling
pathways (see Figure~\ref{fig:sigpath}) is also supported; signaling pathway layout is performed by
the pathway curator.

\blue{The page summary tab includes the pathway diagram and a textual
  description, if available. Other tabs show how the pathway fits into
  the pathway ontology, how the genes of the pathway are organized
  into operons (for single-organism databases only), and a list of
  references for the pathway and all its enzymes and genes.}

\begin{figure}
\begin{center}
\includegraphics[width=5.5in]{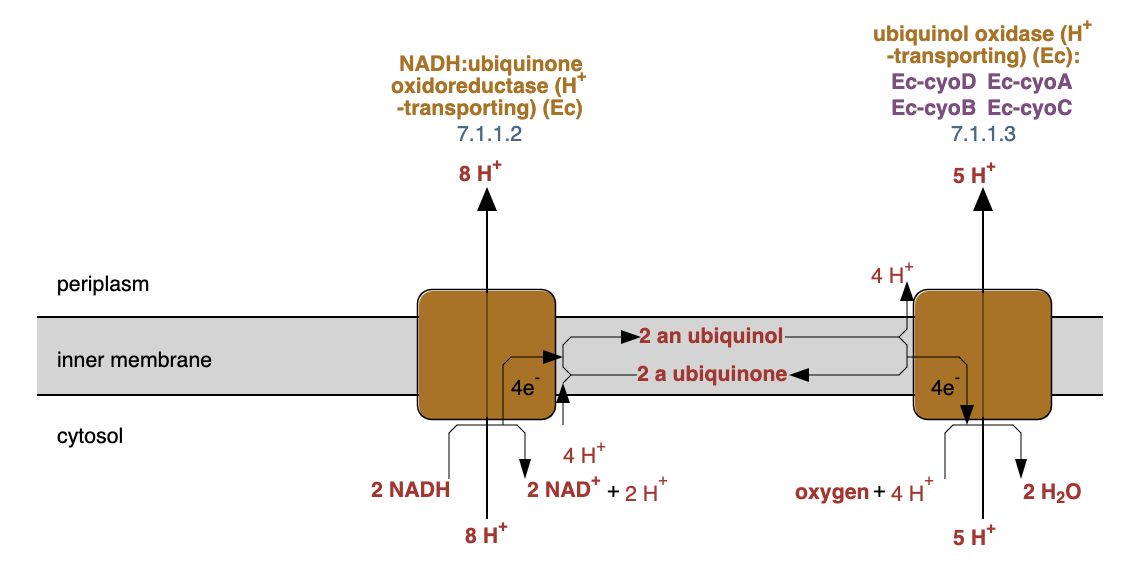}
\caption{An electron transfer pathway diagram.}
\label{fig:etr-pwy}
\end{center}
\end{figure}

{\bf Electron Transfer Reactions and Pathways:} A crucial role in
cellular metabolism is played by electron transfer reactions (ETRs),
which are of key importance in the energy household of a cell. In a
series of redox steps, the high-energy electrons from some
compounds drive the pumping of protons across a cell membrane, to
maintain the proton motive force needed for ATP synthesis.  Because
electrons do not freely exist on their own, these redox reactions seem
less intuitive than other small molecule reactions.  We designed and
implemented drawing code for a special ETR diagram, which shows the
enzyme complex embedded in a membrane, and which schematically depicts
the flow of electrons from one redox half reaction to another.  Inside the
membrane, the quinone/quinol cofactor is shown together with an
indication of the cell compartments that are sources or sinks of the
protons.  An additional directional proton transport reaction can be
added to the diagram.  This results in displaying the flow of all
substrates and products relative to the cellular compartments, in a
similar way to what is customary in the biomedical literature.
Pathways consisting of several ETRs joined together can also be
depicted (see Figure~\ref{fig:etr-pwy}).

{\bf Metabolite Pages:} A metabolite page shows the chemical
structure for the compound (see \href{http://www.ai.sri.com/pkarp/pubs/pt20suppfigs4.pdf}{Supplemental Figure}~8).  It lists all reactions in which the
compound appears, and it lists enzymes whose activity is regulated by
the compound.  These pages now provide a tab-based interface similar
to that of gene/protein pages.

{\bf Transcription Unit Pages:} The display window for transcription units
diagrams the transcription unit and its regulatory sites
including promoters, transcription-factor binding sites, attenuators,
and binding sites for proteins and RNAs that regulate its translation.
The display contains sections describing each site within the
transcription unit.  The promoter section describes which sigma factor
recognizes it.  Sections for a transcription factor binding site
describe its sequence, which transcription factor it binds, ligands that influence
the activity of the transcription factor, and whether the effect of
binding is to activate or inhibit transcription initiation.  Sections
for attenuators describe the signal that the attenuator senses, and
show the sequence regions that form the attenuator.

\blue{{\bf Ontology Viewer:} Pathway, reaction and metabolite pages
  all include an Ontology tab, which contains an interactive ontology
  viewer (see
  \href{http://www.ai.sri.com/pkarp/pubs/pt20suppfigs4.pdf}{Supplemental
    Figure}~28). The ontology viewer is also included in certain other
  page types, including pages for ontology classes and GO terms. A
  graph shows the current object, plus all its parents (and, if it's a
  class, its children) in the appropriate Pathway Tools
  ontology. Classes with other children can be expanded to support
  interactive exploration of the ontology.}

\Ssection{SmartTables: Large-Scale Manipulation of PGDB Object Groups}
\label{sec:smarttables}

SmartTables enable
  users to construct and manipulate groups of PGDB objects through a
  spreadsheet-like user interface \cite{KarpGroups13} (SmartTables
  were previously called Web Groups).  SmartTables provide many
  powerful operations to biologist end users that previously would
  have required assistance from a programmer, and our user surveys
  indicated that SmartTables are reasonably easy for biologists to use
  \cite{KarpGroups13}. To date, more than 10,000 users of BioCyc.org have
  created more than 80,000 SmartTables.

A typical SmartTables use case is for a user to define a SmartTable by
importing a list of PGDB objects from a file.  For example, a user
could define a metabolite SmartTable by importing a list of
metabolites from a metabolomics experiment, where the metabolites are
specified by metabolite name, BioCyc identifier, PubChem identifier,
or KEGG identifier.  (The set of objects in a SmartTable can also be
defined from a query result, from any column of an existing
SmartTable, or from the set of, say, all genes in a
PGDB.)

\begin{figure}
\begin{center}
\includegraphics[width=7in]{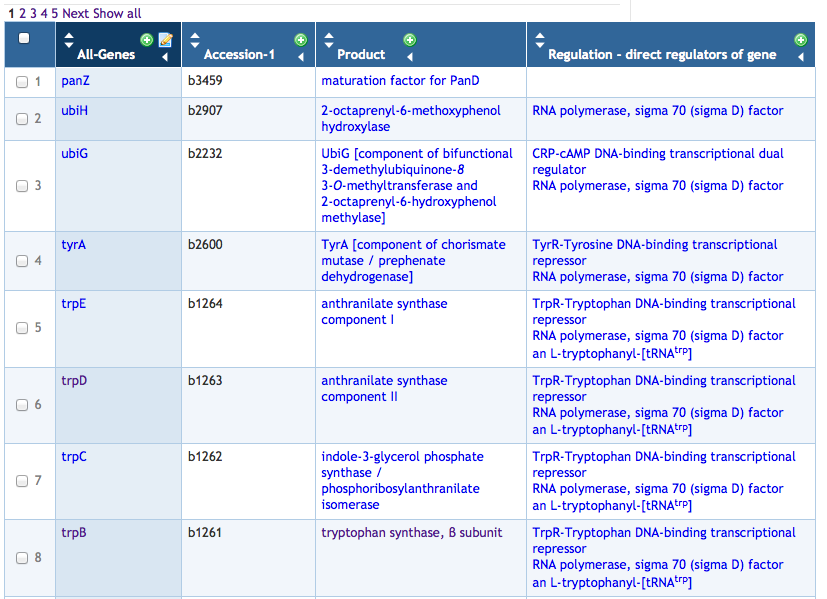}
\caption{A gene SmartTable.  Column 1 shows the gene name, column 2 shows the \coli\ genome
  ``b-number'' accession number for the gene (a property); column 3 shows the gene
  product name (a property).  Column 4 shows the result of a
  transformation in which the regulator(s) of each gene were computed. }
\label{fig:smt}
\end{center}
\end{figure}

The user can browse the set of objects in a SmartTable by paging
  through the table, and can modify the information displayed about
  each object by specifying which table columns to include (see
  Figure~\ref{fig:smt}).  SmartTable columns are derived from the PGDB
  attributes available for each object, and can include information such
  as chemical structures, molecular weights, links to other databases,
  and nucleotide and protein sequence.  A variety of filters and set
  manipulations are provided for SmartTables, such as removing or
  retaining all rows that match a user query; and computing the union,
  intersection, and set difference of two SmartTables.  SmartTables
  are stored in the user's online web account, and a desktop version
  of SmartTables is also provided.  SmartTables are private by
  default, but the user can make them public, share SmartTables with
  selected other users, or archive them in a frozen form in
  conjunction with a publication.

The underlying representation of SmartTables enables them to be
    dynamic, so that if information represented in the SmartTable
    changes (as it might with a new release of a PGDB reflecting newly
    acquired information), those changes will be reflected in the
    SmartTable when it is next examined. For example, if a SmartTable
    is created which includes a column of genes and a column of
    pathways associated with those genes, a newer PGDB might introduce
  new pathways, and the SmartTable would reflect any additional, new
  pathways associated with each gene when next examined. Another
  example, a SmartTable may have
  a column of compounds and an associated column showing their
  structure. If a newer version of the PGDB introduces a different
  structure for a given compound, that would be reflected the next time
  the SmartTable is viewed.

Several more advanced SmartTable operations are provided.  {\em
  Transformations} compute new columns from relationships in a PGDB.
For example, column~4 in Figure~\ref{fig:smt} is a transformation
column that shows one or more regulators for each gene in column~1
that have been computed from PGDB relationships.  Other gene
transformations available include computing the metabolic pathways
in which a gene's product occurs and computing the amino-acid changes
caused by sequence variants.
Different transformations are available for different data types.
For example, the transformations available for a metabolite SmartTable
include computing the reactions in which a metabolite occurs, the
pathways in which a metabolite occurs, the proteins for which the
metabolite is a ligand, and mapping the compounds to their equivalents
in another PGDB.

A user can perform a statistical enrichment analysis on a gene or
metabolite SmartTable to detect over-represented metabolic pathways or
GO terms, or over-represented metabolic pathways, respectively; see
Section~\ref{sec:visenrich} for more details.
In addition, a SmartTable of genes or metabolites can be visualized
on the cellular overview.

 Exporting a SmartTable as a tab-delimited file, for use with a
spreadsheet program, is supported. We also support exporting
FASTA-format files from SmartTables that include sequence information.

 SmartTables can also be used to assist in navigation: a user can put
 the results of a search into a SmartTable, and then navigate the
 objects in a given column from one to the next (without having to
 return to the SmartTable at each navigational step).

 SmartTables are also used for doing sophisticated queries of
 PGDB's. A simple example: create a SmartTable whose initial column
 contains all the genes matching some search criteria; using a
 transformation to show the protein associated with each gene in a new column; using
 another transformation to show the pathways associated with each protein.

\Ssection{System-Level Visualization of Metabolic Networks}
\label{sec:cellov}

Pathway Tools can automatically generate organism-specific metabolic
charts that we call Cellular Overview diagrams
\cite{PToolsOverview06}.  The  Cellular Overview contains all known metabolic
pathways and transporters of an organism (online example: \cite{OverviewURL};
click for example with animated display of omics data:
\cite{OverviewAnimationExampleURL}).  Each node in the diagram represents a
single metabolite, and each line represents a single bioreaction.
The diagram is organized with biosynthetic
pathways on the left, catabolic pathways on the right, energy
metabolism pathways in the middle, and individual reactions not
assigned to any pathway on the far right.  Transport events are
shown in the cell membrane.  Each dot depicts a metabolite and each
line depicts a metabolic reaction.  Reactions are drawn within
metabolic pathways, and pathways are organized into related groups,
such as the group of all cofactor-biosynthesis pathways.
In 2019 the diagram was re-engineered for the web mode of Pathway Tools
so that the diagram can be zoomed in real time within a web browser,
and so that semantic zooming draws additional detail as the diagram is magnified
from the base level to a high level of magnification (see \href{http://www.ai.sri.com/pkarp/pubs/pt20suppfigs4.pdf}{Supplemental Figures}~17--19).

The diagram can be interrogated interactively using an extensive set
of search operations, such as to find an enzyme or metabolite by name,
using a menu in the website sidebar to the right of the diagram.  The
diagram can be generated as a PDF file for printing as a large-format
poster.  Example posters can be downloaded from
\cite{BioCycPostersURL}.

\begin{figure}
\begin{center}
\includegraphics[width=7in]{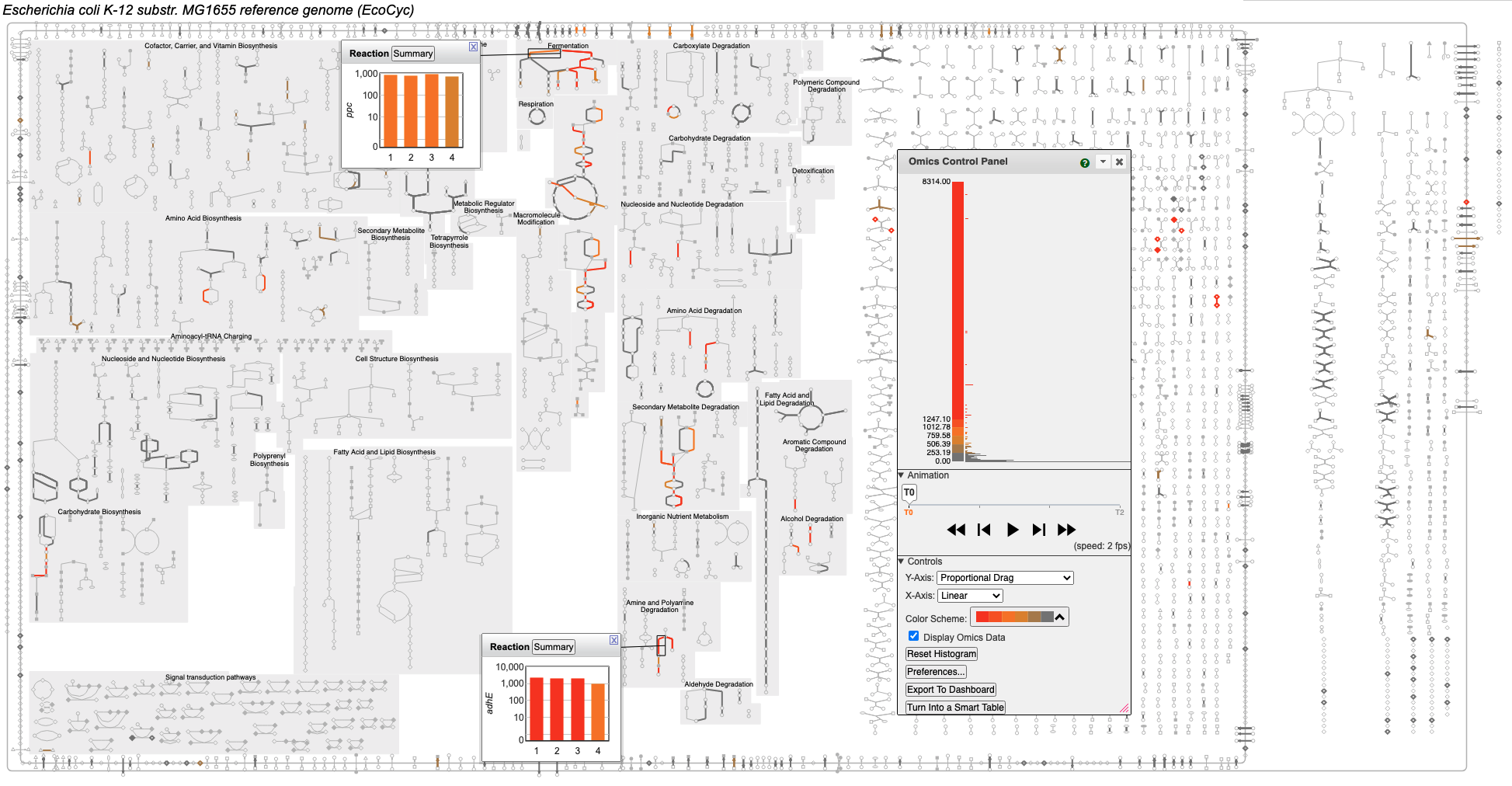}
\caption{The Pathway Tools Cellular Overview diagram for EcoCyc,
painted with a gene-expression dataset measured during shift 
from anaerobic to aerobic growth \cite{vonWulffen17}. Two omics pop-ups are shown that plot
the expression levels of two genes.  The Omics Control panel shows
the mapping of data values to the color scale; users can manipulate
the sliding y-axis labels to dynamically alter the color scale.
\label{fig:cellov}
}
\end{center}
\end{figure}

Omics data (e.g., gene-expression or metabolomics measurements) for a
given organism can be painted onto the cellular overview, as described
in Section~\ref{sec:cellov-omics}.
Figure~\ref{fig:cellov} depicts the Web Cellular Overview at low resolution
painted with gene-expression data.  

Cellular Overview diagrams are generated automatically using an
advanced layout algorithm \cite{PToolsOverview06}.  Automated layout
is essential to enable the diagram to accurately depict the underlying
database content as that content evolves, without requiring
time-consuming manual updates by curators that are bound to overlook
some content changes.  In addition, automated layout enables generation of
organism-specific cellular overviews that reflect the exact pathway
content of each organism-specific PGDB in large PGDB collections such
as BioCyc.

\begin{figure}
\begin{center}
\includegraphics[width=7in]{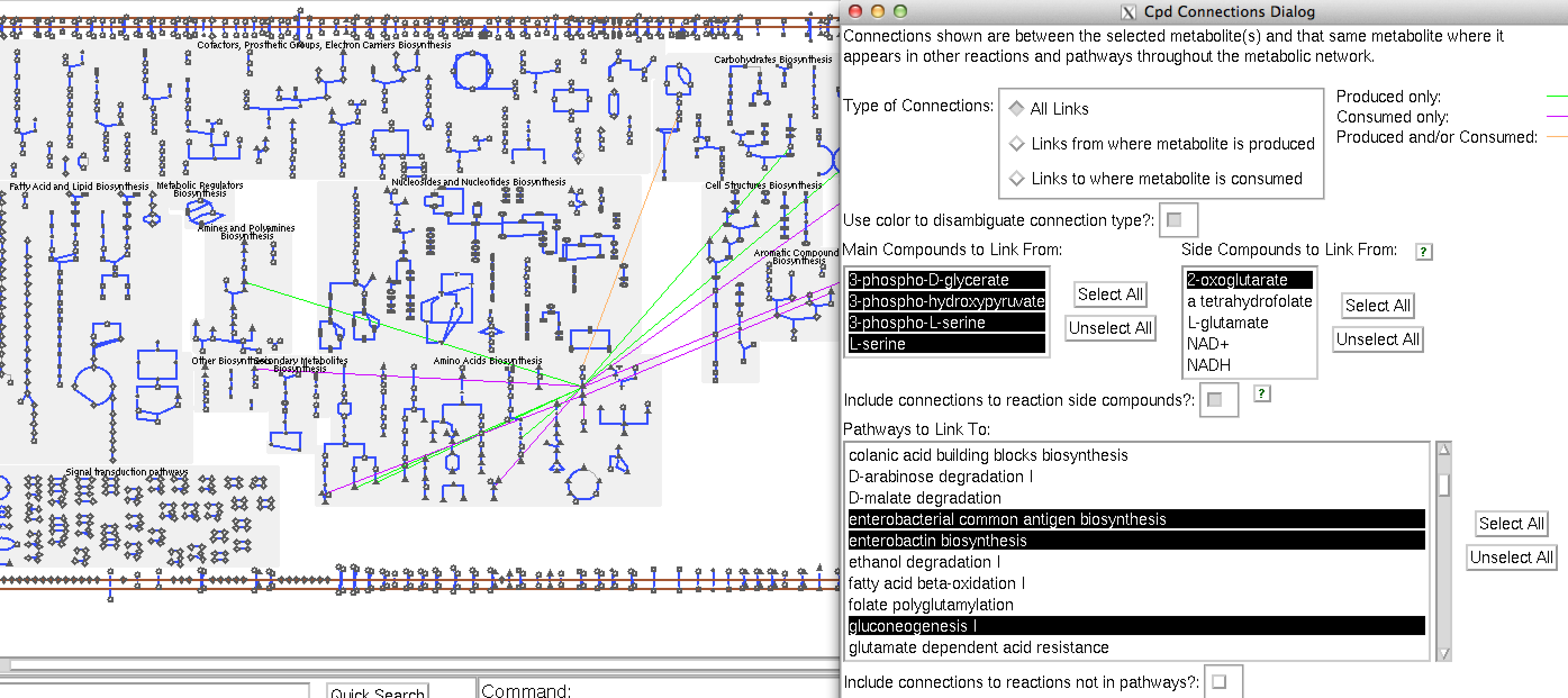}
\caption{The Pathway Tools Cellular Overview diagram for EcoCyc,
  showing connections between metabolites of the serine biosynthesis
  pathway and selected other pathways.}
\label{fig:connections}
\end{center}
\end{figure}

The  Cellular Overview  has many capabilities (described in more
detail in \cite{PToolsOverview06}), including semantic zooming
of the diagram (where the highest magnification corresponds to the
detail shown in the poster version); highlighting of user-requested
elements of the diagram (such as metabolites or pathways);
highlighting large, biologically relevant subnetworks (such as all
reactions regulated by a given transcription factor); and
highlighting comparative analysis results, such as comparison of the 
metabolic networks of two or more  PGDBs.  Because a metabolite
can appear in several different places in the diagram, in
desktop mode, the user can better visualize the flow of material
through the metabolic network by selectively showing connections
between a metabolite of interest, or all the metabolites in a pathway
of interest, and everywhere else those metabolites appear in the
network, as shown in Figure~\ref{fig:connections}.

Cellular Overview diagrams from multiple organisms can be combined to
depict metabolic cross-feeding relationships.  Shared or unique
pathways may be highlighted (see \href{http://www.ai.sri.com/pkarp/pubs/pt20suppfigs4.pdf}{Supplemental Figure}~20).

\Ssection{System-Level Visualization of Regulatory Networks}
\label{sec:regov}

\begin{figure}
\begin{center}
\includegraphics[width=7in]{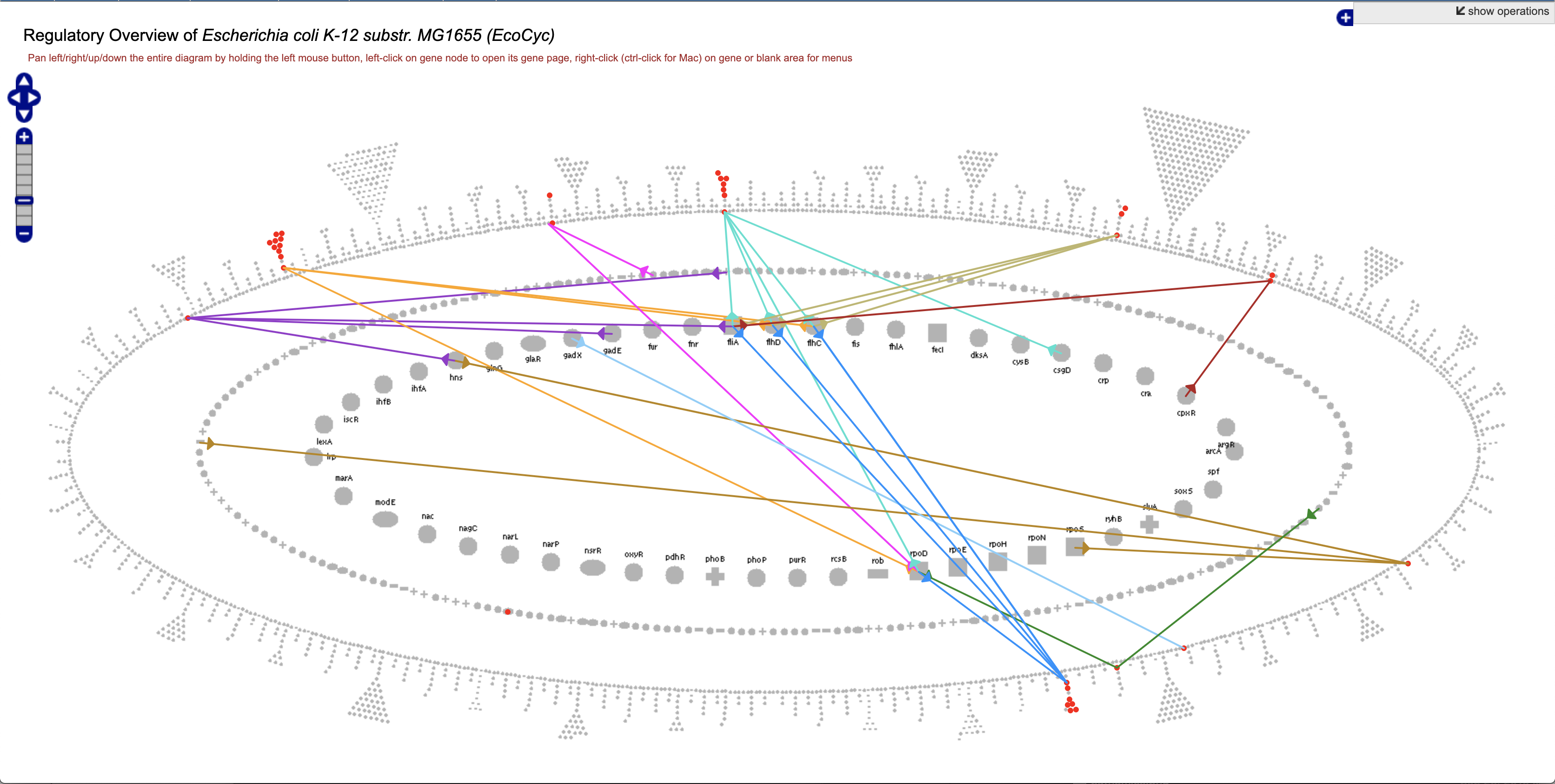}
\caption{The Pathway Tools Regulatory Overview diagram for EcoCyc.
The diagram depicts a full regulatory network as three concentric
rings: the inner ring contains master regulator genes; the middle
ring contains other regulators; the outer ring contains genes that are
not regulators.  An arrow (edge) from gene A to gene B indicates that
gene A regulates gene B.  Initially, no arrows are shown; the user can
interactively add arrows, such as by clicking on a gene and requesting
that arrows are added to genes that it regulates, or from the genes
that regulate it.}
\label{fig:regov}
\end{center}
\end{figure}

The Pathway Tools Regulatory Overview depicts the full transcriptional
regulatory network stored in a PGDB in one screen, and enables the user
to interrogate and explore relationships within the network.
Figure~\ref{fig:regov} shows the Regulatory Overview for EcoCyc, after
the user has asked the system to highlight all genes annotated under
Gene Ontology term GO:0001539 (ciliary or flagellar motility).  We can
see that a few transcription factors control all \coli\
motility genes.  

\Ssection{System-Level Visualization of Genome Maps}
\label{sec:genov}

Pathway Tools includes two different genome browsers: a linear and a
circular genome browser.  The linear browser (named Genome Explorer)
is designed for fluidly zooming
in and out of genome regions to reveal new features, whereas the
circular browser is designed to depict different genome features at
a narrower range of magnifications.

\SSsection{Genome Explorer}

\blue{The web mode Genome Explorer was designed for rapid zooming across the
full scale at which a genome tends to be explored.  In the basic mode,
one replicon (chromosome or plasmid) is selected, yielding a compact,
information dense display, in which the screen real estate is
maximally exploited by showing a section from the replicon, as
multiple line-wrapped horizontal lines at once.  Furthermore, the
functional elements like genes, operons, transcription start sites,
transcription-factor binding sites, and terminators are displayed
in-line on the same line, instead of on many parallel tracks that are
spatially separated from each other, which is what most other genome
browsers do.
}

\blue{Useful semantic zooming levels are provided for covering the large
scale from showing the entire replicon on the screen, all the way down to
the sequence. New graphical features are successively revealed at
different levels, such as gene names, terminators, and binding sites.  
In addition to mouse based navigation, name or coordinate based search
will find specific genes and regions.
}

\begin{figure}
\begin{center}
\includegraphics[width=7in]{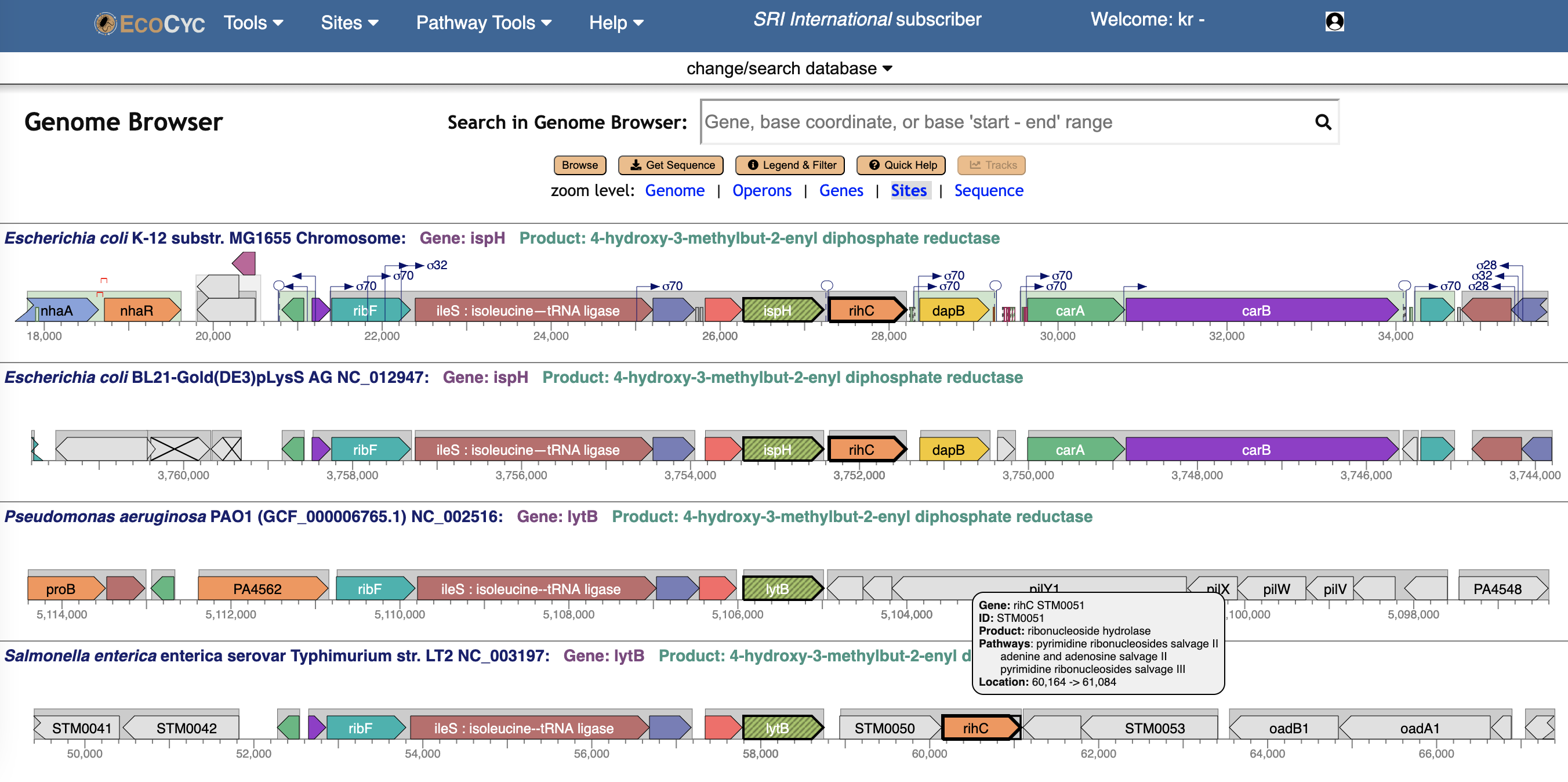}
\caption{Pathway Tools Genome Explorer (web version) in comparative mode.
This screenshot shows 4 genomes aligned and centered by the lead gene
ispH. When mousing over a different gene (rihC in this figure), its
visible orthologs are highlighted by thicker black lines, and an
informative tooltip is popped up.}
\label{fig:comparative-genbro}
\end{center}
\end{figure}

\blue{The comparative mode enables side by side comparisons of many genomes,
to visualize differences in the conservation of genes and other
features.  The genomes are aligned by orthology, starting with the
so-called ``lead gene'' of the lead organism. After selecting other
organisms to compare with, their genes, which are orthologous to the
lead, will be centered in the genome sections of the other organisms,
drawn as horizontal lines, as shown in Figure~\ref{fig:comparative-genbro}.  Genes with the same
colors indicate that they are orthologous to each other.  The caveat
is that only a dozen colors are available, so eventually colors will
get reused for other ortholog pairings.  Navigation like zooming and
translation to the left and right are available.
}

\blue{The tracks mode enables the display of positional data from uploaded, external
datasets. The datapoints are aligned to the genes and other features
of one replicon.  The start and end coordinates of one data point are
aligned underneath the corresponding region of the replicon.  The data
can be shown in 3 different display styles: as a point graph, solid
bar graph, or as simple rectangles.  Intensity values of the data are
converted into colors.
}

\begin{figure}
\begin{center}
\includegraphics[width=7in]{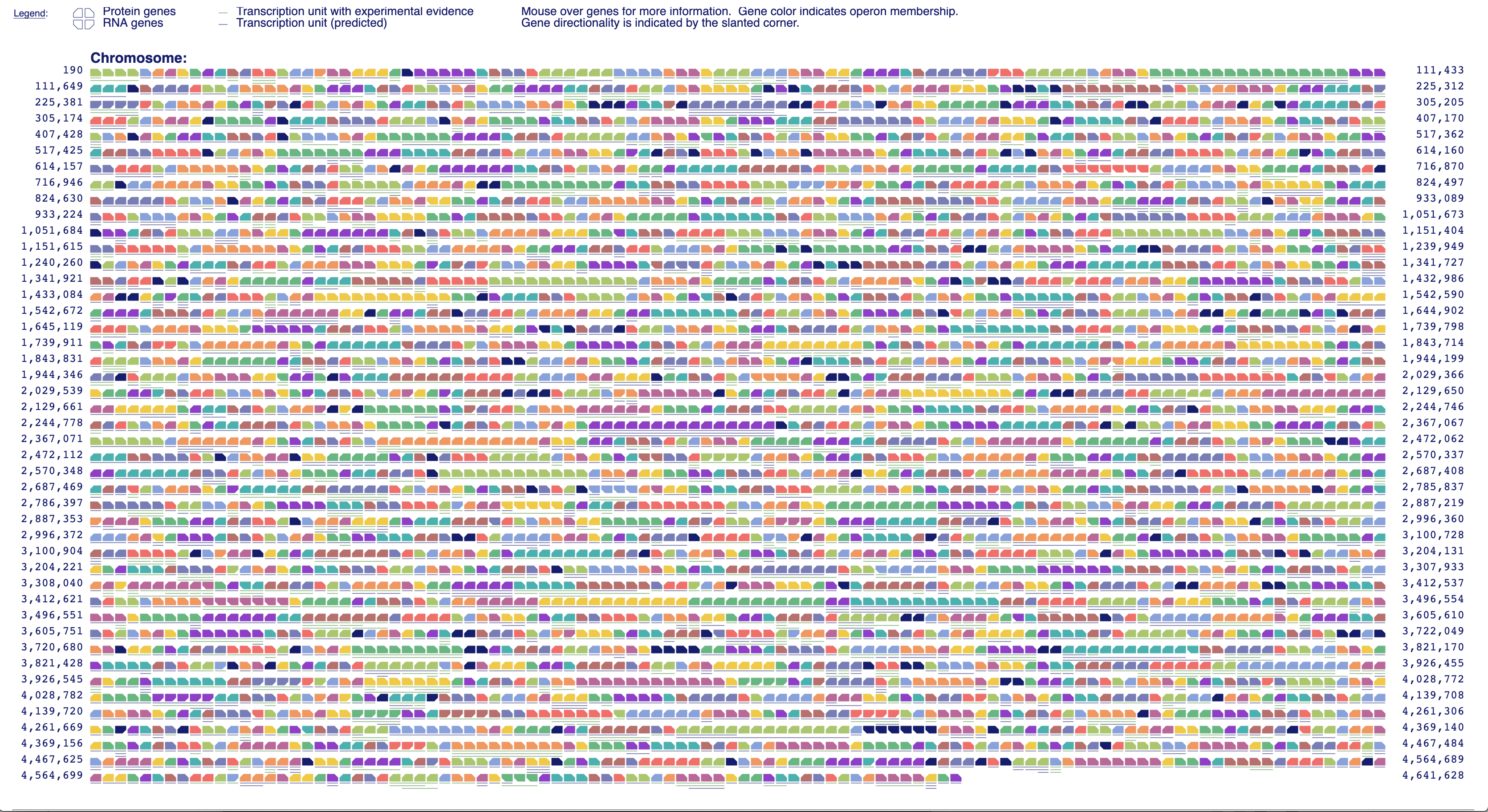}
\caption{Pathway Tools Genome Overview diagram for EcoCyc.
In this diagram genes are not shown to scale.
Adjacent genes drawn in the same color are in the same operon.
Left/right gene direction indicates transcription direction; up/down
gene direction indicates genes coding for proteins versus RNAs.
Horizontal lines under genes indicate transcript extents based on
promoter and terminator information in the PGDB.}
\label{fig:genov}
\end{center}
\end{figure}

\blue{The genome browser can also generate large-format genome posters
  in PDF format; see examples at \cite{BioCycPostersURL}.  The Genome
  Overview for the currently selected organism can be shown by the
  corresponding command in the Tools menu bar. It is a depiction of all the
  genes of the genome, as shown in Figure~\ref{fig:genov}.  This
  diagram can be painted with omics data to provide a global genome
  view of large-scale data sets.}

\SSsection{Circular Genome Browser}

An interactive circular genome display consists of a set of
user-specified concentric circles (tracks) showing features of
interest from one or more circular replicons. A single track shows
elements of a single feature class, such as genes, promoters,
pseudogenes, DNA binding sites, REP elements and others.  Some tracks
can be filtered and/or highlighted based on user-specified
criteria. For example, a gene track can be filtered to only show genes
on one strand or the other, genes in a particular metabolic pathway
class, RNA-coding genes, genes annotated to a specified GO term, genes
regulated by a specified transcription factor, or genes whose names or
whose product names match a specified substring. Alternatively, a
track can include all genes, and then different subsets of genes can
be highlighted in different colors according to one or more of the
above criteria. Users also have the option to upload a file of genes
of interest to filter or highlight a track, \blue{or to generate a
  track from a file of GFF features}. An example circular genome
display with tracks for multiple feature types, and multiple gene
tracks with different sets of highlights is available as
\href{http://www.ai.sri.com/pkarp/pubs/pt20suppfigs4.pdf}{Supplemental
  Figure}~26.

The circular genome viewer can be used to visually compare multiple
related strains. When tracks from multiple strains are included, an
additional highlighting option colors genes that have orthologs
between the selected organisms (See
\href{http://www.ai.sri.com/pkarp/pubs/pt20suppfigs4.pdf}{Supplemental
  Figure}~27). In addition, a highlight that is applied to genes in
one track can also be simultaneously applied to the orthologs of those
genes in other tracks.

\Ssection{Sequence-Based Query and Visualization Tools}
\label{sec:sequencetools}

The following tools support query and visualization of sequence data
from a Pathway Tools web server:

\bitem
\item Nucleotide Sequence Viewer: Gene pages include links to view or
  download the nucleotide or RNA/protein sequence for the gene. When
  viewing the nucleotide sequence, an option is provided to
  include an additional upstream and/or downstream flanking region of
  any desired length.  This option makes it easy to, for example, view
  the sequence of a regulatory region surrounding a gene of interest.
  Alternatively, the user can enter specific start and end coordinates
  and the desired strand to view the sequence of any arbitrary portion
  of the chromosome.

\item BLAST Search: Users can use BLAST to search for all
    sequences within either a single specified genome, \blue{a
      selected set of genomes,} or against all BioCyc
    PGDBs that match a query
    sequence.

\item Sequence Pattern Search (PatMatch): The PatMatch facility
  enables searching within a single genome for all occurrences of a
  specified short nucleotide or peptide sequence (less than about 20
  residues), with the ability to specify degenerate positions.  The
  user can specify the kind and number of allowable mismatches, and
  whether to search coding regions only, intergenic regions, or the
  entire genome.  Examples of situations in which this facility might
  be useful include searching for all occurrences of a particular
  regulatory motif upstream of any gene, or all occurrences of a known
  cofactor binding motif within proteins.

\item Multiple Sequence Alignment: From a gene page, users can request
  a multiple sequence alignment between the nucleotide or amino acid
  sequences of that gene and its orthologs in a user-specified set of
  organisms.  Alignments are displayed using 
  \blue{MSAViewer~\cite{Yachdav16}. Nucleotide sequence alignments can optionally
    include specified upstream and/or downstream offsets. Alignments
    can also be generated for an arbitrary set of genes or sequence
    regions from one or more organisms, and can include sequences
    uploaded by the user.}

\eitem

\Ssection{Pathway Collages}
\label{sec:collages}

The Pathway Collages~\cite{Collages16} facility enables the user to
create multi-pathway diagrams that can be personalized, annotated,
printed, and shared.  Usually the pathways within a collage are of
interest in some biological situation, or are interconnected via
shared metabolites.  The user can specify the pathways of interest in
several ways; the pathways are then exported to the Pathway Collage web
application.  Although the web application requires a web browser,
sets of pathways can be created and exported from the desktop version
of Pathway Tools, which will launch a browser after the export is
complete.

In the web version of Pathway Tools, pathway sets for a collage can be specified
by including a column of pathways in a SmartTable, from data uploaded
into the Cellular Overview, by creating a collage with a single pathway from the
pathway's page, or by selecting pathways from a checklist on the page describing
pathway collages.  In the desktop version of Pathway Tools pathways can be
selected from a SmartTable or from a select pathways tool available in the
Cellular Overview.

\begin{figure*}[!tpb]
\centerline{\includegraphics[width=8cm]{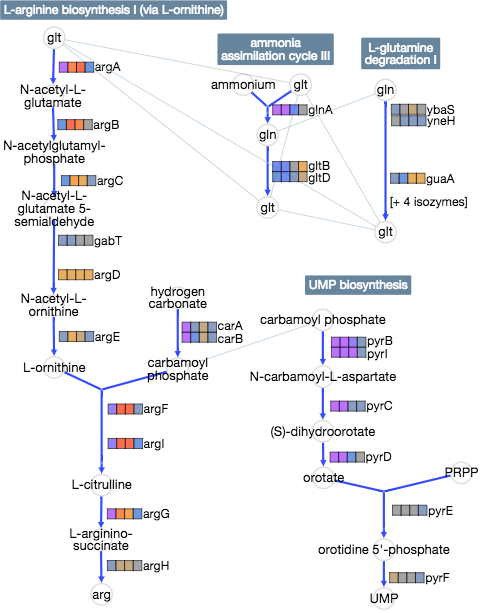}}
\caption{An example pathway collage consisting of four
  pathways. Connections have been drawn between different occurrences
  of the same metabolite.  The collage includes data from a
  four-time-point transcriptomics experiment that is shown in the
  colored grid to the right of each reaction. }
\label{fig:pwycollage}
\end{figure*}

Within the Pathway Collage tool, the user initially sees a
  high-level drawing of the pathways they selected; they can then
  customize the diagram in several ways. The user can zoom
  in (and show more detail), move a pathway by dragging it, and
  perform editing operations (changing drawing properties or labels,
  deleting) on compounds or reactions within the pathway.  Compound
  editing also includes the option to edit all occurrences of the
  compound within the diagram or import other pathways that involve
  the compound.  Compounds can be connected with links, either linking
  all occurrences of a compound, or linking pairs of the users
  choosing.  Alternatively, multiple occurrences of the same compound
  can be merged.  Omics data can also be added to a collage while in
  the collage editor.  An example pathway collage is shown in
  Figure~\ref{fig:pwycollage}.

Finally, the collage can be saved for further editing or
  sharing, or it can be exported as a PNG or SVG file.

\section{Omics Data Analysis}
\label{sec:omics}

Pathway Tools offers a number of tools for analysis of gene
expression, proteomics and metabolomics data.  These include both
visualization tools and analytical tools such as enrichment analysis.

For most of these tools, the input is a tab-delimited file, with the
first column containing gene, protein or metabolite names or
identifiers, followed by one or more columns of numeric data values.
Pathway Tools looks up these names and/or identifiers in its large
tables of synonyms and identifiers.  In many cases, data can also be
imported from a SmartTable or directly from another data repository
such as GEO or Metabolomics Workbench.

The visualization tools span several granularities, from visualizing
data on individual pathways, to visualization on collections of
related pathways (pathway collages), to genome-scale visualization on
a complete metabolic network diagram (Cellular Overview), complete regulatory network
diagram (Regulatory Overview), or Omics Dashboard.  All of these tools can display one type
of omics data at a time, with the exception of 
\blue{the Omics Dashboard and the Cellular Overview,
which can display up to three or four types of omics data at a time, respectively (e.g.,
transcriptomics, proteomics and metabolomics).}
\bitem
\item Omics Dashboard
\item Paint omics data onto whole metabolic network diagram (Cellular Overview)
\item Paint omics data onto multi-pathway diagrams (pathway collages)
\item Paint omics data onto individual pathways
\item Paint transcriptomics data onto whole genome diagram (Genome Overview)
\item Paint transcriptomics data onto whole regulatory network diagram
  (Regulatory Overview)
\eitem

Analytical tools:
\bitem
\item Enrichment analysis
\item Pathway covering analysis
\item Pathway perturbation score analysis
\item MultiOmics Explainer
\item SmartTable analysis
\eitem

\Ssection{Visualization: Omics Dashboard}
\label{sec:omics-dashboard}

\begin{figure*}[!tpb]
\centerline{\includegraphics[width=18cm]{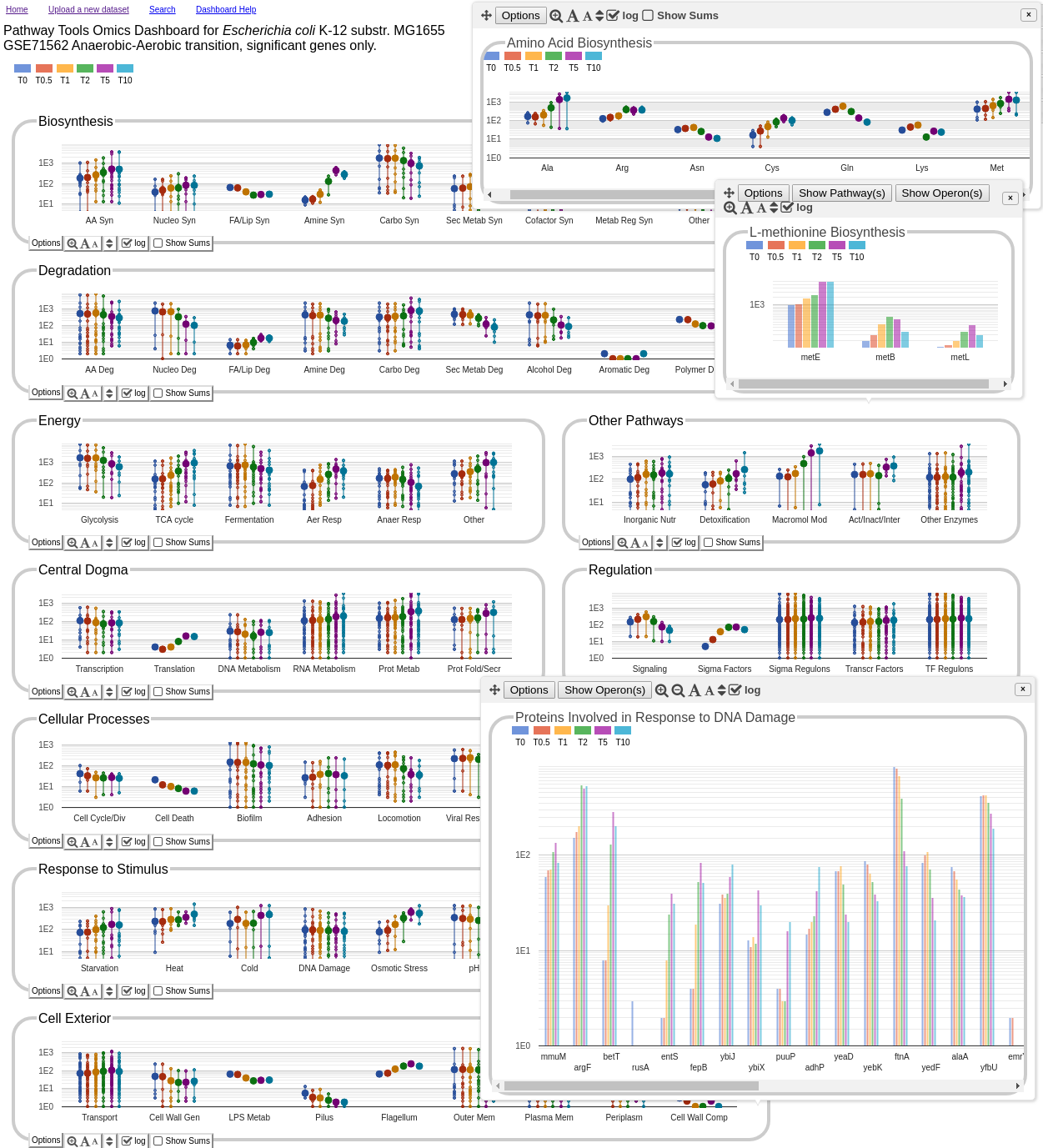}}
\caption{Top-level dashboard view of a gene expression time series
  data set (\cite{vonWulffen17}) for {\em E. coli}, showing changes in
  gene expression after transition to aerobic conditions at T=5.  Two
  different series of drill-down operations are shown.  At top right,
  selecting the AA Syn plot brings up a subsystem panel showing the
  different amino acids.  Drilling down further shows the expression
  levels of the individual genes involved in methionine biosynthesis.
  At bottom right is the panel showing the expression levels of genes
  involved in response to DNA damage.}\label{fig:dash}
\end{figure*}

The Omics Dashboard~\cite{DashGene17,DashMetab24} is a novel tool for
interactive exploration and analysis of omics \blue{and multiomics}
data sets through a hierarchy of cellular systems.  At its highest
level the Dashboard contains panels for cellular systems such as
biosynthesis, energy metabolism, and cellular processes (see
Figure~\ref{fig:dash}).  Each panel contains a series of X--Y plots
depicting the amalgamated expression levels of genes (or quantities of
metabolites, in the case of metabolomics data) within the subsystems
of that panel. For example, the Response to Stimulus panel includes
plots for its component subsystems starvation, DNA damage, osmotic
stress, and others.  Clicking on any plot shows an expanded panel for
that subsystem composed of either (a) plots for its component
subsystems, if component subsystems are defined, or for ``leaf''
subsystems (b) a graph of omics data values for all genes (or
metabolites) in the subsystem.  For example, clicking on the Amino
Acid Biosynthesis plot in the Biosynthesis panel brings up a panel
consisting of plots for each individual amino acid.  Clicking on the
methionine biosynthesis plot then shows a graph of all the genes
involved in methionine biosynthesis.  This organization allows a user
to quickly understand the broad behaviors of systems of interest, and
then successively focus in on more specific areas of function.  The
set of systems and subsystems represented in the Omics Dashboard is
derived from ontologies within Pathway Tools, primarily the pathway
ontology and GO.

In addition to panels and plots, the Omics Dashboard also provides
access to pathway diagrams painted with omics data, diagrams showing
the operon organization of all genes within a given biological system
and plots of all known regulators for the genes within a given system.
It is highly customizable, both in terms of appearance and
content. Enrichment analysis (using the hypergeometric distribution)
is also supported.  Data for the Omics Dashboard and other omics
analysis tools can come from a previously loaded data set, a text
file, or from a SmartTable. \blue{The Omics Dashboard can generate
  combined displays of up to three datasets of potentially different
  types (e.g., transcriptomics, metabolomics and proteomics) to
  support analysis of multiomics data. Data can be filtered by data
  value, significance value, or, for metabolomics data, to exclude
  common metabolites.}

\Ssection{Visualization: Cellular Overview}
\label{sec:cellov-omics}

The Multi Omics Viewer allows you to upload up to four omics datasets
onto the cellular overview, with each dataset available through a distinct 
``visual channel.'' These channels include node (metabolite) colors, 
edge (reaction) colors, node size, and edge thickness. Usually, nodes are
used to visualize metabolomics data and edges are used for visualizing 
transcriptomics, proteomics, and reaction-flux data.

\begin{figure}[h]
  \centerline{\includegraphics[width=14cm]{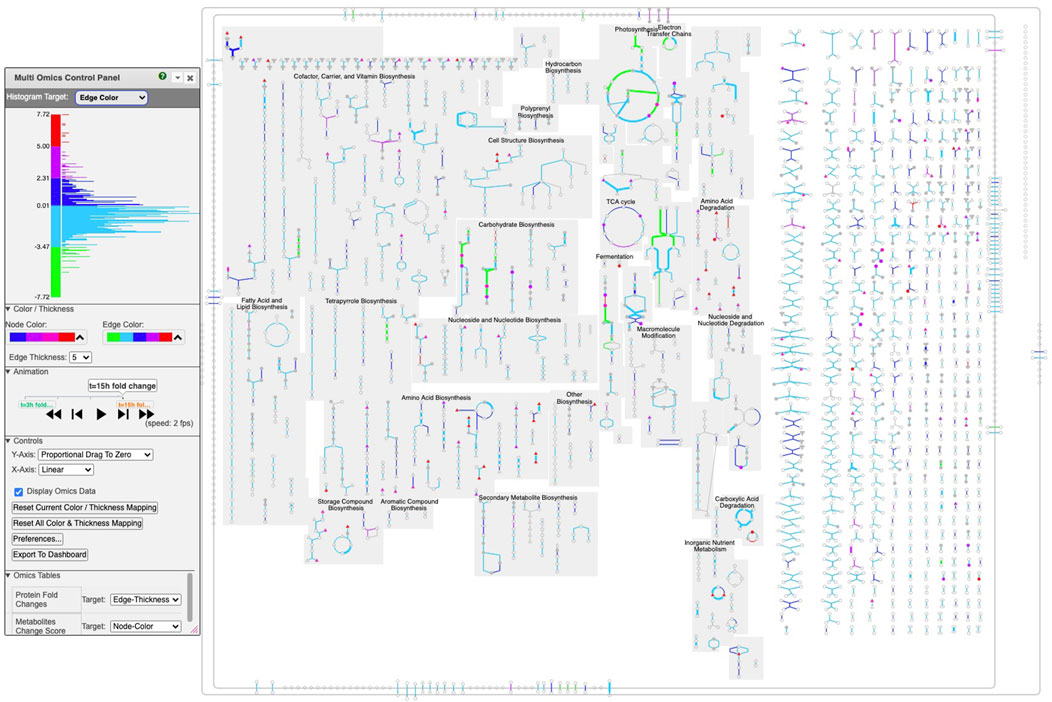}}
  \caption{An example of multi-time point dataset for Synechocystis PCC 6803 shown 
  on the Cellular Overview. The omics viewer control panel is on the left side of 
  the diagram. In this diagram, edge color is assigned to transcriptomics data, 
  edge thickness is assigned to proteomics data, and node color is assigned to 
  metabolomics data. The histogram shown in the image describes the color distribution 
  for depicting transcriptomics data.}
 \end{figure}

The graphical interface of the Multi Omics Viewer enables users to map each
input to its corresponding visual channel. For instance, a user might choose
to map transcriptomics data to the edge-color channel and proteomics data to the
edge-thickness channel, allowing both datasets to be visualized simultaneously.
Additionally, a third dataset, such as metabolomics data, could be assigned to 
the node-color channel.

It is a powerful tool for visual analysis of multi-omics data, enabling the 
simultaneous visualization of up to four types of omics data on an organism metabolic
diagram. The tool supports semantic zooming for detailed exploration and can animate
datasets with multiple time points, providing a comprehensive view of the biological system.
This tool allows the user to interactively adjust the mapping of the data value ranges to 
the displayed colors and thicknesses to provide more informative diagrams.

Transcriptomics, proteomics, metabolomics or reaction flux data can be
painted onto the Cellular Overview diagram (see Section~\ref{sec:cellov}, as shown in
Figure~\ref{fig:cellov}, in order to visualize the results of an omics
experiment on the full metabolic network.  This tool places the data in a
pathway context to enable the user to discern the coordinated
expression of entire pathways (such as the TCA cycle) or of important
steps within a pathway.  When the experiment contains multiple data
columns, such as in a time series experiment, it can be played as an
animation.  Omics pop-ups show all time points for particular
reactions or metabolites of interest.  The display includes histograms
showing the data distribution for each time-point.  The color scheme is
highly customizable, with users able to interactively adjust color bin
cutoffs, and selectively show or hide data ranges of interest.  Omics
data may be loaded from a data file, GEO data set, or SmartTable.
In the desktop software, data may also be imported from
Metabolomics Workbench.

At most two omics data types at a time can be combined in one
invocation of the tool:
\bitem
\item Either transcriptomics, proteomics, or reaction flux data, which
  are mapped to edges of the diagram
\item Metabolomics data, which is mapped to nodes of the diagram
\eitem

\blue{In the web version, an Omics DataTable viewer enables interactive
sorting and prioritizing of the data points in the loaded Omics data,
in a spreadsheet pop-up.  After sorting by names or data values, rows
of interest can be selected and highlighted on the Cellular Overview,
to simplify locating these objects visually.
}

\Ssection{Visualization: Pathway Collages}

The amount of information in the Cellular Overview can be
overwhelming.  To focus on a smaller set of pathways, the user can
build a pathway collage diagram from a specified set of pathways of
interest, and overlay the data on that diagram.  The diagram accepts
transcriptomics, proteomics, reaction flux, or metabolomics data.
See Section~\ref{sec:collages} and Figure~\ref{fig:pwycollage} for
more details.

\Ssection{Visualization: Individual Pathways}

Omics data can be viewed on individual pathway diagrams via omics
pop-ups, as shown in Figure~\ref{fig:pathomics}.  
This tool accepts
transcriptomics, proteomics, reaction flux, or metabolomics data.

\Ssection{Visualization: Genome Overview}

In the desktop Pathway Tools, transcriptomics data, but not
metabolomics data, can be painted onto
the Genome Overview diagram (see Section~\ref{sec:genov} and
Figure~\ref{fig:genov}).  Each gene within the diagram is colored
according to its transcript level, enabling perception of
genome-oriented patterns in gene expression.

\Ssection{Visualization: Regulatory Overview}

\blue{In the desktop Pathway Tools,} transcriptomics data alone can be painted onto the regulatory overview
diagram (see Section~\ref{sec:regov} and Figure~\ref{fig:regov}, to view the system-level impact of different transcription
factors on gene expression.  Each gene within the diagram is colored
according to its transcript level, enabling users to correlate gene
expression patterns with regulatory relationships.

\Ssection{Analysis: Enrichment Analysis}
\label{sec:visenrich}
Enrichment analysis is a statistical technique for analysis of
transcriptomics data and metabolomics data.  For metabolomics data,
the method determines whether the set of input metabolites (e.g.,
those metabolites whose abundances were significantly changed) contains more
metabolites from a given metabolic pathway (meaning the metabolites
are reactants and/or products in the pathway) than one would expect to
occur by chance, i.e., is the metabolite set statistically enriched
for particular metabolic pathways?  

For transcriptomics data the method determines whether the set of
input genes was statistically enriched for particular metabolic
pathways in which the gene products act as enzymes.  Furthermore, the
method can determine whether the input genes were statistically
enriched for genes in particular Gene Ontology categories
\cite{GO2019}, or whether the input genes were statistically
enriched for particular regulators --- for example, did the input genes
contain more genes regulated by a particular transcriptional regulator
than one would expect to find by chance?

\Ssection{Analysis: Pathway Covering}
Pathway Covering is a new approach to interpreting metabolomics
data, available in the web version of Pathway Tools \cite{PathwayCovering19}.
It addresses the same general question as enrichment scores of pathways for compounds,
but takes an approach based on set theory rather than a statistical model
such as the hypergeometric distribution.
It takes a list of metabolites as input, and as output it reports a set of pathways that
includes all metabolites in an input list as substrates (inputs,
outputs or intermediates).  The set returned by Pathway Covering
will always be a minimum cost solution.  Each pathway has a
``cost'' that is calculated by a user selected cost function.
The sum of the costs of each pathway is the cost of a pathway-
covering solution set.
The set of pathways returned will have the lowest total cost
(through other possible sets may have the same cost).

Pathway Tools provides five cost functions for use with Pathway
Covering:
\begin{itemize}
\item{Constant} --- This returns a fixed value for all pathways
\item{Pathway Size} --- This returns the number of reactions in the pathway
\item{Biosynthesis Preferred} --- This starts with the number of reactions in
  the pathway, then reduces the cost by a factor of two if the pathway
  is classified as a biosynthesis pathway in the MetaCyc pathway ontology.
\item{Compound Sparseness} --- This considers all the substrate
  metabolites of a pathway and divides the number of substrate
  metabolites that were not in the input set by the number that were
  in input set.  Because this is a cost function, returning large values for
  low proportions of compounds in the input set is appropriate.
\item{Pathway Harmony} --- This loosely approximates the notion of
  pathway flux. It is sensitive to the direction of change in compound
  abundance.  It takes all the compounds associated with the
  pathway and divides them into three groups: input compounds, output
  compounds, and intermediates.  The function gives low scores if all
  or a majority of compounds in a group are changed in the same direction.
\end{itemize}

Pathway covering typically generates a different set of pathways than an enrichment, which
ignores whether every metabolite is included in the result.  Some weighting functions, e.g.,
Pathway Harmony, will also make use of the direction of change of a metabolite between a pair of
experimental conditions.

\begin{figure*}[!tpb]
\centerline{\includegraphics[width=18cm]{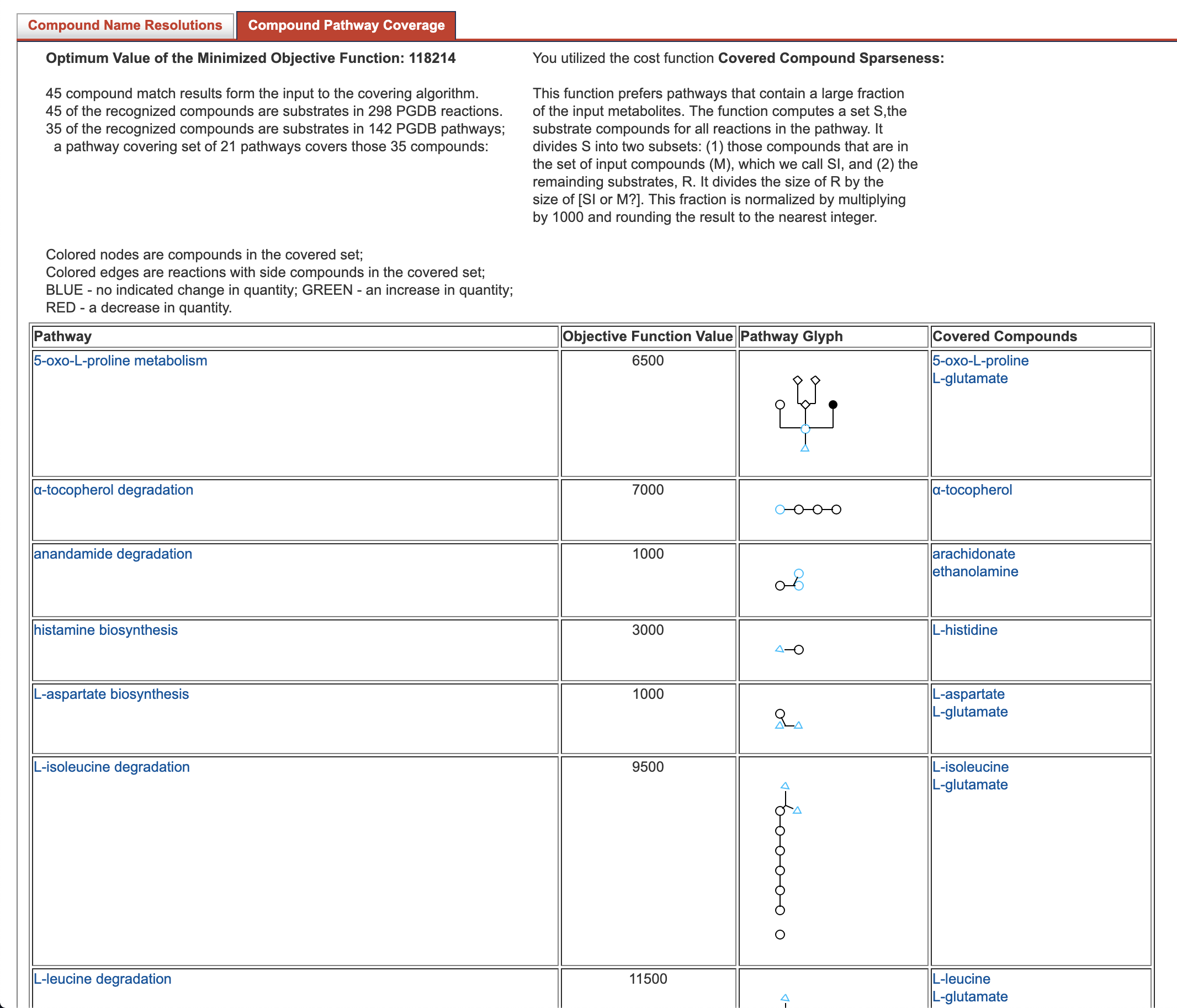}}
\caption{Results of Pathway Covering.}\label{fig:pathway_covering}
\end{figure*}

Pathway Tools reports the set of covering pathways as a table.  The table
additionally includes, the subset of the user supplied compounds that are
covered by each pathway, and a thumbnail showing the location of each
user-supplied compound in the pathway that contains it. An example report
page is shown in Figure~\ref{fig:pathway_covering}.

\Ssection{Analysis: Pathway Perturbation Score}
In additional to visualization, and pathway covering for omics data,
Pathway Tools provides another way to assess which pathways are
affected by an omics experiment.  Pathway Tools can compute a pathway perturbation score (PPS) for
each pathway with omics data at a single time-point.  The PPS attempts to measure
the overall extent to which a pathway is expressed by averaging the level of deviation
from zero (after log transformation) over all the reactions in the pathway.  For omics data
representing a time series, the Differential Pathway Perturbation Score (DPPS) measures the extent
to which a pathway exhibits change between time-points.  Pathway Tools can use omics data from either
gene expression or metabolite abundance to calculate these scores, we
have not attempted to validate that the perturbations scores are consistent between gene and metabolite data in
such mixed sets.

The Pathway Perturbation Score is based on the sum of the perturbation
scores for each reaction in the pathway --- the reaction perturbation
score ($RPS(r)$ for each reaction $r$ in the set of reactions $R$).
The reaction perturbation score in turn looks at each object
associated with the reaction (e.g., genes for gene
expression data, or compounds for metabolomics data). The perturbation score is
a measure of deviation from zero across the participants or genes in the reaction.
For each object, the data value is log transformed (if it was not previously)
and then from the set of values for the reaction, the absolute values
are computed and the largest absolute value is the reaction
perturbation score.

\begin{equation}
  RPS(r) = Max_{gom\,\epsilon\, r}log(gom)
 \end{equation}

Where $gom$ refers to the data representing the level of each
``gene or metabolite'' that is associated with (expressed as membership)
of the reaction $r$.
The pathway perturbation score is the root mean square of the reaction perturbation
scores, or:

\begin{equation}
PPS = \sqrt{\frac{\sum_{r\,\epsilon\,R} RPS(r)^2}{|R|}}.
\end{equation}
Where $|R|$ is the number of reactions in the pathway.

When the data is a time series, the Differential Pathway Perturbation
Score (DPPS) is used to capture the overall
magnitude of pathway activity change across time points; pathways with
high DPPS scores exhibit stronger changes.  The
calculation is similar to the PPS except that the values for each
component (gene, compound) associated with a reaction is the difference of the
maximum and minimum value observed in the time series rather than a deviation from a baseline.

\begin{equation}
DPPS = \sqrt{\frac{\sum_{r\,\epsilon\,R}DRPS(r)^2}{|R|}}.
\end{equation}

Where $DRPS$ for each reaction is 

\begin{equation}
  DRPS(r) = max_{gom\,\epsilon\,r}d(gom)
\end{equation}

and $d$ for each gene or metabolite associated with the reaction is
  
\begin{equation}
d(gom) = Max_{t\,\epsilon\,T}log(data(gom,t)) - Min_{t\,\epsilon\,T}log(data(gom,t)).
\end{equation}

$T$ is the range of time points in the data set and $data(gom,t)$ refers
to the value of a gene or metabolite at a time-point $t$.

\begin{figure*}[!tpb]
\centerline{\includegraphics[width=18cm]{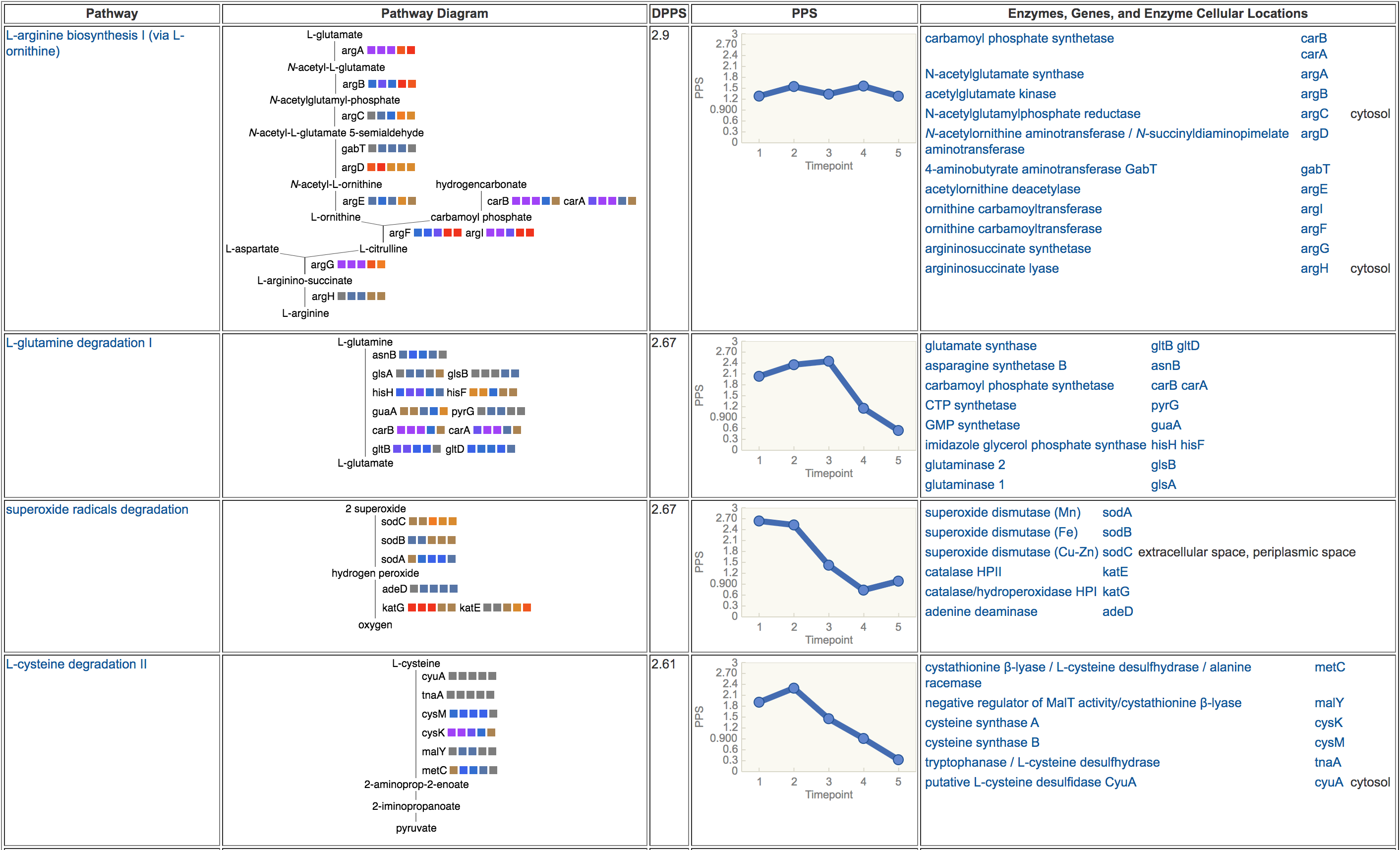}}
\caption{A table of most highly perturbed pathways, computed from an
  {\em E. coli} gene expression time series dataset.}\label{fig:pps-table}
\end{figure*}

Pathway Perturbation Scores are displayed as a table with pathways
sorted by decreasing PPS or DPPS depending on whether the data is a
time series.  The table for time series data includes a graph of the
PPS at each time point as well as the DPPS across all time values.  An
example is shown in Figure~\ref{fig:pps-table}.

A limitation of the PPS and DPPS is that they do not take into
account the consistency of expression changes, e.g., if a regulator
$A$ increased in expression, and the expression of a protein $B$
that was negatively regulated by $A$ also increased, this could be
viewed as an inconsistency.  

\Ssection{Analysis: MultiOmics Explainer} 

\begin{figure*}[!tpb]
\centerline{\includegraphics[width=11cm]{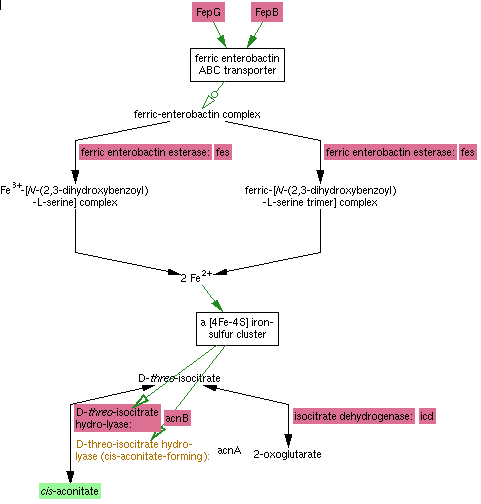}}
\caption{MultiOmics Explainer graph showing genes from \cite{Fuhrer17} that, when knocked out,
  cause levels of cis-aconitate to increase.  In particular, these genes indicate a connection between
  enterobactin-related genes and cis-aconitate metabolism that might not otherwise have been obvious.}
\label{fig:multiomics}
\end{figure*}

High-throughput experiments can
  reveal associations between genes, proteins, and/or metabolites
  whose source is not immediately obvious to researchers, but which
  can be explained by existing knowledge.  The MultiOmics Explainer is
  available only in the desktop software, and leverages
  what is known about an organism's metabolic and regulatory network
  to suggest explanations for some of the results of omics experiments
  (see Figure~\ref{fig:multiomics}).  Its goal is to speed understanding of
  experimental results, find explanations that scientists might
  otherwise overlook, and aid researchers in differentiating which
  effects can and cannot be explained by existing knowledge.

Given a small number of effect entities (genes, proteins or
metabolites), the tool constructs a network of known influences on
those entities.  If one or more condition entities (for example, a
knocked-out gene or a changed metabolite in the medium) are supplied,
the tool attempts to find causal paths in the network that connect the
conditions to the effects.  If no condition entities are supplied,
then the tool attempts to identify common influencers (for example, a
regulatory gene) that can be linked to the effect entities.  The
output of the MultiOmics Explainer is an interactive diagram that
illustrates one or a small number of possible routes by which the
condition entities, or common influencers, influence the effect
entities.

The combined metabolic and regulatory network used by the MultiOmics
Explainer comprises a set of interactions assembled from an organism's:
a) metabolic, transport, and protein modification reactions; b)
enzymes and enzyme activators, inhibitors and cofactors; c)
transcriptional regulators such as sigma factors, transcription
factors, and the small molecules that bind them; and d) translational
regulators such as attenuators, regulatory proteins and RNAs, and
small molecule riboswitches.  The MultiOmics Explainer is unique in
the wide range of potential causal relationships it considers,
reflecting the depth and richness of the Pathway Tools ontology.

\Ssection{Analysis: SmartTables}

As described in Section~\ref{sec:smarttables}, SmartTable
transformations, such as transforming a set of differentially
expressed genes or metabolites to the full set of reactions or
pathways they participate in, are a convenient tool for the analysis
of such data.

\section{Computational Access to PGDB Data}
\label{sec:apis}

\begin{figure}
\begin{center}
\includegraphics{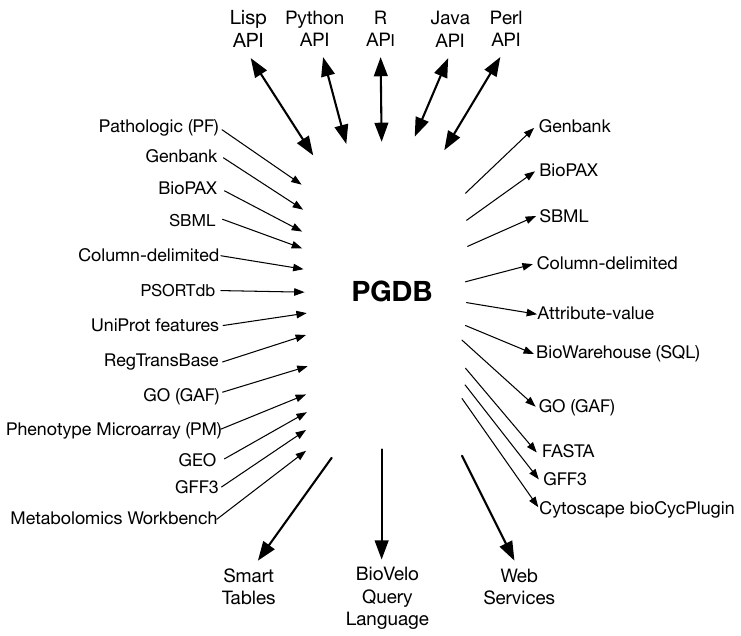}
\caption{Pathway Tools supported formats (left/right) and APIs (top/bottom) for data import and export.}
\label{fig:inputoutput}
\end{center}
\end{figure}

In addition to the user-friendly graphical interfaces to PGDBs
provided through the web and desktop versions of Pathway Tools, the
software supports multiple methods for importing and exporting
data from files and via programmatic interactions, which are
summarized in Figure~\ref{fig:inputoutput}.

{\noindent \bf Programmatic Access through Application Programming
  Interfaces (APIs).} Programmers can access and update PGDB data
directly \cite{PTools05,PToolsQueriesURL} by writing programs in the
Python, R, Java, Perl, and Common Lisp languages.  R, Java, and Perl
queries are executed using systems called RCyc \cite{RCycURL}, JavaCyc
\cite{JavaCycURL}, and PerlCyc \cite{PerlCycURL}.

PythonCyc is a Python package that enables programmatic access to
Pathway Tools. The package provides the basic functions to access and
modify the data of any PGDB. It also exposes more than 150
functions of Pathway Tools, among them, the MetaFlux module. The
PythonCyc package is hosted on GitHub and is a separate installation
from Pathway Tools. A full API documentation and a tutorial is
available online. Please consult the URL~\cite{pythoncycURL} for
access to the package, the API documentation and the tutorial.

{\noindent \bf Data Import Formats.}  Pathway Tools can import data
from many sources into a PGDB.  First of all, PathoLogic can create or
update a PGDB based on a genome in PathoLogic Format (PF)
\cite{PToolsGuide190}, GFF3 format \cite{GFF3URL}, 
or GenBank format \cite{GenbankFormatURL}.  A PGDB can also be populated
from an SBML \cite{Hucka04,SBMLURL} or BioPAX
\cite{BioPAXURL,BioPAX10} file.  The SBML Import tool will read
SBML level 3 files only and is currently limited to SBML files for
prokaryotes.  If the provided file is an FBA SBML file, an option to
write an FBA file after the import is complete will be enabled.  The
resulting \texttt{.fba} file can then be run using MetaFlux.  Bulk
updating of various data items can be accomplished by a simple
tab-delimited table format.

Several specialized data types are also supported.  The following
sources can be imported for proteins.  UniProt sequence annotation
features \cite{UniProtProteinFeatureURL} can be fetched from a
Biowarehouse \cite{Biowarehouse06} server that was loaded with
SwissProt and TrEMBL.  GO (Gene Ontology) annotations can be loaded
from a GAF \cite{GO2000,GOFormatsURL} file.  PSORTdb \cite{Yu11}
cellular localization data can be imported from tab-delimited files.

Gene regulatory data can be imported from RegTransBase
\cite{Cipriano13}, which is a SQL database.  Growth conditions versus
growth media can be imported from Phenotype Microarray (PM)
\cite{Bochner-etal01,Bochner09} files.  High-throughput expression
data can be obtained from NCBI GEO \cite{GEO13}.  Metabolite abundance
data can be obtained from Metabolomics Workbench \cite{Sud15}.

{\noindent \bf File Export Formats.}  Pathway Tools can export PGDBs
into several file formats that we have developed, which include
tab-delimited tables and an attribute-value format (see
\cite{BioCycFlatfilesURL}).  Pathway Tools can also export subsets of
PGDB data to other common formats including SBML
\cite{Hucka04,SBMLURL}, BioPAX \cite{BioPAXURL,BioPAX10}, GO (in the
GAF format) \cite{GO2000,GOFormatsURL}, GenBank
\cite{GenbankFormatURL}, GFF3 \cite{GFF3URL}, and FASTA \cite{FASTAFormatURL}.  The
bioCycPlugin for Cytoscape \cite{bioCycPluginURL} makes use of web
services in conjunction with BioPAX, to select and export pathways into
the Cytoscape environment.  An SBML file for flux balance
  analysis modeling can be generated by providing a MetaFlux FBA
  file.  Generated SBML files are Level 3 version 1 with flux balance
  constraints (fbc) package version 2.

{\noindent \bf Export for Relational Database Access via
  BioWarehouse.}  For scientists who want to query PGDB data through a
relational database system, the attribute-value files exported by
Pathway Tools can be loaded into SRI's BioWarehouse system
\cite{Biowarehouse06}.  BioWarehouse is a MySQL-based
system for integration of multiple public bioinformatics databases.
PGDB data can be queried through BioWarehouse alone or in combination
with other bioinformatics DBs such as UniProt, GenBank, NCBI Taxonomy,
ENZYME, and KEGG.

{\noindent \bf Queries Using the BioVelo Query Language,
  and Web Services.}  
Pathway Tools provides a powerful database query language for PGDBs,
called BioVelo \cite{BioVeloLangURL,BioVeloSAQPURL}.  BioVelo queries
can be issued through an interactive web form, and through APIs.

Pathway Tools web services \cite{PToolsWebServicesURL} enable
programmatic retrieval of numerous data types, based on submitted HTTP
GET or POST commands, and have expanded substantially in recent years.
Users can access BioVelo queries, a Metabolite Translation Service,
and can also invoke omics visualization services and SmartTable
manipulations via web services.  Services for SmartTables include
creation, retrieval, copying, and deletion; applying many
transformations; and changes like adding and deleting rows, columns
and cells.

\section{Graph-Based Metabolic Network Analyses}
\label{sec:analyses}

This section describes Pathway Tools modules for performing
graph-based analyses of metabolic networks.

\Ssection{Metabolic Route Search Tool (Web Only)}
\label{sec:routesearch}

\begin{figure}
\includegraphics[width=6in]{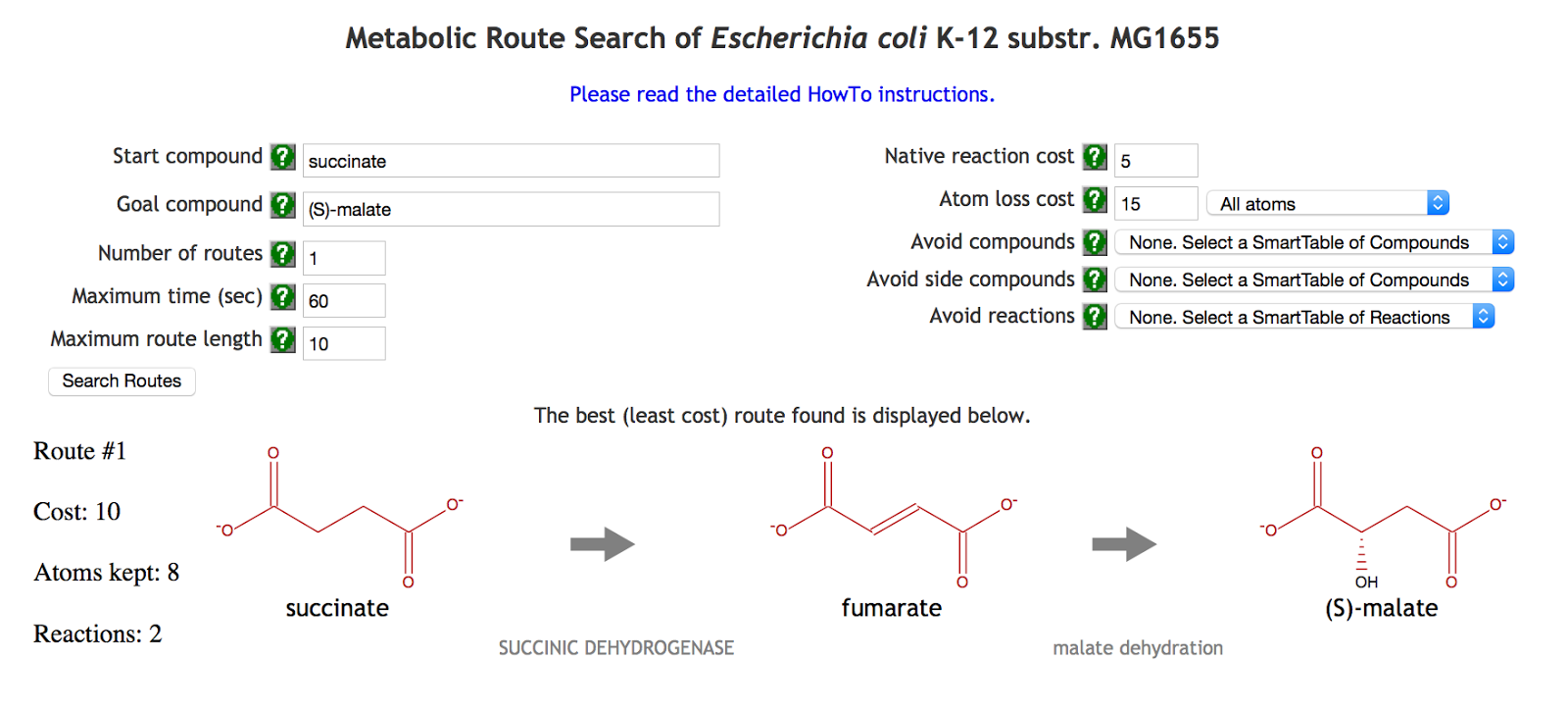}
\caption{The RouteSearch web interface is shown with the result of one short
  pathway found. The arrows represent reactions and are tagged with
  the protein names catalyzing them. All atoms are conserved from the
  start compound (succinate) to the target compound ((S)-Malate) in this
  simplified example.}
\label{fig:routesearch}
\end{figure}

RouteSearch~\cite{Latendresse14} is a Pathway Tools component that
enables the exploration of the reaction network of a PGDB, and
engineering of
new metabolic pathways. RouteSearch computes optimal metabolic routes (that is,
an optimal series of biochemical reactions that connect start and
goal compounds), given various cost parameters to control the
optimality of the routes found.  RouteSearch can display several of
the best routes it finds using an interactive graphical web page (see
Figure~\ref{fig:routesearch} for an example).

When RouteSearch is used in metabolic engineering, it uses the
MetaCyc database as its external reaction database for new reactions
to include in an organism. The cost for adding one reaction is selected
by the user. Typically this cost, an integer, is larger than the cost of
using one reaction from the organism. In such a case, a new reaction
would be added for a route if it conserve more atoms from the start
compound to the target compound. The cost of losing one atom from the
start compound is also selected by the user, and it is typically
larger than the cost of one reaction from the organism or the external
library of new reactions.

In computing optimality, RouteSearch takes into account the
conservation of non-hydrogen atoms from the start compound to the goal
compound. The more atoms that are conserved, the more efficient the
transformation from start to goal becomes, therefore resulting
in a lower cost. To compute the number of conserved atoms,
RouteSearch uses pre-computed atom mappings of reactions that
are available in MetaCyc. An atom mapping of a reaction gives a
one-to-one correspondence of each non-hydrogen atom from reactants to
products. RouteSearch is available only in web mode in Pathway Tools.

\Ssection{Multi-Organism Route Search (Web Only)}
Multi-Organism Route Search (MORS) \cite{MORS19} extends our
previous single-organism 
Metabolic Route Search to accept arbitrary sets of organisms,
simultaneously, for searching across the union of the reactions in the
selected organisms.  A typical use case is searching HumanCyc plus the
organisms in a microbiome body site, such as the gut, to
investigate how a combination of organisms might synthesize a toxic compound,
and to see which specific organisms are participating.

Originally, the single-organism Route Search tool would seek the best (least cost) route 
between user-specified start and goal compounds.  A route is a linear series of
reactions.  The cost of a route involves a weighted combination of the
length of the route (number of reactions) and the number of atoms lost
from the start to the goal compound (computed using atom mapping information).

The MORS mode adds a multi-organism selector for selecting the set of
BioCyc organisms to be searched; the set of reactions searched by
MORS will be  the union of reactions from that organism set.
Additionally, a cost for ``organism switching'' can be set.  A switch
occurs when the two organism sets of two consecutive reactions in a
route have no overlap.  In other words, if the first reaction is known
to occur in one set of organisms and the second reaction is occurring
in a different organism set, but there is no organism that contains
both reactions simultaneously, then the route has to switch organisms
by transferring the compound connecting both reactions, from one
organism to another (by unspecified transport mechanisms).  An
organism switch is depicted in a route with a red vertical line.  A
SmartTable containing the route can be generated, which shows the 
organism sets containing enzymes that catalyze  each reaction along
the route.

\begin{figure*}[!tpb]
\centerline{\includegraphics[width=18cm]{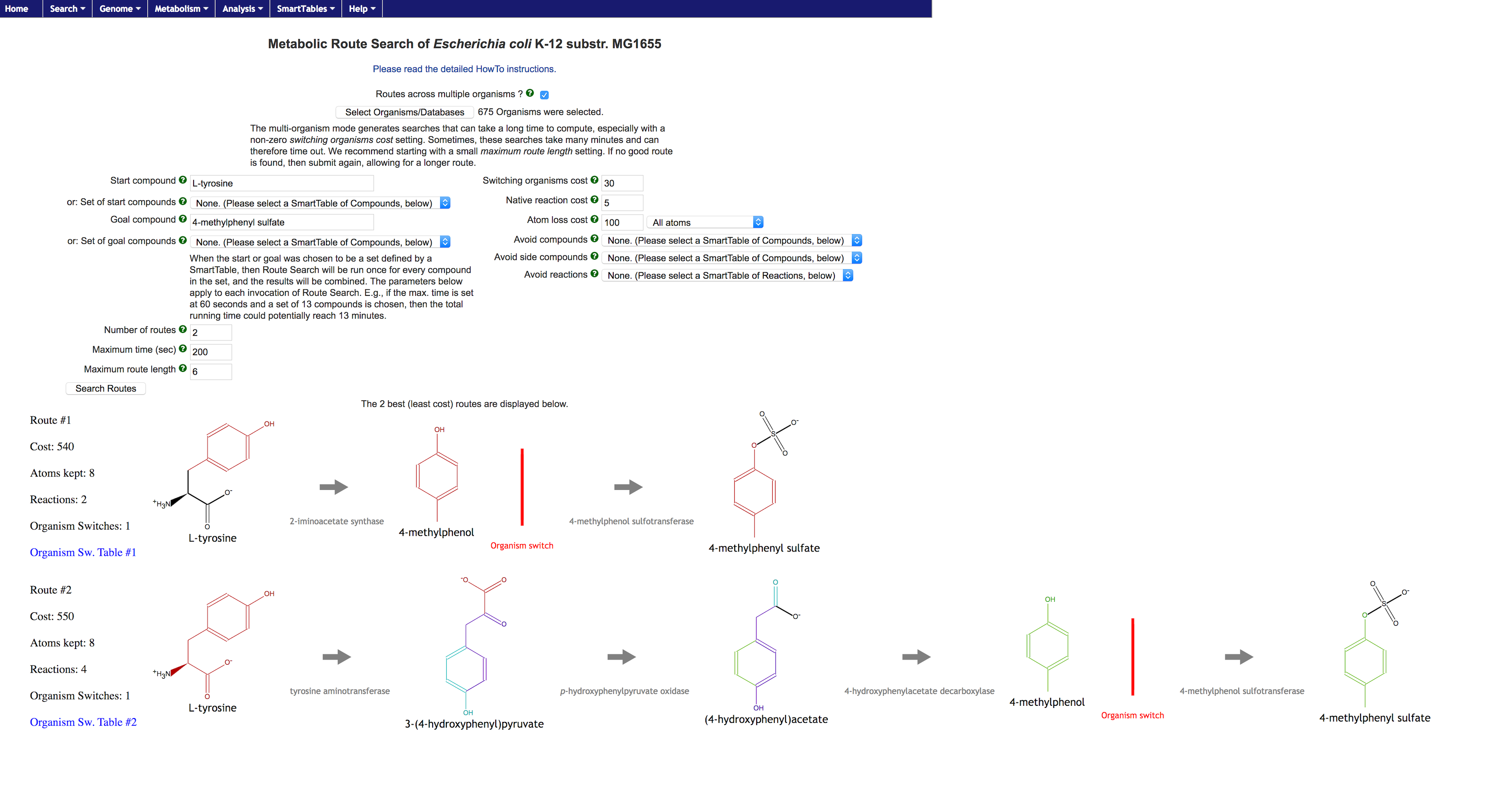}}
\caption{Result of MORS routes to 4-methylphenyl sulfate.}\label{fig:mors}
\end{figure*}

As an example, let us examine how dietary L-tyrosine is transformed
into toxic 4-methylphenyl sulfate, which is a protein fermentation
product that has been modified in the liver and is implicated in
kidney problems.  As it is known that this toxin originates from
L-tyrosine \cite{Selmer01}, the start compound was set to L-tyrosine
and the goal compound to 4-methylphenyl sulfate.  We selected all
organisms in the human microbiome body site called
``gastrointestinal-tract'' plus {\em Homo sapiens}.  The total count
of organisms was 675.  The resulting top two routes are shown in
Figure~\ref{fig:mors}.  Both routes retain eight atoms.  The first route consists
of two reactions, and the second of four reactions.  The last
reaction, after the organism switch, is only found in {\em Homo
  sapiens}.  However, the reaction immediately before the switch
occurs in 412 organisms in the first route and in 80 organisms in the
second.

\Ssection{\blue{Metabolic Network Explorer (Web Only)}}
\label{sec:metnetexplorer}

\begin{figure*}[!tpb]
\centerline{\includegraphics[width=18cm]{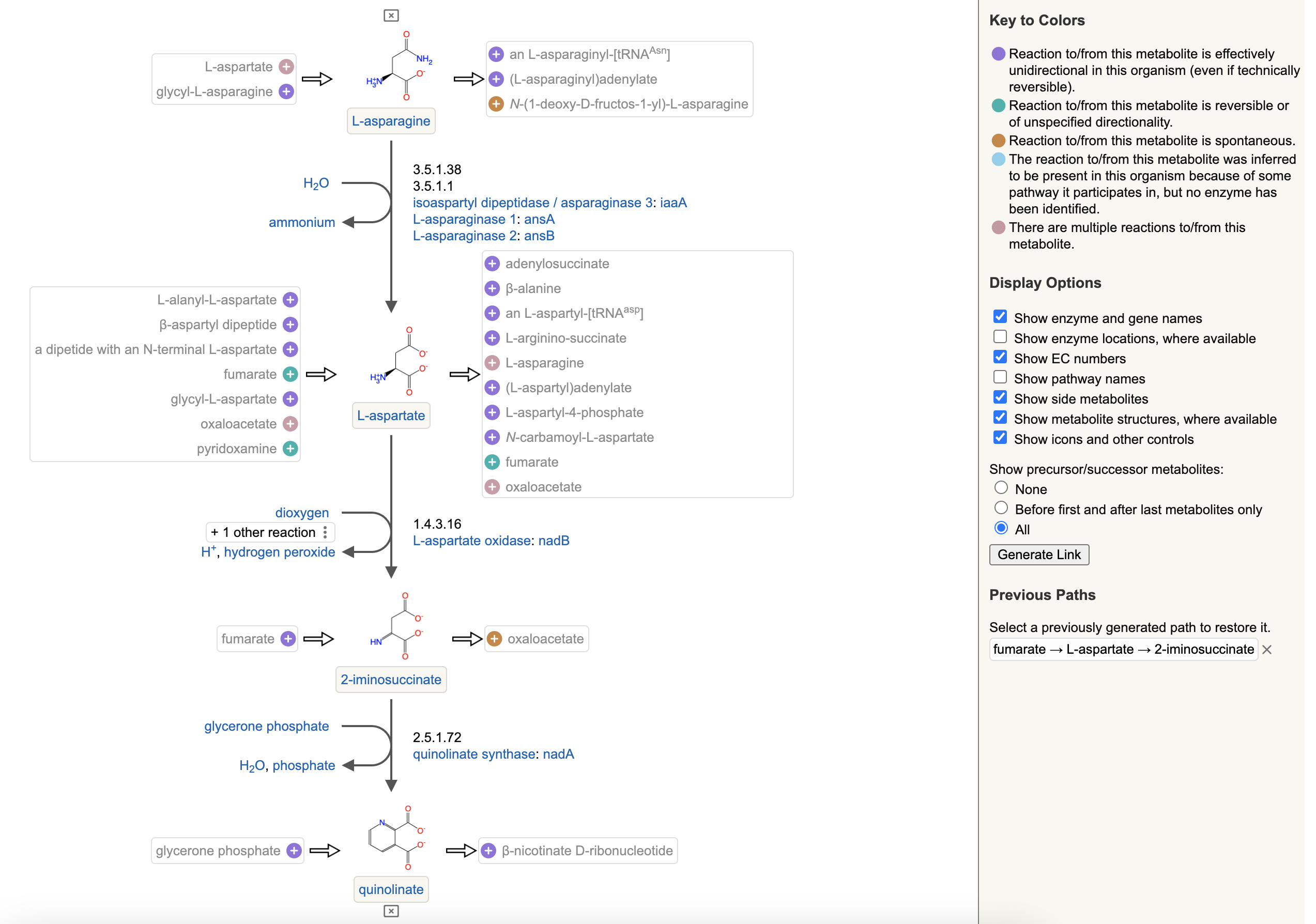}}
\caption{\blue{The Metabolic Network Explorer, showing an exploration that
  started at L-aspartate, and expanded backward one step to
  L-asparagine and forward two steps to quinolinate. The lists of
  metabolites to the left and right of each metabolite along the
  central path suggest alternative directions of exploration.}}
\label{fig:metnetexp}
\end{figure*}

\blue{The Metabolic Network Explorer~\cite{Paley21a}, shown in
  Figure~\ref{fig:metnetexp}, is a web-based tool for interactive
  exploration of small subsets of a metabolic network. From a
  starting metabolite, the user is shown a list of potential
  precursors and successor metabolites. When one is selected, it is
  added to a linear path, along with its potential precursors and
  successor metabolites. The user can interactively expand in either
  direction to build up a linear route through the metabolic network.
  A tooltip for each potential expansion metabolite lists the relevant
  connecting reactions and enzymes.  Expanding from an intermediate
  metabolite will replace the portion of the path previously connected
  to that metabolite so the path remains linear, but the a link to the
  previous path is kept, making it easy for the user to switch back to
  it.}

\Ssection{Dead-End Metabolite Analysis}
\label{sec:reachability}

Dead-end metabolites are metabolites that are only produced by the metabolic
network or only consumed by the network.  
More precisely, the tool for identifying dead-end metabolites will
report a small-molecule metabolite $M$ as a dead-end metabolite in the cellular compartment
$C$ if and only if
one of the following conditions is true:

\benum

\item $M$ or parent classes of $M$ are only {\bf consumed} by small-molecule
reactions occurring in $C$, and $M$ or parent classes of $M$ are not transported into $C$.

\item $M$ or parent classes of $M$ are only {\bf produced} by small-molecule
reactions occurring in $C$, and $M$ or parent classes of $M$ are not
transported out of $C$, and no enzyme in the PGDB uses $M$ as a cofactor

\eenum

\Ssection{Computation of Blocked Reactions}
\label{sec:blockedreactions}

The MetaFlux component of Pathway Tools computes the blocked reactions
in a reaction network (see~\ref{sec:logfile})---reactions that can
never carry flux because of blockages in the network.

\Ssection{Prediction of Network Choke Points}

One application of a metabolic network model is to find network bottlenecks,
which, if blocked, could kill the cell.  Such bottlenecks could
constitute antimicrobial drug targets.  We have developed a
tool for predicting these so-called choke points.

The Pathway Tools choke point detection algorithm examines the reactions attached to a given
metabolite, and processes one metabolite at a time.  The first step is
to assemble the list of metabolites to examine.  This is done by
collecting (1) all reactions that are in pathways, plus (2) reactions
that stand alone, but which use only small molecule metabolites.  The
reactions that came from pathways may use some macromolecular
substrates, such as proteins that are modified by the reaction.  From
this list of reactions, the algorithm collects all of their substrate
metabolites (meaning their reactant or product metabolites).

Definition \cite{Yeh04}: A ``choke point reaction'' is a reaction that
either uniquely consumes a specific substrate or uniquely produces a
specific product in a metabolic network, and is also balanced by at
least one reaction that respectively produces or consumes that
substrate.  Specifically, the algorithm searches for two types of
choke point reactions:  (a) Reactions $R_1$ such that only a single
reaction $R_1$ produces metabolite $M,$ and at least one reaction
consumes $M.$ (b) Reactions $R_2$ such that only a single reaction
$R_2$ consumes metabolite $M,$ and at least one reaction produces $M.$
These definitions imply that to find a choke point, all reactions
involving $M$ must be unidirectional.  These choke point reactions are
collected and returned as the result.  Note that the definition
excludes reactions directly connected to dead-end metabolites.

The resulting candidate choke point reactions can be painted onto the
cellular overview to facilitate further analysis.

\section{Comparative Tools}
\label{sec:comparative}

Pathway Tools contains a rich set of operations for comparing the
information in two or more PGDBs.  These operations range from
comparison of genome-related information to comparison of pathway
information.  These comparisons are of several types.

\Ssection{Genome Comparison}

The comparative genome browser discussed in Section~\ref{sec:genov}
displays replicon regions centered on orthologous genes across a
set of genomes (see \href{http://www.ai.sri.com/pkarp/pubs/pt20suppfigs4.pdf}{Supplemental Figure}~9).

\Ssection{Pathway Comparison}
\label{sec:pwy-comparison}

The user can generate a comparative table for \blue{one or more} given
metabolic pathways across a specified set of organisms.  For each
organism the table shows the presence of pathway enzymes and operon
structures of genes within each pathway. \blue{If no enzyme exists for
  some reaction in one organism, the software will search for
  orthologs of the genes for that step in the other organisms in the
  set, and will list any that are found.}

\Ssection{Metabolic Network Comparison}

A global comparison of the metabolic networks of multiple PGDBs can
be performed by highlighting on the Cellular Overview diagram (see
Section~\ref{sec:cellov}).  This tool enables the user to highlight
in the Cellular Overview reactions that are shared, or not shared,
among a specified set of organisms.

\Ssection{\blue{Comparative Genome Dashboard}}

\blue{The Comparative Genome Dashboard is a tool for visualizing the
  overall biological capabilities of an organism or set of organisms,
  as encoded by their respective genome and pathway annotations. It
  facilitates a rapid user survey of all cellular systems and enables
  the user to quickly identify similarities and differences between
  organisms. In addition to reflecting the biology of an organism, the
  capabilities it depicts will depend on the quality of the genome
  annotation.}

\begin{figure*}[!tpb]
\centerline{\includegraphics[width=18cm]{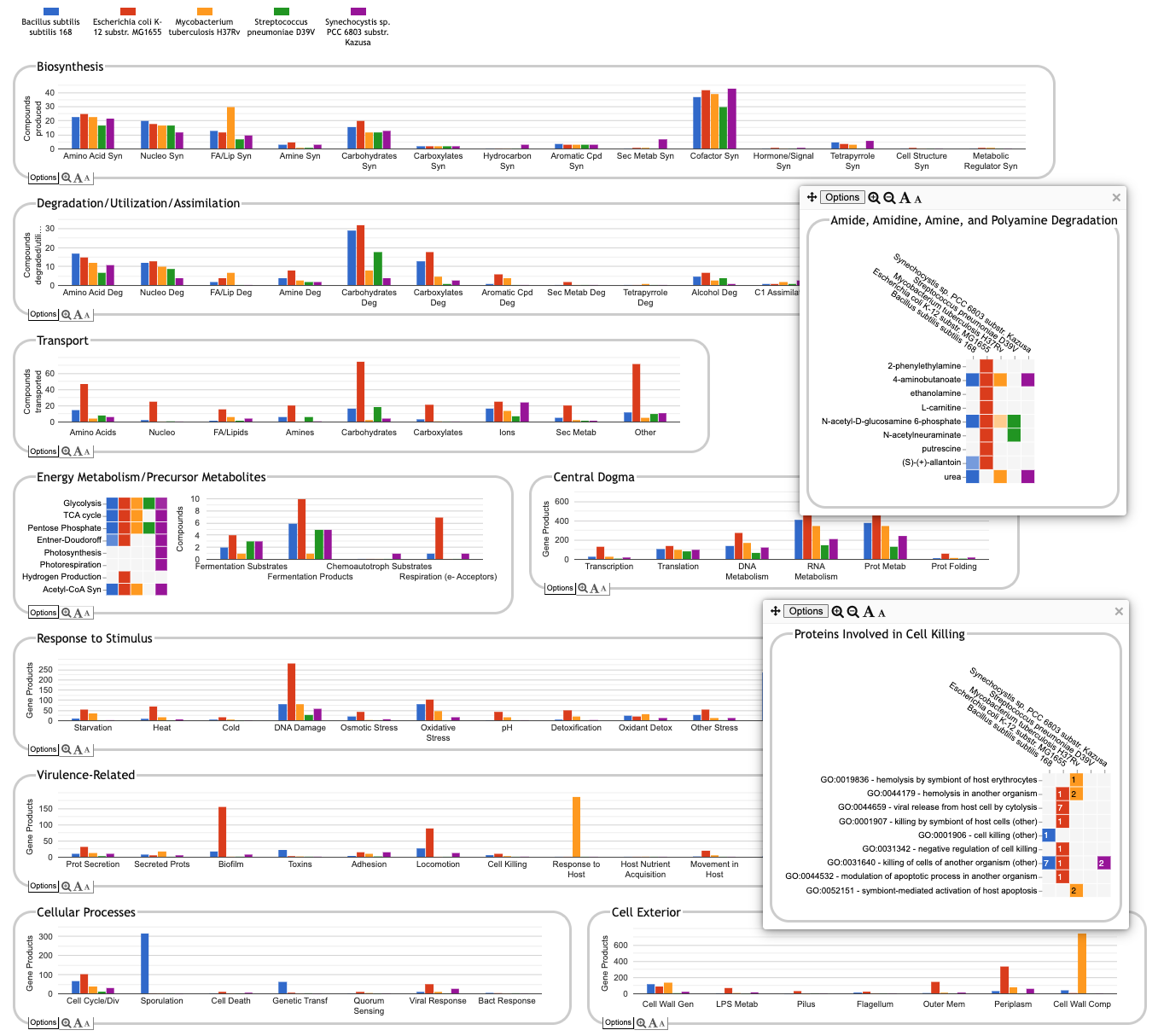}}
\caption{\blue{A comparison of five organisms using the Comparative Genome
  Dashboard. Clicking on the Amine Deg plot in the Degradation panel
  brings up a detail panel indicating which amines can be degraded by
  which organisms. Clicking on the Cell Killing plot in the Virulence
  panel brings up a detail panel listing specific GO terms and the
  numbers of proteins annotated with each term in each organism.}}
\label{fig:gdash}
\end{figure*}

\blue{The Comparative Genome Dashboard, shown in
  Figure~\ref{fig:gdash}, consists of a set of panels, each
  representing a cellular system, e.g.,
  Degradation/Utilization/Assimilation. For each metabolic panel we
  show a series of bar charts (plots) depicting the number of
  compounds of various types predicted to be produced or consumed,
  based on the metabolic pathways present in the PGDB. The transport
  panel shows the numbers of compounds of different types transported,
  based on the transport reactions defined in the PGDB. For
  non-metabolic panels the bar charts depict the number of genes
  annotated to each of a set of subsystems, based on GO term
  annotations within the PGDB.}

\blue{Clicking on a plot brings up a detail panel for that
  subsystem. The lowest level detail panels for the metabolic and
  transport subsystems show the list of compounds produced, consumed
  or transported by each organism. The lowest level detail panels for
  the non-metabolic subsystems list specific GO terms and the number
  of associated proteins for each organism. Tooltips provide links to
  data pages for compounds, pathways, and proteins, as well as to more
  detailed object-specific comparison pages, such as the pathway
  comparison tables described in Section~\ref{sec:pwy-comparison}.}

\Ssection{Comparison Report Tables}

Finally, a general comparative analysis facility enables the user to
generate comparative report tables for many aspects of a PGDB.  As
well as being used for comparative analyses, these tools can be used
to generate statistics regarding the content of a single PGDB.  These
tools are general in that they present their results in a standard
format, and they enable the user to drill down to specific results in a
consistent fashion.  The initial report page shows summary statistics,
but the user can drill down to compare all instances of a category
by clicking on elements of a report table.

For example, consider the transporter report page
in \href{http://www.ai.sri.com/pkarp/pubs/pt20suppfigs4.pdf}{Supplemental Figure}~10.  Table~2 within that report
summarizes the number of uptake transporters found in two organisms.
A user who wants to see the actual transported substrates
clicks on the text ``Compounds transported into the cell''
to generate a new report page containing a table listing the union of all substrates
imported by both organisms, along with an indication of which
organisms transport each substrate, and which transporter is utilized.
If the user clicks on a data cell within Table~2, such as the
number of imported substrates in \coli\ K-12 (156), a page is
generated that lists those substrates only.  Similar functionality
applies to most tables in these reports.

The following report types are provided.  An example comparative
report is available at URL \cite{BioCycCompGenomicsExample1URL}.

\bitem

\item Reaction report includes the following statistics for each
selected organism:
\bitem
\item Number of reactions containing substrates of different types
(e.g., reactions for which all substrates are small molecules, and for
which some substrate is a protein or a tRNA)
\item Number of reactions in each Enzyme Commission (EC) category
\item Number of reactions containing different numbers of isozymes
\eitem

\item Pathway report includes these statistics:
\bitem
\item Number of pathways in each category within the MetaCyc pathway
ontology
\item Number of pathways with different numbers of pathway holes
\eitem

\item Compound  report includes these statistics:
\bitem
\item Frequency with which different compounds appear in different
metabolic roles (substrate, cofactor, inhibitor, activator)
\eitem

\item Protein report includes these statistics:
\bitem
\item General statistics on number of monomers versus multimeric
complexes, breakdown of multimers into heteromultimers and
homomultimers
\item Statistics on multifunctional enzymes
\eitem

\item Transporter report includes these statistics:
\bitem
\item Number of efflux versus influx transporters
\item Number of genes whose products are transporters
\item Number of unique transported substrates, both overall and broken
down by efflux versus influx
\item Number of transported substrates that are substrates in
metabolic pathways or are enzyme cofactors
\item Transporters with multiple substrates, and substrates with
multiple transporters
\item Operon organization of transporters
\eitem

\item Ortholog report includes these statistics:
\bitem
\item List of all orthologous proteins across the selected organisms
\item Proteins that are shared in all selected organisms, or unique to
one organism
\eitem

\item Transcription Unit report includes these statistics:
\bitem
\item Distribution of number of genes per transcription unit
\item Distribution of number of operons into which metabolic pathway genes
are distributed
\eitem

\eitem

\section{Software and Database Architecture}
\label{sec:arch}

Pathway Tools is mostly implemented in the Common Lisp programming
language,\footnote{We use the Allegro Common Lisp implementation from
Franz Inc., Oakland, CA.} with some components implemented in
JavaScript and MySQL.
We chose Common Lisp because it is a
high-productivity programming environment.  Because Lisp is a very
high-level language, one line of Lisp code is equivalent to several
lines of code in a language such as Java or C++. Therefore, the same
program can be written more quickly in Lisp, with fewer bugs.  A study
by Gat found that compiled Lisp programs generally run faster than
Java programs, and that a given program can be developed 2--7 times
faster in Lisp than in Java \cite{Gat00}.  Common Lisp also has a very
powerful interactive debugging environment.

Lisp has powerful dynamic capabilities that are illustrated
by a Pathway Tools feature called auto-patch.  Imagine that a Pathway
Tools user site has reported a bug in the software.  Once our group
has found a fix for the bug, we put a patch file that redefines the
offending Lisp function(s) on the SRI website.  The next time Pathway
Tools is started at remote sites, it automatically downloads the patch
(in compiled form) from the SRI website, puts the patch in an
appropriate directory, and dynamically loads the patch file into the
running Pathway Tools to redefine the altered function(s).

Pathway Tools consists of \nLinesCodePTools\ lines of Common Lisp code, organized
into 20 subsystems.  In addition, \nLinesJavascriptPTools\ lines of JavaScript code are
used within the Pathway Tools web interface.  Pathway Tools runs on
the Macintosh, Linux, and Microsoft Windows platforms.  Pathway Tools was ported
to 64-bit architectures several years ago.

The main bioinformatics modules of Pathway
Tools are the Navigator, Editors, PathoLogic, and MetaFlux.
PTools also includes a
chemoinformatics subsystem that includes tools such as SMILES
\cite{Smiles1} generation and parsing, a chemical substructure
matcher, and a large set of shared utilities that we call the Pathway
Tools core.  Pathway Tools uses an object-oriented database system
called Ocelot.  The Pathway Tools user interface relies on a graph layout
and display package called Grasper \cite{GRASPER-CL}, and web and
desktop graphics packages called CWEST and CLIM
(the Common Lisp Interface Manager).
Other software used by (and often included with) Pathway Tools are:

\bitem
\item Bioinformatics: Textpresso \cite{Textpresso04}; \blue{Clustal Omega
  \cite{Sievers11}; MSAViewer \cite{Yachdav16};}
PatMatch \cite{PatMatch05}; BLAST \cite{BLAST}; \blue{DIAMOND
\cite{Diamond21};}
cytoscape.js \cite{Cytoscapejs15}

\item Chemoinformatics: MarvinJS \cite{MarvinURL}; GlycanBuilder
  \cite{Damerell12}; InChI \cite{InChI03}

\item Lisp: ARNESI \cite{ARNESIURL}; alexandria \cite{AlexandriaURL}; 5am \cite{5AMURL};
  cl-json \cite{CLJSONURL}; cl-store \cite{CLStoreURL}; SKIPPY \cite{SKIPPYURL}

\item Other: SCIP \cite{SCIPURL}; Ghostscript; 
SOLR \cite{SOLRURL}; MySQL; 
Yahoo User Interface library (YUI) \cite{YUIURL}; \blue{jQuery UI; jsTree;
jBox; tippy; Google Charts; VEGA \cite{VEGAURL} }
\eitem


Ocelot is an object/relational database management system (DBMS)
developed by our group at SRI \cite{EcoCycJCB96,Karp-JIIS-97a}.  
Ocelot combines the expressive power of frame knowledge representation
systems \cite{KarpFreview} developed within the AI community (whose
object data model is far superior to the relational data model for
representing biological data\footnote{Superior aspects of the object data model
include the following.  The object data model is better at managing very complex
schemas.  That is, if the same domain is represented within the object data and within
the relational model, the object schema is usually much more compact and easier to
comprehend.  One reason is that inheritance enables the object data model to define
subclasses by extending existing classes (e.g., the class Polypeptides is a subclass
of the class Proteins), whereas the relational model would force attributes shared between
the two tables to be duplicated in each, which both obscures the fact that the two
tables are related, and complicates schema evolution.  Relational normalization also
increases the size of the schema by forcing the creation of new tables for every
multi-valued attribute, which is not required in the object data model.  The object data model
used by Ocelot is particularly flexible in supporting any type of schema evolution without
forcing the entire database to be reloaded (unlike relational DBMSs), which is important in bioinformatics because
the complexity of biological data forces never-ending enhancements to the schema (note
that not every object DBMS provides such flexibility).}) with the scalability of relational
database management systems (RDBMSs).  Ocelot DBs are persistently
stored within a MySQL RDBMS.  Ocelot objects are faulted on
demand from the RDBMS, and in addition are faulted by a background
process during idle time.  Objects that were modified during a user
session are tracked and saved to the RDBMS during a save operation.
Ocelot uses optimistic concurrency control
\cite{Karp-JIIS-97a}---during
a save operation it checks for conflicts between the updates
made by the user and updates saved by other users since the saving
user began their session or last made a save operation.  This approach
avoids the overhead of locking that becomes problematic in object
databases because modifications to one object often cascade to related
objects and could require a large number of lock operations.  The
optimistic concurrency control works well in practice because curators
tend to focus in different biological areas and therefore rarely
update the same objects at the same time.

\begin{figure}
\begin{center}
\includegraphics[height=3in]{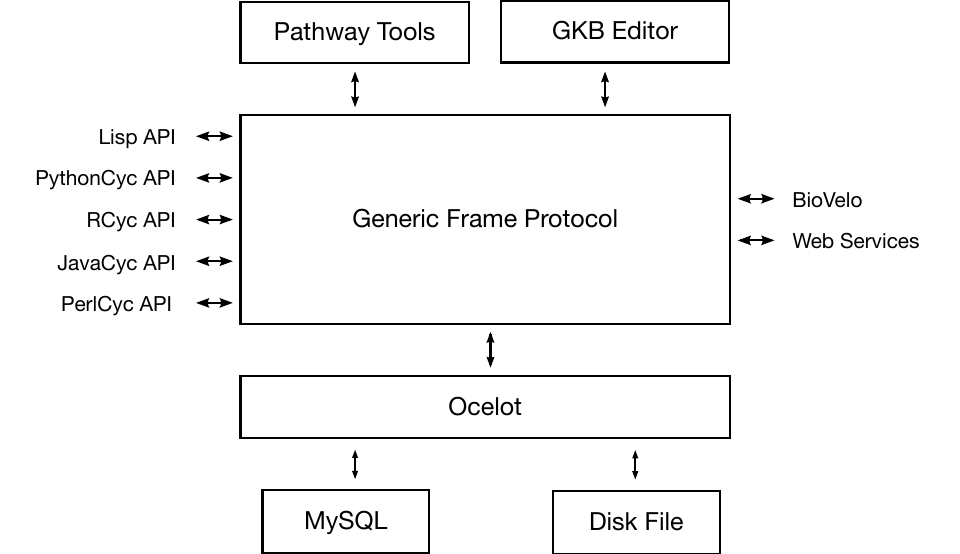}
\caption{Database architecture and APIs for Pathway Tools.}
\label{fig:arch-storage}
\end{center}
\end{figure}

Ocelot DBs can also be saved to disk files, in which case the RDBMS is
not needed (see Figure~\ref{fig:arch-storage}).  The file persistence
configuration is simpler to use since it does not require 
installation of an RDBMS.  It provides an easy and low-cost way to
begin a PGDB project; a project can switch to an RDBMS configuration
as its complexity grows.  The advantage of an RDBMS configuration is
that it provides Ocelot with multi-user update capabilities, and it
permits incremental (and therefore faster) saving of DB updates.  The
RDBMS configuration also enables Ocelot to maintain a history of all
DB transactions --- DB curators can examine the history of all updates
to a given object to determine when a given change was made, and by
whom.  This functionality is very useful when diagnosing mistakes
within a PGDB.

\begin{figure}
\begin{center}
\includegraphics[height=3in]{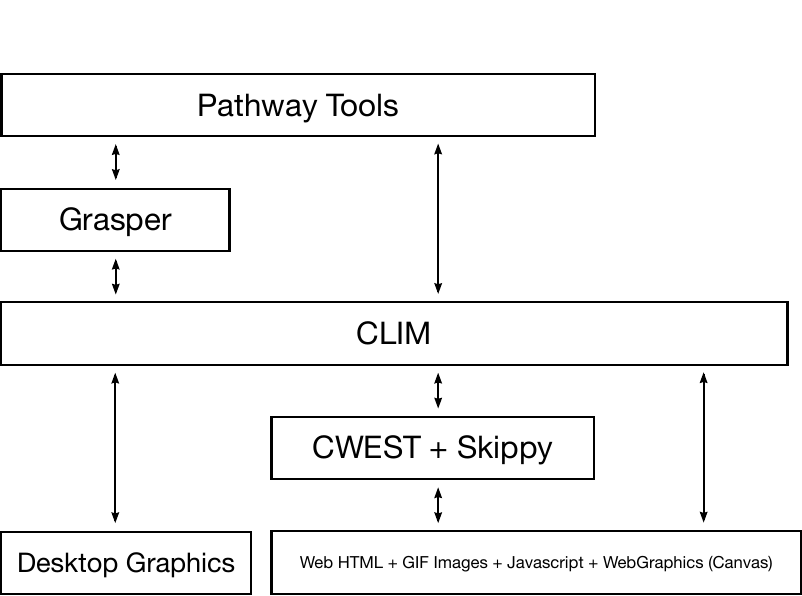}
\caption{Graphics architecture of Pathway Tools.}
\label{fig:arch-graphics}
\end{center}
\end{figure}

Figure~\ref{fig:arch-graphics} shows the graphics architecture of Pathway Tools.
The Grasper graph toolkit is used in pathway layouts, and in the
cellular overview and regulatory overview.  Grasper graphics, and
all other graphics generated by Pathway Tools, are rendered using the
Allegro CLIM Common Lisp graphics system, which is implemented using the X window system on Linux and Mac, and the native Windows API on Windows.
When Pathway Tools runs as a desktop application, CLIM graphics 
directly update the user's screen.  

\Ssection{Web Mode}

Pathway Tools can also run as a web server, which is how it powers
websites such as BioCyc.org.  Pathway Tools does not run with an
associated HTTP server such as Apache.  Instead, Pathway Tools provides a
fully functional AllegroServe web server that includes services such as compression
and connection keep alive.  
HTTP servers typically start a new operating-system process for each incoming web
request that terminates after the request has been serviced.  In
contrast, Pathway Tools starts one long-lived web server process that
can service many thousands of web requests; that process forks an internal thread
to service each incoming request.  

Pathway Tools processes an incoming web request in the following
manner.  The top-level directory name within an incoming URL indicates
whether the operation is requesting a static file or dynamically
generated page.

\bitem
\item {\bf Static files:} 
A small number of web pages, such as the home page and informational
pages for BioCyc.org, are implemented as disk files.  Pathway Tools
can serve file-based web pages like a traditional web server.

\item {\bf Dynamically generated pages:} Most Pathway Tools pages are
  generated dynamically by processing that typically includes querying
  PGDBs within the Pathway Tools virtual memory and generating query
  outputs and visualizations, often using the same code as for desktop
  mode.  The CLIM graphics generated by software such as the pathway
  layout code are dynamically converted to HTML and images using CWEST
  \cite{CWEST2}.  The resulting HTML and images are returned to the
  user's web browser, and the Pathway Tools web server awaits the next
  query.  \blue{Images are generated as JSON objects that are rendered
    by the browser on HTML5 canvas using our internal web graphics
    protocol, which includes specification of mouse-sensitive regions
    and of what operations should be invoked when the user clicks on
    such a region. Most such images are individually zoomable and can
    be output to a printer or postscript file.}

\eitem

\Ssection{\blue{Access Control}}

\blue{Pathway Tools in web mode supports both coarse- and fine-grained
  PGDB access control mechanisms. Coarse-grained access control simply
  allows a specified subset of PGDBs on a system to be accessed from
  the web, but does not support differential access for different
  users. The fine-grained access control system uses a Web Accounts
  database stored in MySQL to customize access on a per-user, per-PGDB
  basis, as follows:}

\begin{itemize}
  \item \blue{A PGDB can be designated either public or private. A public
    PGDB is accessible to all users of the Pathway Tools website. A
    private PGDB is inaccessible to users who are not logged in, and
    logged-in users who lack access privileges.}
  \item \blue{By default, a user has no access privileges to any private
    PGDBs. A user account can be granted either a blanket privilege to
    access all private PGDBs, or access to a specific, enumerated set
    of PGDBs.}
\end{itemize}

\section{Metabolic Modeling with MetaFlux}
\label{sec:metaflux}

The MetaFlux component of Pathway Tools is used to develop and execute
quantitative metabolic flux models for individual organisms and for
organism communities.  MetaFlux uses the steady-state modeling
technique of flux balance analysis, which can be used to predict the
phenotypes of an organism, or a community of organisms, based on   
a specification of available nutrients in the growth environment.

MetaFlux offers several modes of operation:

\begin{enumerate}
\item Solving mode: execute a metabolic model for a single organism or
  for a community of organisms
\item Development mode: generate hypotheses on how to fill gaps in a
  developing metabolic model
\item Knockout mode: run metabolic models under gene knockout
  scenarios
\item Flux variability analysis mode: perform flux variability
  analysis on a metabolic model and generate a robustness report
\end{enumerate}

MetaFlux can be controlled via a graphical user interface and via a Python API.

The next section provides an overview of metabolic model development
using MetaFlux.

\subsection{The MetaFlux Model Development Process}
\label{sec:metaflux-dev}

Pathway Tools provides a unique environment for the development of
metabolic flux models for several reasons.  First, it includes a range
of tools that support fast and accurate development of metabolic
models from annotated genomes.  Second, metabolic models developed
with MetaFlux are highly accessible to the user, and are coupled with
extensive enriching information, resulting in models that are easier
to understand and reuse.

The high-level steps for developing a metabolic model from an annotated genome
using Pathway Tools are as follows.  For more information on the 
genome-scale metabolic reconstruction process, we suggest the 
comprehensive COBRA-based \cite{COBRA2011} overview published by 
Thiele and Palsson \cite{Thiele10}.

\benum
\item The PathoLogic tool computes a metabolic
reconstruction by inferring the reactome of an organism from its
annotated genome.  By combining the enzyme-name matching tool with the
extensive reaction information in the MetaCyc database, we obtain
a very complete mapping of annotated enzymes to metabolic reactions.  

\item Subsequent inference of metabolic pathways by PathoLogic 
fills a significant number of missing (gap) reactions, because as well
as inferring the presence of pathways,
pathway inference also infers the presence of pathway
reactions that were not initially identified.

\item The pathway hole filler identifies enzymes that catalyze those pathway-hole
reactions. (Note this step is optional and is informative in nature,
because it does not modify the set of reactions in the model.)

\item The user can request that MetaFlux compute an initial
set of biomass metabolites for the organism.  MetaFlux will do so if
the organism falls within the twelve taxonomic groups for which
MetaFlux has defined biomass compositions, obtained from the
experimental literature.

\item The user supplies an objective function and constraints 
on metabolite uptake and secretion.  These can be based on experimental
observations of the organism under study, or can be set to arbitrary values 
to explore the theoretical behavior of metabolism.

\item The MetaFlux gap-filler identifies missing reactions, nutrients,
and secretions that enable a model to be solved.  It can be run on one
compartment at a time to simplify the gap-filling process for
eukaryotic organisms.  MetaFlux tools for computing blocked reactions and dead-end metabolites
identify potential errors and omissions in the metabolic network definition.

\item The reaction network, objective function, and constraints 
of the metabolic model are adjusted by the user until its predictions
match experimental results.
\eenum

When developing a metabolic model with MetaFlux, the reactions and
metabolites within the model are derived from (and stored in) the PGDB.  MetaFlux
automatically generates the system of linear equations for the model
from the PGDB.  Thus, to modify the reactions within a model, the user
edits the PGDB; to inspect the reactions and metabolites within a
model, the user can query the PGDB using the plethora of Pathway Tools query and
visualization operations.  The entire PGDB/model can be published on a
website using Pathway Tools, where, all reactions and pathways
that utilize a given metabolite are listed on the Pathway Tools
metabolite page for that compound; all model metabolites within a given molecular weight
range can be found using metabolite searches; all reactions within a
given cellular location or using a given set of reactants and products
can be found using reaction searches.

Furthermore, compared to other modeling environments, a metabolic
model stored within a PGDB contains extensive additional enriching
information.  Chemical structures within a PGDB enable reaction mass
and charge balancing.  Chemical structures and reaction atom mappings
aid users in understanding the chemistry of reaction transformations.
Pathways arrange reactions into biologically meaningful groupings.
Couplings between reactions, enzymes, and genes enable reasoning about
the roles of multi-subunit complexes, isozymes, and gene knockouts.
Regulatory information supports inferences about metabolic regulation.
Model testing and validation are facilitated by PGDB storage of growth
media and growth experiment results, and of gene knockout experiment
results.

Taken together, the Pathway Tools modeling environment---with its
extensive tools for inspecting metabolic model content and its
enriching information---renders MetaFlux models
significantly easier to understand, learn from, validate through
inspection, reuse, and extend than models produced with other
metabolic modeling software environments.

\subsection{Description of an FBA Model}

The description of a flux balance analysis (FBA) model is provided to
MetaFlux partly via a text file called an FBA input file and partly
via a PGDB. Typically, the FBA file specifies many parameters, but we
will describe only the most important ones. Some parameters are only
relevant for specific modes of MetaFlux, so that we will present these
parameters when describing that mode. All parameter names end with a
colon, whereas keyword options for parameters start with a colon. In
the following explanation, we omit the colon character to reduce
clutter.

\subsection{Solving Mode}

Solving mode computes flux values for the reactions in the metabolic
model given four inputs: a set of nutrient compounds, a set of
secreted compounds, a set of biomass metabolites that are synthesized
by the cell, and a set of metabolic reactions. The first three inputs
are supplied by the FBA file.  The set of metabolic reactions are
provided by the PGDB, but may be altered by the FBA file.
For example, the following file describes a (very simple) FBA model for
\coli: 

\begin{verbatim}
pgdb: ecoli

reactions: 
metab-all           # Include all metabolic reactions of PGDB ecoli.
FUMHYDR-RXN         # Example of including a reaction according to its 
                    # PGDB unique identifier.


mal -> fum + water  # Example of including a reaction by reaction equation.

biomass:

CYS[CCO-CYTOSOL] 0.0054
GLN[CCO-CYTOSOL] 0.2987
GLT[CCO-CYTOSOL] 0.2987
GLY[CCO-CYTOSOL] 0.3431

nutrients:

GLC[CCO-EXTRACELLULAR]             :upper-bound 10.0
OXYGEN-MOLECULE[CCO-EXTRACELLULAR] :upper-bound 1.0

secretions:
CARBON-DIOXIDE[CCO-EXTRACELLULAR]
WATER[CCO-EXTRACELLULAR]


\end{verbatim}

Note that this example is meant to show the syntax of the file
describing a model, it is not meant to show a working
model. The~\texttt{pgdb} parameter specifies the PGDB to be used in this
model.  The set of model reactions is specified by using the
\texttt{reactions} parameter. The keyword \texttt{metab-all}
specifies all metabolic reactions from the PGDB, whereas
\texttt{FUMHYDR-RXN} specifies one reaction using a unique
identifier.  A reaction equation can be provided to describe a
reaction that is to be present in the model, but is not present in the PGDB. 

The set of biomass metabolites are specified by parameter
\texttt{biomass}; each metabolite is specified either by metabolite
name or unique identifier, plus a compartment identifier in square
brackets, and an optional coefficient. Nutrients and secretions are
provided in the same manner, but no coefficients are allowed. Upper
and lower bounds can be provided to constrain nutrient uptake rates or
secretion production rates.

Specifying the metabolites of the biomass
reaction using groups is also possible. The use of groups present a better
structure of the biomass reaction. For example,

\begin{verbatim}
biomass:

:group val
Charged-VAL-tRNAs[CCO-CYTOSOL]   0.423162
VAL-tRNAs[CCO-CYTOSOL]          -0.423162
:end-group

:group thr
Charged-THR-tRNAs[CCO-CYTOSOL]   0.253687
THR-tRNAs[CCO-CYTOSOL]          -0.253687
:end-group
\end{verbatim}

These two groups, one named {\tt val}, the other named {\tt thr},
specify a relationship between a charged and an uncharged tRNA, for
valine and threonine. Essentially, the groups enable gathering the
compounds where one or more compounds, with negative coefficients, are
needed to produce the other(s) with positive coefficients. In other
words, these groups describe partial sub-reactions of the biomass
reaction because the metabolites with negative coefficients can be
considered as reactants, and the metabolites with positive coefficients
can be considered as products of a reaction, which is not necessarily
mass balanced. The negative and positive values do not need to be the
same absolute values because these sub-reactions are allowed not to be
mass balanced.  Metabolites can be repeated in different groups, but
repetition of a metabolite is not allowed outside groups.  Groups have
another use with the \texttt{try-biomass} parameter in development
mode (see subsection \ref{sec:developmentMode}).

\subsubsection{Modeling a Community of Organisms using Dynamic FBA}

The organisms within a community model share the same physical
space, which can now be specified by the user as a rectangular grid
such as to simulate an animal digestive tract.  At the start of the
simulation, the user seeds the grid squares with different starting biomasses of
organisms and metabolite concentrations.  Organisms and metabolites
may also be introduced at specified grid squares at specified time
points as the model runs.  At each dynamic FBA time step,
metabolite concentrations are updated to reflect metabolites
secreted by each organism, and by computed diffusion of
metabolites across the grid (diffusion of organisms is also computed). This
computation emulates the dispersion of metabolites and organisms due
to Brownian collision.  The diffusion coefficients of the metabolites and organisms are set to
$5x10^{-6}$ $cm^2/s$ and $3x10^{-9}$ $cm^2/s$, respectively.  At each
step, the concentrations of nutrients in each grid
square are decreased and the concentrations of secreted metabolites
are changed based on the rate of nutrient uptake and rate of
secretion by each model.  Each model is run independently in each time
step to compute growth rates for each organism, and then the organism
biomass is updated with respect to both that growth rate and a
specified death rate.  A death rate can be specified as a percentage
(between 0 and 100) of organisms dying at each step using the
parameter \texttt{organism-death-rate}.  The same organism death rate is
applied to all models. The order in which organism models are solved
is chosen at random at each step.

For example, the following COM file specifies two {\em E. coli}
models with some starting biomass at opposite locations on a $5x5$
grid.  The colony in the upper right is the wild type, but
the colony in the lower left is missing a single reaction
(NADH:ubiquinone reductase) which reduces its ability to use oxygen to
speed growth, resulting in somewhat lower population and less CO2
production.  The simulation shown here is anaerobic for the first
five time steps and aerobic for the remaining time.  The simulation is
run for 12 steps where each step represents one hour.

\begin{verbatim}
community-name: two-ecoli

fba-files: 
ecoli-strain-A.fba   :biomass 0.01 :locations (0 0) :steps 1
ecoli-strain-B.fba   :biomass 0.01 :locations (4 4) :steps 1

organism-death-rate: 0

grid-dimensions: 5 5
grid-real-dimensions: .5 .5 .5
nb-steps: 12       # default is 24
time-step: 3600    # seconds, default is 3600

exchange-compartments: [CCO-EXTRACELLULAR]

community-nutrients:
GLC[CCO-EXTRACELLULAR]             :supply 10 :locations (* *) :steps 1
OXYGEN-MOLECULE[CCO-EXTRACELLULAR] :supply 10 :locations (* *) :steps 5
\end{verbatim}

After the simulation run is complete, a solution file will be produced
that describes the nutrients used, secretions produced, accumulated
biomasses and metabolites used or produced for each FBA model and for
each grid.  These results can be visualized using 
generated figures and MPEG movie files with an animation of the
simulation results over time.

\subsubsection{Visualization of MetaFlux Community Model Outputs}
\begin{figure}
\begin{center}
\includegraphics[width=5in]{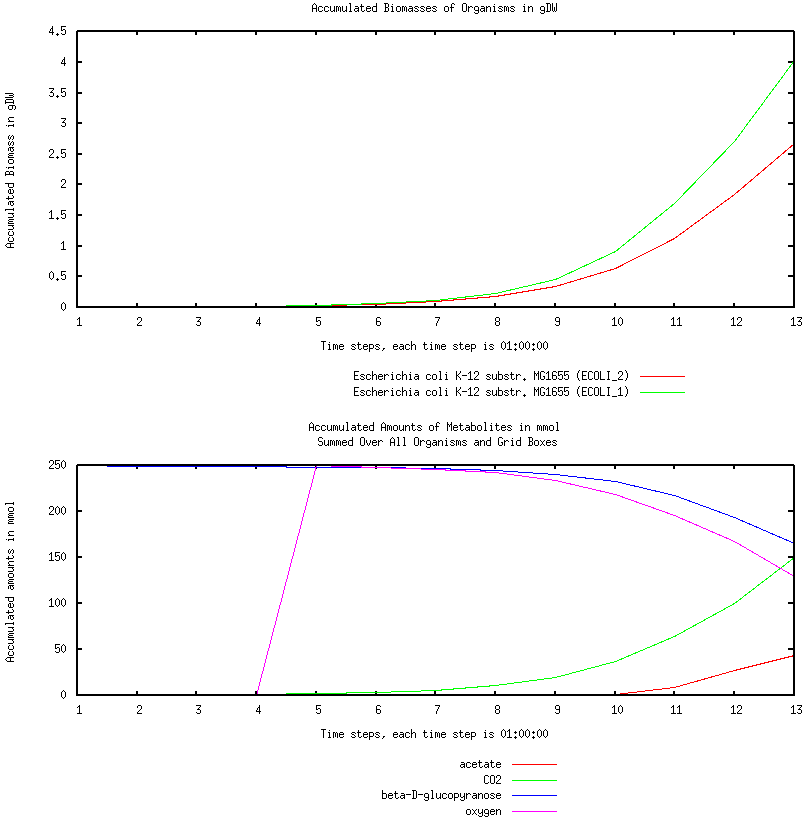}
\caption{Example X-Y Plots for aggregated biomass and metabolites for
  a dynamic FBA simulation of two {\em E. coli} models. The top figure shows the accumulated
  biomasses of each {\em E. coli} model over the entire simulation. The
  bottom figure shows the accumulated acetate, carbon dioxide,
  glucose, and oxygen over the entire simulation.}
\label{fig:dfba-aggregated}
\end{center}
\end{figure}

\begin{figure}
\begin{center}
\includegraphics[width=7in]{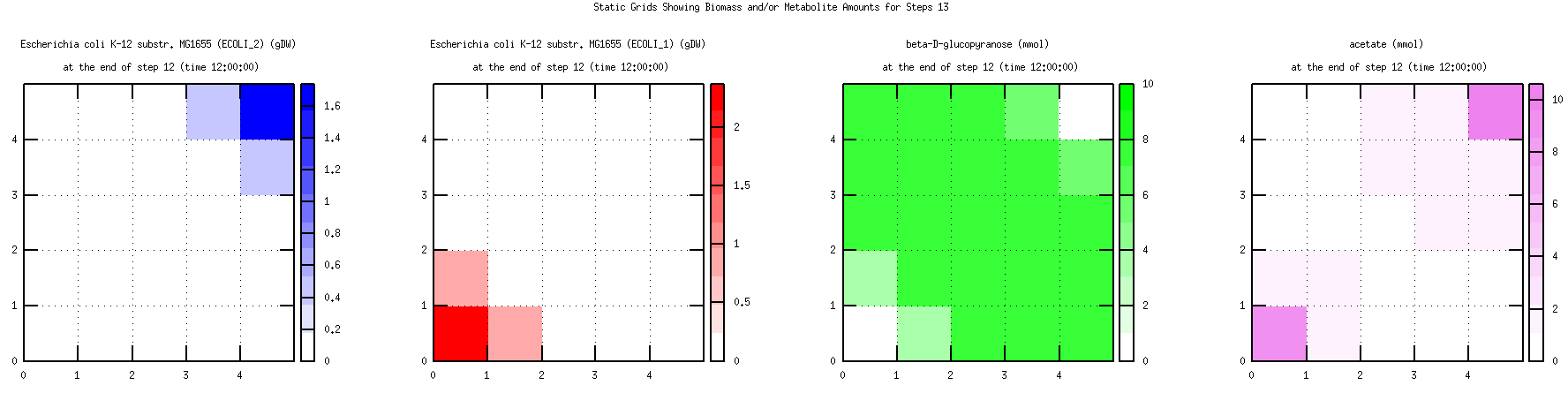}
\caption{Example spatial static grid output for biomass and
  metabolites for a dynamic FBA simulation of two {\em E. coli}
  models. Shown here are the final accumulated biomass (for both
  models) and metabolites (for glucose and acetate) at the end of the
  simulation.}
\label{fig:dfba-static-grids}
\end{center}
\end{figure}

When MetaFlux is used to model a community of organisms, 
its output results consist of organism biomass values for each time point in the simulation, and metabolite
concentrations for each time point for those
metabolites supplied as nutrients or secreted into the extracellular
space.  If the simulation takes place across a spatial grid, then the output
consists of organism biomass values and metabolite concentrations at each
square within the grid, for each time point in the simulation.  We developed
several new visualization tools that aid the user in digesting this large
amount of output data:

\begin{itemize}
\item {\bf X-Y Plot: Aggregated Biomasses/Metabolites.} This tool
  produces two separate X-Y plots.  One graphs the organism biomasses
  (in gDW) accumulated in the grid as a function of time.  The second
  graphs aggregate metabolite amounts (in mmol) across the grid as a
  function of time. The metabolites shown are for the extracellular compartment.
  The metabolites and organisms shown are selected by the user (see
  Figure~\ref{fig:dfba-aggregated}).
  
\item {\bf X-Y Plot: Metabolites Used/Produced per Organism.}
  The metabolites used and produced by each organism at each time step are
  shown in a separate X-Y plot for each organism. The metabolites are selected
  by the user. (The X-axis is time; the Y-axis shows aggregate metabolite
  concentrations across the grid for that organism.)

\item {\bf Static Grids: Biomasses/Metabolites.}
  A series of static 2D heatmaps depict the spatial grid in which the
  organism community grows, showing the organism biomasses (in gDW
  per grid box) and/or
  metabolites (in mmol per grid box) at several time steps. Each image
  of the grid depicts one organism and one metabolite among those
  selected by the user, at one time point (see Figure~\ref{fig:dfba-static-grids}).

\item {\bf Dynamic Grids: Biomasses/Metabolites.} This option is similar to the
  preceding static grids display, but an MPEG movie is created, in which
  each simulation time point corresponds to one step in the animation.
  An animation step is shown every three seconds. The FFmpeg
  program is used to create this MPEG movie, which will be shown
  using the default web browser installed on the user's computer.

\end{itemize}

\subsection{Development Mode}
\label{sec:developmentMode}

The development mode can be used to create an FBA model or to discover
what is wrong with a model that does not grow when growth is expected.
The main parameters used in development mode are the four parameters
\texttt{try-biomass}, \texttt{try-reactions}, \texttt{try-secretions}
and \texttt{try-nutrients}.

When MetaFlux is provided with a list of metabolites for the \texttt{try-biomass} in development mode,
the software tries adding these metabolites to the biomass reaction. 
That is, in development mode, MetaFlux computes
the maximum number of \texttt{try-biomass} metabolites
that can be produced by the model. In other words,
MetaFlux will output as a solution the largest subset of 
metabolites it can produce as biomass. 
In the early phases of developing a model, typically the entire biomass reaction is specified by
\texttt{try-biomass} and no metabolites are specified for the
\texttt{biomass} parameter.

The \texttt{try-biomass} parameter cannot specify any metabolite with
a negative coefficient not included in a group. This is required
because any such metabolite could be used as a ``free'' nutrient
for the organism. On the other hand, inside a group, negative
coefficients are allowed because MetaFlux tries to produce the entire
group of metabolites, not any one of them independently. A group of metabolites
is supposed to form a cohesive unit where any metabolite specified
with a negative coefficient is used to produce some other metabolites
of that group. Therefore, declaring all metabolites that have negative
coefficients in some groups enables the entire biomass reaction to be used as
a try-biomass set, indicating for all metabolites which ones can or
cannot be produced.

The \texttt{try-reactions} parameter can be used to try to add
candidate reactions to the model to increase  the number of
\texttt{try-biomass} metabolites produced by the model. Therefore, the
\texttt{try-reactions} parameters is typically used when at least one
biomass metabolite is not produced. The candidate reactions are selected from MetaCyc.
A single keyword, \texttt{metacyc-metab-all}, instructs MetaFlux to try all
the metabolic reactions of MetaCyc.  Alternatively, a list of candidate reactions
can be specified by their unique identifiers. 
MetaFlux tries to produce as many 
biomass metabolites as possible from the \texttt{try-biomass} section, by
adding as few reactions  as possible from the \texttt{try-reactions} set.
This computation is performed using optimization as a
Mixed-Integer Linear Program (MILP), which can be computationally
expensive to solve. The amount of time to compute an optimal solution
varies depending on the number of candidate reactions. It can take a
few seconds to several hours, or even days.

MetaFlux also has a fast development mode that can be used with the
\texttt{try-reactions} parameter only. It may run much faster than the
general development mode and may provide different solutions. It uses
a heuristic that does not necessarily provide an optimal solution. But
optimality is not always the best possible biological solution. Please
consult~\cite{LatendresseGapFill14} for a more complete
description of the advantages and disadvantages of fast
development mode.

\subsection{Knockout Mode}

Knockout mode is used to computationally evaluate the impact of
removing genes or reactions from an FBA model. This mode is used to
predict essential genes of an organism for a given growth
environment, and can also be used to evaluate the accuracy of a model if
experimental gene knockout data are available. MetaFlux can compute
single, double, or higher numbers of simultaneous knockouts.


When run, MetaFlux solves the model without any knockouts, and
then solves for each gene to knockout by deactivating the reactions associated
with that gene. Note that a given gene may deactivate none, one, or several
reactions, because some genes may have isozymes, or catalyze several
reactions. Requesting a summary only of the results of
the knockout or requesting that, beside the summary, a complete solution
file be produced for each gene knockout is also possible. See the subsection~\ref{sec:sol}
for more details on the solution files produced.

\subsection{Flux Variability Analysis Mode}
Flux variability analysis (FVA) can be used to determine the robustness of 
metabolic models under different simulated growth conditions by
computing the minimum and maximum possible flux values of each
reaction for a given growth condition.  After FVA is run, a report is generated
showing the reactions which are blocked (these reactions cannot
carry flux), fixed (these reactions have identical,
non-zero minimum and maximum flux values), or free (these reactions
have flux values that freely fluctuate within some range between the
lower and upper bounds).  Other uses of FVA
may include classifying reactions as to whether they are fully-coupled,
partially coupled, or not coupled to biomass flux or investigating 
alternate solutions by fixing relevant reaction flux values.

\subsection{Outputs Generated by MetaFlux} 

Whichever mode is used to execute MetaFlux, the following output files are
generated: the solution file that contains a description of the active
reactions and metabolites used and produced, the log file to
describe issues that may exist in the model, and a data file
for the fluxes that can be displayed with the Cellular Overview map of
Pathway Tools.

In solving mode, the main output is the computed optimal assignments
of reaction fluxes.
For an organism-community model, solution, log, and data files are produced
for each organism.

In development mode, the outputs are the set of biomass metabolites
that can be produced; the try nutrients utilized (if any); the try
secretions produced (if any); the reactions that actively carry flux;
and a minimal list of suggested reactions to be added to the model,
and reactions whose directions are reversed (if any) to produce
otherwise unproducible biomass metabolites.  Development mode also
identifies which biomass metabolites could not be produced by the
model (if any), despite the additions from the try sets.

\subsubsection{Reports Generated by MetaFlux}
\label{sec:logfile}

The set of reactions specified by the PGDB and the FBA input file may contain
reactions that cannot, or that should not, be used in a model. MetaFlux
checks each reaction  to ensure that its inclusion in the model
will not invalidate the model.  In
particular, all reactions are verified to be mass and
charge balanced. Reactions that are unbalanced, or cannot be shown to be balanced, are not included in
the model, but are listed in the log file.

Another step in model execution is to instantiate the
generic reactions of a model according to the compounds available in the PGDB. A
generic reaction has some compound classes as products or reactants (e.g.,
``sugars'').  Each computed reaction instantiation
is added to the model. If a generic reaction instantiation
was not possible (for example, when which compound
instances should be used for classes on each side of a reaction is ambiguous)
then such generic reactions are listed in the log file.

{\noindent \bf Blocked reactions.}  MetaFlux analyzes the network of
reactions specified by the PGDB plus the FBA input file to detect if
some reactions are blocked. A blocked reaction is a reaction that can
never carry flux, given the nutrients, secretions, biomass
metabolites, and reactions specified for a particular model
execution. That is, blocked reactions are a function not only of the
network, but of the cellular growth conditions.  A blocked reaction
$R$ has at least one reactant $M$ that is not produced by any reaction
in the model, and that is not provided as a nutrient; or $R$ has a
product $M$ that is not used as a reactant by any other reaction and
that is not secreted, and is not specified as a biomass metabolite
with a positive coefficient.  These reactions are the {\it basic
  blocked reactions}, and the metabolites that caused basic blocked
reactions to be blocked are called {\it basic blocking metabolites}.

Blocked reactions can never carry flux because in steady-state
modeling, the producing and consuming fluxes for every metabolite must
be balanced.  But the preceding metabolites $M$  could not have
balanced fluxes if a blocked reaction $R$ carried flux, because
according to the preceding definition, $M$ must have either zero
reactions that produce it  or zero reactions that consume it.

Additional blocked reactions can be identified by eliminating basic
blocked reactions from the model, causing more reactions to become
basic blocked reactions. This process of removing basic blocked
reactions from the model is repeated until no more reactions become
blocked. The detection of blocked reactions is done before the linear
solver is called (that is, this detection does not depend on the
fluxes of reactions, but is a static evaluation of the model).

The set of blocked reactions is listed in the log file, grouped by
basic blocking metabolite (the reactions are also grouped by
pathways). The basic blocking metabolites are, in a sense, the root
causes of blocked reactions, thus their identification is quite
valuable for model debugging.

\subsubsection{The MetaFlux Solution File}
\label{sec:sol}

The content of the solution file generated by MetaFlux 
depends on the mode of MetaFlux. In solving mode, the solution file
contains the uptake rates of the nutrients, the production rates of
the biomass metabolites and secretions, and the fluxes of the model
reactions. If no growth is obtained (that is, the flux of the biomass reaction is
zero, or very near zero), this condition is stated in the solution file. 

When solving a community model, a community solution file is generated
that contains a summary of the result, and a complete solution file is
produced for each organism in the community.  The summary
lists many values: the growth rate of each model, the compartments
that were involved in the exchange of metabolites between the
organisms; the secretions used as nutrients by some organisms with
their flux; the secretions that were not used by any organisms; and the
secretions that were not produced by any organism of the community. In
essence, the summary gives the overall view of the exchange of
metabolites between the organisms and the rate of growth of each
organism.

In development mode, the solution file contains essentially what is
given in solving mode, but the flux values are only meaningful as to
whether they are zero or non-zero.  For example, if a
\texttt{try-biomass} parameter was specified with some metabolites,
the metabolites that could or could not be produced are
identified. This is different from the solving mode, where only one
biomass metabolite that cannot be produced will result in no growth,
but with no indication of which metabolites could not be produced.

In knockout mode, a solution file with a summary of the results is
produced. The summary includes the growth of the model with no knockouts,
followed by a list of every knockout that was requested, based
on genes and/or reactions to knockout, and the resulting flux for the biomass
reaction. Requesting a complete solution file 
for each knockout performed is also possible.

\subsubsection{Painting MetaFlux Fluxes on the Cellular Overview and
  Omics Dashboard}

MetaFlux also generates a data file that can be used with the Cellular
Overview of Pathway Tools (see Section~\ref{sec:cellov}). The graphical user interface of MetaFlux
enables the user to click one button to invoke the Cellular Overview
with the data from that file. The Cellular Overview displays all the
reactions, grouped by pathways, of the organisms, and highlights, with
colors indicative of flux values, the reactions that have positive fluxes. This overview
enables the user to visually assess the metabolic activities of a
model solution.

For a community of organisms, a data file is generated for each
individual organism, and the graphical user interface gives direct
access to each individual Cellular Overview of the organisms involved in
the community.

MetaFlux has been extended so that the reaction fluxes computed by
solving mode can be displayed on the Omics Dashboard \cite{DashGene17}
(see Section \ref{sec:omics-dashboard}) to speed user interpretation of fluxes.  The Dashboard is an
interactive tool for hierarchical visualization of large-scale
data sets.  The top level of the Dashboard graphs the aggregate
behavior of every metabolic subsystem of the cell, from nucleotide
biosynthesis to aromatic compound degradation to aerobic respiration.
The user can drill down into any of these graphs to view the fluxes of
individual pathways.

%
%

\section{Survey of Pathway Tools Compatible Databases}
\label{sec:pgdbs}

According to user surveys, Pathway Tools users have created 9,800
PGDBs for organisms from all domains of life, in addition to the
\NbiocycPGDBs\ PGDBs available from BioCyc.org.  This section
summarizes sites that provide public-access user-generated PGDBs.

With highly curated PGDBs available for many important
organisms, it is not clear why users would consider using the
uncurated (and therefore lower quality) pathway DBs available for
these same organisms from other pathway DB providers.  For example,
consider the highly curated AraCyc pathway database for {\em Arabidopsis thaliana}
\cite{Mueller03,AraCycURL}. AraCyc contains mini-review summaries for enzymes
and metabolic pathways; thousands of literature references; evidence
codes for enzyme functions and metabolic pathways (indicating which pathways are supported
by experimental evidence); and information on
enzyme subunit structure, activators, inhibitors, and cofactors.  In
contrast, KEGG data on {\em Arabidopsis}
contains none of the preceding information.  In addition, AraCyc
curators have carefully refined the metabolic reactions and pathways
present in AraCyc, such as to remove false-positive computational
predictions, and to add {\em Arabidopsis} reactions and pathways from
the biomedical literature to AraCyc.  Although KEGG updates its
reference pathway map diagrams periodically to contain new pathways
and reactions from different organisms, the KEGG approach of
computationally coloring reactions within pathway maps based on the
presence of enzymes for those reactions within a genome results in
significant ambiguity.  If AraCyc curators are reasonably certain that
a reaction or pathway is absent from {\em Arabidopsis}, they remove it
from the database.  The KEGG model does not allow such removal, so whether an uncolored reaction is truly
absent from an organism or whether the gene for its enzyme has not
yet been identified in the genome
is never clear within KEGG.  This situation results in a real
conundrum for a scientist who wishes to assemble the list of reactions
likely to be present in {\em Arabidopsis} from KEGG, because is no
way exists to distinguish the many uncolored reactions that are likely
present but for which no gene has been identified, from the many
uncolored reactions that are clearly known to be absent from {\em
Arabidopsis} (which curators have deleted from AraCyc).

Table \ref{tab:pgdb-collections} lists PTools-based websites,
the number of PGDBs they contain, and their
supporting institution.  This table is based on the web page
\cite{OtherPGDBsURL} which is updated on a semi-regular
basis.

\begin{table}[!h]
  \footnotesize
  \centerline{
\begin{tabular}{|l|l|l|l|} \hline
{\bf Database or Collection}	&{\bf Scope}	&{\bf Institution}	&{\bf URLs} \\ \hline \hline
AcypiCyc	&{\it Acyrthosiphon pisum}	&Universite
de Lyon	& \cite{AcypiCycURL}\\
	&{\it Aphis glycines}	&	& \\
	&{\it Daktulosphaira vitifoliae}	&	& \\
	&{\it Diuraphis noxia}	&	& \\
	&{\it Myzus cerasi}	&	& \\
	&{\it Myzus persicae}	&	& \\
	&{\it Rhoplalossiphum padi}	&	& \\
\hline
ArthropodaCyc	&PGDBs for 32 arthropods	&ArthropodaCyc group, France
& \cite{ArthropodaCycURL} \\ \hline
CalbiCyc	&{\it Candida albicans}	&Department of Genetics,
Stanford U., USA	&  \cite{CandidaCycURL}  \\ \hline
CottonCyc	&{\it Gossypium raimondii}	&CottonGen,	& \cite{CottonCycURL}\\
	&{\it Gossypium hirsutum}	& Washington State University	& \\
	&{\it Gossypium arboreum}	&	& \\
	&{\it Gossypium barbadense}	&	& \\
	&{\it Gossypium turneri}	&	& \\ \hline
DiatomCyc	&{\it Phaeodactylum tricornutum}	&Center for
Plant Systems Biology,	& \cite{DiatomCycURL}\\
	&	&University of Ghent, Belgium	&\\
\hline
Gramene	&{\it Brachypodium distachyon}	&Gramene curators
(CSHL, OSU, EMBL-EBI)	& \cite{GramenePathwayURL} \\
	&{\it Oryza sativa japonica} Nipponbare	&	&\\
	&{\it Sorghum bicolor} BTX623	&	&\\
	&{\it Zea mays}	&	&\\
	&and mirrored PGDBs for 9 organisms	&	&\\
\hline
GutCyc	&418 gut bacteria available as flat files only	&University of
British Columbia,	& \cite{GutCycURL}\\
	&	&	Stanford University	& \\
\hline
MicroScope	&PGDBs for 3749 organisms	&C. Medigue,
Genoscope, France	& \cite{MicroScopeURL}\\
\hline
PlantCyc	&PGDBs for 101 plants	&Plant Metabolic
Network, 	& \cite{PlantCycURL}\\
	&	&Carnegie Institution, USA	&\\
\hline
PseudoCyc	&{\it Pseudomonas aeruginosa}	&F. Brinkman, Pseudomonas
Genome Project,	& \cite{PseudoCycURL} \\
	&	& Simon Fraser U., Canada	&\\
\hline
SoyCyc	&{\it Glycine max}	&SoyBase	& \cite{SoyBasePGDBURL}\\
\hline
Trypanocyc	&{\it Trypanosoma brucei}	&INRA Toulouse	& \\
	&{\it Leishmania major}	& France	& \cite{TrypanoCycURL}\\
\hline
USDA Bacterial Genomes	&{\it Mannheimia haemolytica} Strain	&USDA
SciNet	& \cite{USDACycPathwaysURL} \\
	&USDA-ARS-USMARC-183	&	& \\
	&{\it Salmonella enterica} Strain	&	& \\
	&serovar Montevideo str. CDC 2012K-1544	&	& \\
	&{\it Escherichia coli} O157:H7 350	&	& \\
\hline
VitisCyc and others	&{\it Vitis vinifera}	&Center for Genome Research
and Biocomputing,	& \cite{VitisCycURL} \\
	&{\it Eucalyptus grandis}	&Oregon State University	& \\
	&{\it Fragaria vesca}	&	& \\
	&and mirrored PGDBs for 10 organisms	&	& \\
\hline
Yeast Biochemical Pathways	&{\it Saccharomyces cerevisiae}	&SGD
Curators, & \cite{SGDURL} \\
	&	&Stanford University, USA	& \\
\hline
\end{tabular}
}
\caption{\label{tab:pgdb-collections}
{\bf Main collections of PGDBs outside of BioCyc.}
}
\end{table}

To facilitate sharing of PGDBs among multiple users, we have created a
PGDB Registry that enables peer-to-peer sharing.  PGDB sharing is
desirable because a user whose own computer has a copy of a PGDB can
use Pathway Tools functionality that would not be available through a
remote Pathway Tools web server, such as functionality that exists in
desktop mode only, and such as comparative operations.  Comparative
analysis of two or more PGDBs is possible only when they are loaded
into the same instance of Pathway Tools.

The PGDB Registry uses a server maintained by SRI that tracks the
locations of available PGDBs that PGDB authors have registered for
downloading.  The author of a PGDB can register that PGDB by
using a command within Pathway Tools that creates an entry for the
PGDB in the Registry server, and places the PGDB on an FTP or HTTP
server of the author's choosing.  Users who want to download a PGDB
from the Registry can view available PGDBs by using a web browser (see
URL \cite{PToolsRegistryURL}) or using Pathway Tools itself.  With a
few mouse clicks, a user can download a PGDB from the registry using
Pathway Tools.

\section{Comparison of Pathway Tools with Related Software Environments}
\label{sec:related}

Pathway Tools stands out with respect to related software tools in the
breadth of the functionality and the high level of integration that it
provides.  It addresses a very large number of use cases.  And it
provides schema, visualization, and editing support for an unusually
large number of data types in addition to metabolic and signaling pathways, including
chromosomes, genes, enzymes, transporters, regulatory networks, and compounds.
The following comparison is organized roughly according to the use
cases presented in Section~\ref{sec:use-cases}.

\Ssection{Development, Visualization, and Web Publishing of Organism Specific Databases}

\SSsection{Metabolic Pathway Information}

Other software systems for managing metabolic pathway information are
KEGG \cite{KEGG06,KEGG08}; The SEED \cite{Overbeek14};
GenMAPP \cite{GENMAPP07,PathVisio08};
PathCase \cite{PathCase03,PathCase06}; VisANT \cite{VisANT07}; and Reactome
\cite{Reactome14}.  KEGG, Reactome, and GenMAPP employ static, pre-drawn pathway diagrams, a model that does not
scale to produce custom pathway diagrams for thousands of
different pathways in different organisms.  Nor can the static approach produce
multiple views of a given pathway at different levels of
detail, as can the Customize Pathway option in Pathway Tools that enables
the user to choose exactly which graphical elements (e.g., gene names, EC
numbers, metabolite structures, activators and inhibitors)
appear in the pathway diagram, and to color a pathway diagram with
omics data.

PathCase and VisANT do have pathway layout capabilities,
but the resulting diagrams bear little resemblance to those found in
the biomedical literature, nor are they particularly compelling
visually.  Their diagrams can be customized in various ways
(e.g., choice of layout algorithm, data overlays, and selective
compaction of nodes), but they do
not offer the same types of customization or multiple-detail views offered by
Pathway Tools.

Cytoscape \cite{Cytoscape03} is a general tool for displaying
biological networks that embodies the philosophy that general graph
layout techniques can satisfactorily depict any biological network.
Although the Cytoscape layout algorithms are a terrific fit for
display of protein interaction maps, we assert that the general layout
algorithms do not
produce useful results for metabolic pathways.  We believe that
superior visualization results are obtained when layout algorithms
are specifically tailored to metabolic pathways.  For example, Pathway
Tools provides separate layout algorithms for circular, linear, and
tree-structured pathways to make the structure of those pathways stand
out prominently to the biologist.  Biologists developed their
pre-computer depictions of metabolic pathways for important reasons,
namely to clearly present the topology of a pathway.

SEED models lack a unifying pathway visualization framework---they
use KEGG pathway diagrams, often supplemented with simple
text-based diagrams and/or images imported from other sources, with
limited interactivity beyond hyperlinks.

GenMAPP, Reactome, VisANT, KGML-ED \cite{Klukas07}, and PathCase have
pathway-editing capabilities.  KEGG and SEED lack pathway editing,
meaning that users cannot introduce new organism-specific pathways,
nor modify a reference pathway to customize it to a specific organism,
thus eliminating the possibility of removing erroneous reaction steps
from a pathway, or of adding missing reactions to a pathway.

KEGG Atlas \cite{KEGGatlas08} has a comparable
analog of our Cellular Overview diagram (which we introduced in 1999
\cite{Cyclone99,OverviewURL}).  The KEGG Atlas diagram was constructed
by hand and provides a single set of overview metabolic maps for all
organisms in KEGG, as opposed to the organism-specific overviews that
Pathway Tools generates through advanced layout algorithms.  The KEGG
diagram is not as flexibly queryable or interactive as the Pathway
Tools diagram, and it also lacks semantic zooming for adding increasing
details. A notable improvement occurred with Pathway Projector
\cite{Tomita09}.  It relies directly on KEGG for the overall
overview diagram and other data, but graphically, it is enhanced,
adding EC numbers and genes to reaction edges.

Reactome includes a different type of high-level overview diagram
called Fireworks, which shows a hierarchical organization of
nodes representing
metabolic and signaling pathways, mostly the latter.  Users can navigate to pathways of interest with this
diagram, and highlight pathways that contain objects such as compounds
or genes that are supplied in an analysis interface.  Fireworks can
zoom in until pathway names are shown.  From there, one can click to a
specific pathway display for the details.  However, Fireworks does not
depict individual reactions or metabolites; therefore, it is not a
metabolic map diagram.

Tools such as VisANT and
Cytoscape can be used to display very large networks, including the
entire metabolic network for an organism, and these diagrams can be
used to visualize omics data and interacted with in other ways, but
given that such diagrams are laid out using general-purpose network
layout algorithms, individual pathways are not likely to be
recognizable.

No other general tool can show reactions with atom-mapping
information.

Computationally generated diagrams for electron transfer reactions and
pathways, as well as for transport reactions, with their depiction of
compartments at a cellular membrane, are unique features of Pathway
Tools.

\SSsection{Genome and Proteome Information}
\label{sec:gen}

Most model-organism DB tools include genome browsers and gene pages.  A
representative sample of such systems includes GBrowse
\cite{GBrowse02,GBrowse07}; JBrowse \cite{Skinner09}; IMG
\cite{Markowitz14}; Entrez Genome \cite{EntrezGenomeURL}; the UC Santa
Cruz (UCSC) Genome Browser \cite{UCSC15}; Ensembl \cite{Ensembl15};
and PATRIC \cite{Wattam14}.

The Pathway Tools genome browser uses a different approach than the others.
First, all the other genome browsers depict the chromosome as a single horizontal
line, with tracks information below it.  Pathway Tools uses this approach
when the user explicitly enables tracks, but when tracks are not
enabled, a multi-line wrapped display is used to present the most
possible information in the available display window.  Second, all the
other genome browsers use tracks to depict coding regions, promoters,
and other genome features.  Although Pathway Tools does provide
a tracks capability, its assumption is that consensus curated
information on coding region extents, promoters, terminators, and
transcription-factor binding sites should be depicted ``inline'' on
the multi-line chromosome.  This approach results in a much more
information-dense display.

GBrowse is highly customizable through the addition of data tracks.
Wrapped multiline displays that will make full use of screen real
estate are not available.  In GBrowse, the semantics of semantic
zooming has to be configured but provides flexibility.  GBrowse can
provide comparative genomics with additional tools like SynView
\cite{Wang06} or GBrowse\textunderscore syn \cite{McKay10}.

JBrowse is a JavaScript-based genome browser that could become a
replacement for GBrowse.  Its main innovation is to rely on the web
browser to perform most of the processing and drawing, making it fast
and smooth.  PATRIC uses JBrowse.

The IMG genome browser (or ``chromosome viewer'', in their parlance)
is very basic, and does not support zooming.  Some views are shown as
wrapped multiline displays, and several gene coloring schemes can be
selected, but no data tracks are offered.  For comparative genomics,
three different, unrelated synteny viewers are available.

The Entrez Genome sequence viewer has no real semantic viewing other
than showing the DNA sequence at maximum zoom level.  It offers support
for adding extra tracks and an alignment view for comparative genomics.

The UCSC Genome Browser is mainly geared towards eukaryotes and has an
extensive support for addition of data tracks.  But no real semantic
viewing is available other than showing the DNA sequence at maximum zoom level.

Ensembl is focused on eukaryotes, similarly to UCSC, and also has
extensive support for extra tracks.  It offers no real semantic
viewing, other than showing the DNA sequence at maximum zoom level.

Most sequenced genomes are circular, but Pathway Tools seems to have
the only genome browser that can seamlessly depict the junction between
the first and last base-pair to show a contiguous view.  All other
genome browsers seem to artificially linearize all chromosomes.  Also,
Pathway Tools seems to be unique in providing a compact overview of
the entire genome (the ``genome overview''), which enables coloring each gene with omics data.



\SSsection{Regulatory Networks}

A number of bioinformatics databases include regulatory network
information; however, the majority of these databases and their
associated software environments can represent information on
transcription-factor-based regulation only, such as RegTransBase
\cite{RegTransBase07}; TRANSFAC \cite{Transfac06}; CoryneRegNet \cite{CoryneRegNet07}; ProdoNet
\cite{ProdoNet08}; and DBTBS \cite{DBTBS08}.  The exception is
RegulonDB \cite{RegulonDB08}, which can also capture RNA-based
regulation, including riboswitches, attenuators, and small-RNA
regulators.

We are not aware of tools comparable to the Regulatory Overview in
being able to display and interrogate large, complete cellular
regulatory networks, although CoryneRegNet and ProdoNet
display smaller regulatory networks.  CoryneRegNet 
also displays omics data onto its regulatory network diagrams.  


\SSsection{Query Tools}

Other organism DBs provide a subset of the three tiers of
queries provided by Pathway Tools (quick search, object-specific
searches, and Structured Advanced Query Page).  Virtually all provide
a quick search.  Sites providing particularly extensive
object-specific searches are FlyBase \cite{Flybase05}; Mouse Genome
Informatics \cite{MGD08}; EuPathDB \cite{PlasmoDB09}; and BioMart
\cite{Biomart09}.  BioMart is used by bioinformatics DBs, including
WormBase, Rat Genome Database, UniProt, Reactome, and Galaxy. Its
underlying query language is Perl, using the BioMart libraries.
However, none of the preceding systems provides the query power of the
Pathway Tools SAQP.  For example, BioMart does not allow the user to
construct arbitrary queries that perform joins (queries that combine
multiple data types); it provides only the ``and'' logical operator
(the ``or'' operator is not available); and it includes only a limited
form of ``not.''  Biozon \cite{Biozon06} (\verb+biozon.org+) provides
a web interface that provides fairly complex querying, including join
operations, over several biological DBs.


\Ssection{Extend Genome Annotations with Additional Computational Inferences}

KEGG, Model SEED, and Reactome are the only other tools that can
predict pathways from genome data.  The pathway hole filler and
transport inference parser tools are unique to Pathway Tools.  Many
genome-annotation pipelines include operon predictors.

\Ssection{Analysis of Omics Data}

Kono \etal\ introduced a tool based on the Google Maps API for
painting omics data onto an enhanced KEGG Atlas map, called Pathway
Projector \cite{Tomita09}.  This tool does not produce animations as
our omics viewers do.  However, it can depict time-series expression
data as small histograms that are reminiscent of Pathway Tools' omics pop-ups.
It also provides more powerful highlighting options than KEGG Atlas.
Sequence-based search and highlighting is available via KEGG BLAST,
although it can take minutes.  Users can manually annotate the map
with custom markers and lines, which can be exported via XML and
shared with others.

KEGG Atlas \cite{KEGGatlas08} enables some limited highlighting, but
mostly by using KEGG identifiers.  No animation or omics pop-ups seem
to be available.

Reactome \cite{Reactome14} can paint omics data onto the 
Fireworks hierarchical pathway overview.  Zooming to high
detail reveals the pathway names, but zooming does not  go
to the reaction level.  Showing time-series
data as animations is possible.  No omics pop-ups seem to be available.

GenMapp \cite{GENMAPP02}; VitaPad \cite{Holford04}; VisANT; and ArrayXPath
\cite{Chung04} paint omics data onto single pathways, rather than onto
a full metabolic overview.

\Ssection{Metabolic Modeling}

Other well-known tools providing constraint-based metabolic modeling
include COBRA \cite{COBRA2011} and Model SEED \cite{Henry10}.
Many other FBA tools exist, but a detailed comparison with all
of them is beyond the scope of this paper. Lakshmanan \etal\ provide a
comparison among 19 FBA tools, including MetaFlux (see their Tables 2,
3, and 4) \cite{Lakshmanan14}. This paper was first published in 2012
and therefore covers an old version of MetaFlux. Additions to MetaFlux
since that time include expansion to the Windows platform, import of
SBML files, solving of models for organism communities, and support
for dynamic FBA.  Other tools  have probably developed new capabilities as
well.  We now examine some of the issues raised by Lakshmanan \etal\ 
in more detail.

Our design of MetaFlux has emphasized ease of use and acceleration of
the very slow model development times of 12--18 months stated in some
FBA publications.  At SRI we have reliably been able to create FBA
models for bacteria and fungi in 3--4 weeks (albeit to a lower level
of validation than some published models).
We note that in another recent survey of 
metabolic modeling software \cite{Hamilton14}, Pathway Tools
was the only software package recommended for users new to modeling.

Elements of the MetaFlux approach that speed model
development are as follows.  
(a) Use of MetaFlux does not require
programming ability as some other tools do. 
(b) Lakshmanan mentions the collection of metabolic reactions for an
organism at the start of all metabolic modeling projects in passing,
without noting the strong link between model accuracy
and the completeness of the initial reaction set.  The
PathoLogic enzyme name matcher (see Section~\ref{sec:pwy-inference})
combined with the extensive reaction
database of MetaCyc \cite{MetaCycNAR18} provide a powerful resource for
reactome prediction.  Furthermore, when metabolic pathway prediction
follows reactome prediction, it produces a more complete metabolic
network, because it imports all reactions in a
pathway, even those that lack enzyme assignments, thus simplifying the
later gap-filling step.  Furthermore, the fact that Model SEED contains
approximately $1/3$ as many metabolic pathways as does MetaCyc raises
a major question in the use of Model SEED for pathway
reconstruction.\footnote{This number was obtained by manually
  reviewing the 1,320 subsystems present in the February 2015 version
  of Model SEED, then removing all subsystems not related to metabolism,
then removing subsystems for individual metabolic enzymes.  This
process yielded 730
subsystems that correspond to metabolic pathways, compared to the 2,300
pathways in the February 2015 version of MetaCyc. }
(c) MetaFlux has a more
powerful gap-filler than any other tool: it can gap-fill not
only reactions but also nutrients and secretions, and it can
identify which biomass metabolites cannot be synthesized by
the model, a critical step in helping the user discover which aspect of the
network is incomplete.  This is a feature not found in other tools to our knowledge.
(d) MetaFlux computes blocked reactions and metabolites, reporting to the user the basic blocking
metabolites that are the root causes of model blockages.
(e) MetaFlux computes the reaction balance from the reaction equations plus chemical
structure data to ensure that unbalanced reactions are prohibited from
inclusion in a model.

Many other tools use the GLPK solver (some not exclusively).  MetaFlux
uses the SCIP solver \cite{SCIPURL}, which has a much faster
Mixed-Integer Linear Programming (MILP) solver, needed for gap-filling,
and is free for research use by academics.  The SCIP MILP solver is fast,
although not as fast as commercial solvers.

The comment by Lakshmanan \etal\ on p8 that ``none of
the [network visualization] tools can handle large-scale models''
clearly does not apply to MetaFlux, as its cellular overview diagram
does handle genome-scale models.

Lakshmanan \etal\ suggest (p10) that more tools should consider linking with
a biological model storage DB.  Pathway Tools has taken this approach
to an extreme, as described in Section~\ref{sec:metaflux-dev}.

\section{Limitations and Future Work}
\label{sec:limitations}

Here, we summarize limitations of Pathway Tools, organized by use
case.  Some of these limitations are being addressed in current
research; many of the others will be addressed in future work.

{\noindent \bf Development of Organism Specific Databases.}
Pathway Tools has an emphasis on prokaryotic biology, although over
time we have added, and plan to add, more support for eukaryotic biology.
One remaining limitation is that although the software can capture many types of prokaryotic
regulation, we have not attempted comprehensive coverage of eukaryotic regulation.
Another limitation is that the editing tools within Pathway Tools are not web based, but require
installation of Pathway Tools on every computer that will be used for editing.

{\noindent \bf Visualization and Web Publishing of Organism Specific Databases.}  
In recent years we have made significant progress in making the
capabilities of the web and desktop modes of the Pathway Tools
Navigator as similar as possible.  For example, many cellular overview
capabilities previously present only in desktop mode are now also available
in web mode.  However, 
not all capabilities of Pathway Tools are available in
both the web and desktop modes.  For example, many comparative tools
function in web mode only, whereas all aspects of PathoLogic and the
editing tools are available in desktop mode only.

{\noindent \bf Analysis of Omics Data.}
Pathway Tools is not a general-purpose environment for analysis of
omics data.  Our assumption is that scientists will use one of the many
other software packages for the early stages of omics data analysis
(such as normalization), and will provide the output of those analyses to
Pathway Tools for display with the omics viewers.

{\noindent \bf Analysis of Biological Networks.}
We would like to see many additional network analysis tools present
within Pathway Tools, such for computing the scaling properties of
metabolic networks \cite{Jeong00}, and functional modules within metabolic
networks \cite{Ma04}.


\section{Summary}

Pathway Tools treats a genome as far more than a sequence and a set of
annotations.  Instead, it links the molecular parts list of the cell
both to the genome and to a carefully constructed web of functional
interactions.  The Pathway Tools ontology defines an extensive set of
object attributes and object relations that enables representing a rich
conceptualization of biology within a PGDB, along with enabling
querying and manipulation by the user.  Furthermore, a PGDB can be
transformed into a quantitative metabolic model for the organism.

Pathway Tools provides a broad range of functionality.  It can
manipulate genome data, metabolic networks, and regulatory networks.
For each data type, it provides query, visualization, editing, and
analysis functions.  It provides model-organism database development
capabilities, including computational inferences that support fast
generation of comprehensive databases, editors that enable
refinement of a PGDB, web publishing, and comparative analysis.  A
family of curated PGDBs has been developed using these tools for
important model organisms.

The software also provides visual tools for analysis of omics
data sets, and tools for the analysis of biological networks.

\section{Software Availability}
 
Pathway Tools runs on Macintosh, Windows, and Linux.  It is freely
available to academic and government researchers; a license fee
applies to commercial use.  See \url{http://BioCyc.org/download.shtml}.

\section*{Funding}

This work was supported by grants R01AI160719, 1R24GM150703, R01
LM013229, GM75742, GM080746,
and GM077678 from the National Institutes of Health; and by grant
NSF2109898 from the National Science Foundation.  The
contents of this article are solely the responsibility of the authors
and do not necessarily represent the official views of the National
Institutes of Health or the National Science Foundation.

\section*{Acknowledgments}

Pathway Tools has benefited from advice, input, and contributions from
many scientists during its lifetime.  We particularly wish to
recognize contributions from Ian Paulsen, Robert Gunsalus, Monica
Riley, John Ingraham, Jean-Francois Tomb, and Peifen Zhang.  
Lukas Mueller developed PerlCyc and
has provided many helpful suggestions.  Thomas Yan developed JavaCyc.
Tomer Altman developed RCyc.  Jeremy Zucker developed the first SBML
generation module, and contributed many other ideas.  Christos
Ouzounis was a co-developer of the original metabolic pathway
prediction algorithm, contributed an early version of the
import/export system, and has been a source of much sound advice.

\section*{Key Points}

\bitem
\item The Pathway Tools software is a comprehensive 
environment for creating model organism DBs that span
genome information, metabolic pathways, and regulatory networks.

\item Pathway Tools inference capabilities include 
prediction of metabolic pathways, prediction of metabolic pathway hole
fillers, inference of transport reactions from transporter functions,
and prediction of operons.

\item  Its metabolic modeling capabilities include flux balance
  analysis modeling for individual organisms and organism
  communities, with model gap-filling and the ability to model
  gene knockouts.

\item Pathway Tools provides interactive editing tools for use by
database curators.

\item Omics data analysis tools paint genome-scale data sets onto a
complete genome diagram, complete metabolic network diagram, and
complete regulatory network diagram.  Pathway Tools also computes
enrichment analysis, and provides an Omics Dashboard tool for visual,
hierarchical, interactive analysis of omics data sets.

\item Other tools include comparative analysis operations,
dead-end metabolite and blocked-reaction analysis of metabolic networks,
and metabolic route searching.
\eitem


\end{document}